\documentclass[a4paper,fleqn,usenatbib]{mnras}

\usepackage{newtxtext,newtxmath}
\usepackage{multicol}        

\usepackage[T1]{fontenc}
\usepackage{ae,aecompl}

\usepackage{graphicx}	
\usepackage{amsmath}	
\usepackage{amssymb}	

\title[Modern stellar spectroscopy caveats]{Modern stellar spectroscopy caveats}

\author[S. Blanco-Cuaresma]{
Sergi Blanco-Cuaresma$^{1}$\thanks{E-mail: sblancocuaresma@cfa.harvard.edu}
\\
$^{1}$Harvard-Smithsonian Center for Astrophysics, 60 Garden Street, Cambridge, MA 02138, USA\\
}

\date{Accepted XXX. Received YYY; in original form ZZZ}

\pubyear{2018}

\begin{document}
\label{firstpage}
\pagerange{\pageref{firstpage}--\pageref{lastpage}}
\maketitle

\begin{abstract}
Multiple codes are available to derive atmospheric parameters and individual chemical abundances from high-resolution spectra of AFGKM stars. Almost all spectroscopists have their own preferences regarding which code and method to use. But the intrinsic differences between codes and methods lead to complex systematics that depend on multiple variables such as the selected spectral regions and the radiative transfer code used. I expand iSpec, a popular open-source spectroscopic tool, to support the most well-known radiative transfer codes and assess their similarities and biases when using multiple set-ups based on the equivalent-width method and the synthetic spectral-fitting technique (interpolating from a pre-computed grid of spectra or synthesizing with interpolated model atmospheres). This work shows that systematic differences on atmospheric parameters and abundances between most of the codes can be reduced when using the same method and executing a careful spectral feature selection. However, it may not be possible to ignore the remaining differences, depending on the particular case and the required precision. Regarding methods, equivalent-width-based and spectrum-fitting analyses exhibit large differences that are caused by their intrinsic differences, which is significant given the popularity of these two methods. The results help to identify the key caveats of modern spectroscopy that all scientists should be aware of before trusting their own results or being tempted to combine atmospheric parameters and abundances from the literature.
\end{abstract}




\section{Introduction}

The automation of high-resolution stellar spectral analysis for AFGKM stars has become a necessity in recent years owing to the enormous increase of publicly available observations. Large surveys such as APOGEE \citep{2011AJ....142...72E, 2017AJ....154...94M} or the Gaia-ESO Public Spectroscopic Survey \citep[GES; ][]{2012Msngr.147...25G, 2013Msngr.154...47R}, complemented by smaller surveys like OCCASO \citep{2016MNRAS.458.3150C, 2017MNRAS.470.4363C} plus other independent studies and observational proposals, have contributed to this golden period of stellar spectroscopy.

Several research groups have developed codes to analyse all these data, and some have made their work openly available (e.g., SME \citealt{1996A&AS..118..595V}, GALA \citealt{2013ApJ...766...78M}, FAMA \citealt{2013ApJ...766...78M}, StePar \citealt{2013hsa7.conf..673T}, iSpec \citealt{2014A&A...569A.111B}, The Cannon \citealt{2015ApJ...808...16N}, ZASPE \citealt{2017MNRAS.467..971B}, FASMA \citealt{2018MNRAS.473.5066T}). Except those methods that use the whole spectrum (or a single continuous spectral region) for their analysis (e.g., MATISSE \citealt{2006MNRAS.370..141R}, FERRE \citealt{2006ApJ...636..804A}, ULySS \citealt{2009A&A...501.1269K}, Starfish \citealt{2015ApJ...812..128C}, sick \citealt{2016ApJS..223....8C}), most of them base the analysis on certain spectral features.

The derivation of atmospheric parameters and abundances from stellar spectra can generally be accomplished by following either of two possible strategies: the equivalent-width method or the synthetic spectral-fitting technique. The equivalent-width method first requires the measurement of the equivalent width of a selection of neutral and ionized iron absorption lines. This is generally achieved by fitting a Gaussian profile and then computing the width of the spectral continuum that has the same area as the absorption line. Next, a radiative transfer code is used to derive the individual line abundances for a given set of initial atmospheric parameters. The stellar parameters are found by flattening the abundance trends with respect to the reduced equivalent width, the lower excitation potential, and the ionization stage. In the case of the synthetic spectral-fitting technique, the observed spectrum is compared with theoretical spectra that are synthesized on-the-fly or interpolated from pre-computed grids (in both cases, a radiative transfer code is also necessary), and a minimization algorithm is executed. Frequently, a selection of spectral features is used instead of the full spectrum in order to reduce the computation time and to concentrate on the more informative spectral regions. The above-mentioned codes generally follow one of these two strategies.

In addition to the significant diversity of codes available, all of them can be set up with many different combinations of the necessary ingredients, such as the grid of model atmospheres (where interpolations are needed to derive the right model for the desired atmospheric parameters), the reference solar abundances (these can be scaled up or down following the desired metallicity, or certain elements can be enhanced/depleted to follow certain patterns such as the enhancement of alpha elements observed for metal-poor stars), the radiative transfer code, the atomic data (e.g. wavelengths, oscillator strengths, line-broadening parameters such as the radiative/Stark/van der Waals damping parameters), the selection of spectral features used (e.g. some regions may carry more information than others, or it may be that some codes and models are better at reproducing certain regions), or the continuum normalization procedure. These inhomogeneities have led to large discrepancies in the atmospheric parameters and abundances present in the literature \citep{2014AJ....148...54H}.

The source of this problem was first explored in \cite{2016ApJS..226....4H}, where four spectra were analysed using six different codes with a subsequent re-analysis with common atmospheric parameters and atomic line lists to determine chemical abundances. The study showed that homogenizing the atomic data and atmospheric parameters led to an improvement in the agreement between abundances derived by each method, although the dispersion remained high for several elements. The authors concluded that it is necessary to investigate further the inherent different results between spectroscopic techniques.

The same problem was tackled again even more thoroughly by \cite{2017A&A...601A..38J}, who determined four abundances for four Gaia Benchmark Stars \citep{2014A&A...564A.133J, 2015A&A...582A..81J, 2015A&A...582A..49H, 2016A&A...592A..70H} using six different methods and one representative line for each of the four elements with fixed atmospheric parameters (i.e. effective temperature, surface gravity and metallicity). The study showed that equivalent-width methods are less affected by shifted absorption lines than synthesis methods, which need to implement mechanisms to detect the shift and perform a correction. The agreement between methods improved when a common normalization was applied (see also the brief discussion about normalization effects in section 4.1 of \cite{2015A&A...577A..47B}) and when the same microturbulence was used. This parameter is very sensitive to the method used, however, which makes it impossible of have a good criterion to define a good common value for all the techniques. The neglect of hyperfine structure (i.e. shifts and splittings in the energy levels of atoms, molecules and ions owing to the interaction between the state of the nucleus and the state of the electron clouds), and different assumptions on the abundances of blending elements contribute to increasing the discrepancies between analysis. Different atmospheric model interpolation methods lead to very small differences, which are only relevant if very high-precision abundances are required. For equivalent-width methods, differences arise from the use of different radiative transfer codes. For synthetic methods, different line masks (i.e. the spectral region that includes the target absorption lines) do not seem to lead to any significant differences, but differences in broadening parameters do.

These studies have provided the first glimpses into the discrepancies currently found in the literature for spectroscopic analysis. However, they are limited to a very low number of spectra, and the execution involved several different research groups using their own codes with, sometimes, manual operations. Despite the excellent coordination, it is easy to make mistakes when the analysis is not fully automatic or when a completely homogeneous analysis cannot be guaranteed. In order to reveal the essential caveats of modern spectroscopy it is necessary to have fully automatic and reliable tests that analyse homogeneously a higher number of spectra covering a wider range of stars.

In this work, I have extended iSpec\footnote{The iSpec version used in this work was released as v2019.03.02} \citep{2014A&A...569A.111B} by: (1) including a large number of radiative transfers codes widely used for spectral synthesis and equivalent-width analysis \citep{2017hsa9.conf..334B, 2017ewas.confE...7B}; (2) adding spectral interpolation capabilities, which allows the user to use/compare a different spectroscopic approach (i.e. spectral interpolation instead of model atmosphere interpolation). iSpec has become a useful tool for spectroscopic analysis and also a very convenient framework in which to discover and assess the caveats of modern stellar spectroscopy. Using this tool, I designed several fully automatic experiments that compare the impact of using distinct radiative transfer codes, and different set-ups and spectroscopic techniques.

\section{Data}
\label{s:data}

The Gaia FGKM Benchmark Stars \citep{2014A&A...564A.133J, 2015A&A...582A..81J, 2015A&A...582A..49H, 2016A&A...592A..70H} constitute a set of very well-known stars covering a wide range in effective temperature (3\,500 to 6\,600~K), surface gravity (0.50 to 4.60~dex) and metallicity (-2.70 to 0.30~dex). They are especially convenient for spectral analysis assessments because they are accompanied by reference atmospheric parameters obtained from methods independent of spectroscopy.

For this work, I used the high-resolution spectra provided by the public library\footnote{\url{http://www.blancocuaresma.com/s/}} \citep{2014A&A...566A..98B}. The original non-normalized spectra came from different instruments with different resolutions, spectral ranges and signal-to-noise ratios (S/Ns). Some pre-processing was executed to homogenize the data set, including merging separate wavelength regions from the same observation, co-adding spectra to increase the S/N, cleaning areas affected by telluric lines (i.e. setting fluxes to zero), correcting radial velocities, estimating the S/N and error fluxes (a procedure that strongly influences the derived parameter errors), selecting the spectral range of 480 to 680 nm (optical range) and degrading the resolution to 47000, which matches the resolution and range of the UVES set-up used in the Gaia-ESO Survey.

\section{Pipeline}

\subsection{Methods}
\label{s:methods}

iSpec can derive atmospheric parameters using the synthetic spectral-fitting technique and the equivalent-width method. The former compares the observed fluxes (weighted by the flux errors if present) with synthetic spectra for a selected set of spectral features, and then a least-squares algorithm minimizes the differences (i.e., by computing the $\chi^2$) by varying the atmospheric parameters until convergence is reached. The spectral features can be absorption lines or any other spectral region. For instance, it is common to use the wings of the H-$\alpha$/$\beta$ and Mg triplet to help break degeneracies, given that these regions are highly sensitive to the effective temperature and surface gravity, respectively. The synthetic spectra can be computed on demand by interpolating from a grid of model atmospheres and using a radiative transfer code, or now also by interpolating from a grid of pre-computed spectra with iSpec or another tool (this also gives the possibility of using grids of synthetic spectra for stars cooler or hotter than AFGKM, or even grids of observed spectra). In both cases, the input grid is used to construct convex hulls, and a linear barycentric interpolation is executed at each necessary triangle. No specific code was written for this: I used widely tested methods present in the Qhull and SciPy packages \citep{Barber:1996:QAC:235815.235821, SciPy2001}.

In the case of the equivalent-width method (which also requires model atmosphere interpolations), the analysis starts with a selection of absorption lines produced by neutral and ionized iron, for which their equivalent width is measured. Usually this is done by fitting Gaussian profiles and determining the area of each absorption line. Then the equivalent width can be transformed to abundances by using a radiative transfer code, and the atmospheric parameters are varied until there is no correlation between abundances and equivalent widths, and excitation equilibrium plus ionization balance is reached (i.e. there is no correlation with excitation potential, and the average iron abundances from neutral and ionized lines are equal). In this case, no spectral features other than iron absorption lines are used, and the analysis is fast because the amount of information to be computed is small compared with that in synthesis methods (i.e. the full line profiles are not considered - only their area).

In this work I compare: (1) the synthetic spectral-fitting technique using a grid of atmospheric models; (2) the synthetic spectral-fitting technique using a grid of pre-computed synthetic spectra; (3) the equivalent-width method.

\subsection{Radiative transfer codes}
\label{s:radiative_transfer_codes}

For all the method described in Section~\ref{s:methods}, iSpec offers a broad variety of radiative transfer codes (a summary can be found in Table 1). It is worth noting that all of these radiative transfer codes assume local thermodynamic equilibrium (LTE), which means that the mean free path of photons is smaller than the scale over which thermodynamic quantities vary, and thus the atmospheric state (e.g. temperature) at a given depth is affected by radiation below or above that point. This approximation is not valid for OB stars or extremely metal-poor stars, which have optically thin layers where non-local, non-thermal influences (i.e. radiation) overcome local thermalizing ones (i.e. collisions). Nevertheless, the approximation is good enough for AFGKM stars, depending on the type of analysis and scientific goals.

Most of the codes use model atmospheres calculated assuming plane-parallel geometry, and hence the radiative transfer is solved while neglecting the curvature of the atmosphere and only considering one depth variable. This is a valid approximation for most stars but it breaks down when the size of the stellar atmosphere starts to be relevant compared with the stellar radius (e.g. cold giants and supergiants). Only some codes can consider the curvature when providing model atmospheres that were calculated assuming a spherical geometry.

All the codes were integrated in iSpec in the most homogeneous way possible; this implies that certain default behaviours were overriden. For instance, codes such as SYNTHE extract the abundances from the model atmosphere input file, while others, such as MOOG, have default hard-coded values that can be modified if the user issues the right commands. In addition, certain functions provided by these codes were not used, such as resolution degradation, macroturbulence and rotational effects. To guarantee comparable results, these effects are directly implemented in iSpec and they are homogeneously applied to all synthetic spectra independently of what radiative transfer code is used. Researchers that make use of these tools outside the iSpec framework should expect differences that were minimized for this work.

\begin{table*}
    \begin{center}
        \begin{tabular}{l|c|c|c|c|c}

    &   SPECTRUM    &   Turbospectrum   &   SME &   MOOG    &   WIDTH9/SYNTHE   \\
\hline
1D plane-parallel model atmosphere geometry                 &   \checkmark   &   \checkmark  &   \checkmark  &   \checkmark  &  \checkmark     \\
\hline
1D spherical model atmosphere geometry                      &                &   \checkmark  &   \checkmark  &               &                 \\
\hline
Non-LTE                                                     &                &               &               &               &                 \\
\hline
Grids of Non-LTE departure coefficient                      &                &               &   \checkmark  &               &                 \\
\hline
Customizable chemical abundances                            &   \checkmark   &   \checkmark  &   \checkmark  &   \checkmark  &  \checkmark     \\
\hline
Customizable isotopes                                       &   \checkmark   &   \checkmark  &               &   \checkmark  &                 \\
\hline
Customizable molecular dissociation constants               &   \checkmark   &   \checkmark  &               &   \checkmark  &                 \\
\hline
Re-computed model atmosphere electron density               &   \checkmark   &   \checkmark  &   \checkmark  &               &  \checkmark     \\
\hline
Continuum scattering                                        &                &   \checkmark  &   \checkmark  &               &                 \\
\hline
Radiative damping parameter due to natural broadening       &   \checkmark   &   \checkmark  &   \checkmark  &   \checkmark  &  \checkmark     \\
\hline
Stark broadening due to collisions with charged particles   &   \checkmark   &               &   \checkmark  &               &  \checkmark     \\
\hline
Classical van der Waals damping parameter                   &   \checkmark   &   \checkmark  &               &   \checkmark  &  \checkmark     \\
\hline
Anstee and O'Mara van der Waals broadening theory           &   \checkmark   &   \checkmark  &   \checkmark  &   \checkmark  &  \checkmark     \\
\hline
Hydrogen broadening                                         &   AG           &   BPO         &   BPO         &   BPO         &  AG             \\
\hline
Customizable hydrogen lines parameters                      &                &   \checkmark  &   \checkmark  &   \checkmark  &  \checkmark     \\
\hline
Base line profiles                                          &   Voigt        &   Voigt       &   Voigt       &   Voigt       &  Voigt          \\
\hline
Average synthesis time in seconds (480 - 680 nm)            &   $\sim$123     &   $\sim$56   &   $\sim$222   &   $\sim$68    &  $\sim$360       \\
        \end{tabular}
    \end{center}
    \caption{
        Summary of radiative transfer code features. 'Customizable' denotes the possibility of changing values without recompiling the program; 'BPO' stands for Barklem-Piskunov-O'Mara \citep{2000A&A...363.1091B}; and AG stands for Ali-Griem \citep{1965PhRv..140.1044A, 1966PhRv..144..366A}.
    }
    \label{tab:codes_summary}
\end{table*}

\subsubsection{SPECTRUM}

SPECTRUM version 2.76e\footnote{\href{http://www.appstate.edu/~grayro/spectrum/spectrum.html}{http://www.appstate.edu/~grayro/spectrum/spectrum.html}} \citep{1994AJ....107..742G} is a radiative transfer code written in C (compatible with the gcc compiler) that can synthesize spectra and derive abundances from equivalent widths. However, the latter functionality is done by fully synthesizing each absorption line and it is computationally expensive (i.e. significantly slower) compared with MOOG or WIDTH9, which use a faster direct computational analysis \citep[][chapter 16]{2008oasp.book.....G}. I did not use SPECTRUM for the tests based on the equivalent-width method, but it is an ideal code for the synthetic spectral-fitting technique because it is one of the fastest.

\subsubsection{Turbospectrum}

Turbospectrum version 15.1\footnote{\href{http://www.pages-perso-bertrand-plez.univ-montp2.fr/}{http://www.pages-perso-bertrand-plez.univ-montp2.fr/}} \citep[written in Fortran and compatible with the gfortran compiler][]{1998A&A...330.1109A, 2012ascl.soft05004P} is similar to SPECTRUM in terms of usage, and I again excluded it from tests based on the equivalent-width method. In contrast to SPECTRUM, which only works with plane-parallel model atmospheres, Turbospectrum can use spherical models (which offer a better approximation for giant stars), in which the stellar radius and the depth of each layer have to be provided (as shown in Table~\ref{tab:input_values_model_atmospheres}).

\subsubsection{SME}

SME version 4.23\footnote{\href{http://www.stsci.edu/~valenti/sme.html}{http://www.stsci.edu/~valenti/sme.html}} \citep{1996A&AS..118..595V} is the only radiative code considered that is closed-source, which makes its debugging and scientific assessment more difficult. It is distributed with IDL scripts that call a pre-compiled binary library that performs the spectral synthesis; iSpec only uses this library. SME only does synthesis and, equivalent to Turbospectrum, it can work with spherical model atmospheres. The code is ready to consider departure coefficients for non-LTE effects, although this has not been considered for this work.

\subsubsection{MOOG}

MOOG version February 2017\footnote{\href{http://www.as.utexas.edu/~chris/moog.html}{http://www.as.utexas.edu/~chris/moog.html}} \citep{2012ascl.soft02009S} is a radiative transfer code written in Fortran (compatible with the gfortran compiler) that can synthesize spectra and derive abundances from equivalent widths efficiently. Unfortunately, MOOG depends on the non-free SM package (formerly SuperMongo) for plotting results. Given that iSpec already has its own free python interface, I developed a SM package mock with the same functions but empty implementation that allows MOOG to be compiled without the official non-free SM package. This is the only code that does not recompute electron densities, but it keeps them fixed them as provided by the input model atmosphere.

\subsubsection{WIDTH9/SYNTHE}

WIDTH9 version 9 March 1993 and SYNTHE version 20 July 2001\footnote{\href{http://atmos.obspm.fr/}{http://atmos.obspm.fr/}} \citep{1993KurCD..18.....K, 2004MSAIS...5...93S} are programs that share the same radiative transfer code, but the former is used to transform equivalent widths into abundances, while the latter computes synthetic spectra. Both are written in Fortran and require the Intel compiler, which is not open-source (in contrast to gfortran), and thus pre-compiled executables are included in iSpec by default. More details about this code can be found in \cite{2002A&A...387..595C}.

\subsection{Model atmosphere}
\label{s:model_atmosphere}

For the grid of model atmosphere grid I used MARCS\footnote{\url{http://marcs.astro.uu.se/}} \citep{2008A&A...486..951G}, which was computed with solar abundances from \cite{2007SSRv..130..105G}. Note that iSpec also support ATLAS/Kurucz and many other solar abundances, but MARCS includes models computed with plane-parallel and spherical geometries. The latter allow for spherical radiative transfer (although only with codes that support it, as explained in Section~\ref{s:radiative_transfer_codes}) and ensure a more realistic temperature structure of the model atmospheres because the spherically symmetric radiative transfer scheme takes into account the geometric dilution of flux. The spherical radiative transfer is generally not important for line formation, but is very important for the model atmosphere structures \citep{2006A&A...452.1039H}. iSpec uses the model atmosphere grid to construct convex hulls, where linear barycentric interpolations at each necessary triangle can be executed to generate the required model with the necessary atmospheric parameters (always within the grid ranges). Regarding the radiative transfer codes, not all of them require the same model atmosphere input values: the differences are shown in Table~\ref{tab:input_values_model_atmospheres}.

\subsection{Atomic data}
\label{s:atomic_data}

The atomic data used in this work corresponds to the line list version~5 from the Gaia-ESO Survey line list \citep{2015PhyS...90e4010H}. The line list format is automatically transformed by iSpec to fit the requirements from every radiative transfer code. Furthermore, some lines are not used for certain codes if they are not compatible or necessary. For instance, SPECTRUM has several hard-coded strong absorption lines (which should not be included in the atomic line list or SPECTRUM would generate strong lines twice): hydrogen-line series (Lyman, Balmer, Paschen, Brackett, Pfund and Humphreys), 31 helium I lines, 11 iron II lines, 1 magnesium I line and 2 magnesium I lines, 1 calcium I line and 2 calcium II lines, 1 scandium II line, and 1 strontium II line. Turbospectrum and MOOG also include their own data for hydrogen and helium lines. All these included lines will not be taken from the GES line list when using these codes.

In terms of isotopes, the SPECTRUM documentation indicates that it supports 311 atomic isotopes plus 40 molecular isotopes, and their relative abundances can be fine-tuned using an input file. The rest of the codes do not seem to offer this possibility (they have hard-coded values) and are less well documented, which makes homogenization and comparison difficult. Apart from filtering out isotopes not supported by SPECTRUM, no other atomic data selection has been performed based on isotopes.

Regarding molecules, not all the codes support the same molecules, and SPECTRUM is again the best-documented code, while Turbospectrum seems to be the code that supports the most molecules. Furthermore, every code includes its own dissociation energies for molecules, and only SPECTRUM (via the input solar abundance), MOOG (via the input line list) and Turbospectrum (via specific molecule input files) allow the user to override them without modifying the source code. In any case, the public version of the GES line list does not include molecules (which are relevant only for the coolest Benchmark Stars).

In all the cases, I discarded atomic lines for second or higher ionized atoms (e.g. Fe III), lines with a lower state excitation potential higher than 15 eV (corresponding to only 20 lines) and auto-ionizing transitions for metals (corresponding to only 17 lines) to reduce computation time because their contribution is small for FGKM stars and furthermore they are not supported by all the codes.

Regardless of starting with a common atomic line list, there are differences that arise owing to the intrinsic functioning of each code. Moreover, not all codes use the same input values, as shown in Table~\ref{tab:input_values_atomic_linelist}. Sometimes the differences are just a matter of units or format, but in other cases there are values that are not required at all by some codes.

\subsection{Line selection}
\label{s:line_selection}

No matter whether the analysis method is based on the equivalent-width technique or on the synthetic spectral-fitting technique, the spectral ranges used in the study are going to have an impact on the final derived atmospheric parameters. Given the nature of the methods used in this work, most spectral ranges will correspond to absorption lines because they carry key information related to the atmospheric parameters of the star.

It is common to find studies in the literature in which authors use a line selection carried out by other authors. But this approach carries risks. A line might be reliable when using a specific spectroscopic pipeline with a concrete set-up (e.g. normalization process, atmospheric models, radiative transfer codes, atomic data) and observed spectra with a particular resolution, but very bad when any of these components change. For instance, a line selection carried out with high-resolution spectra might not be convenient for lower resolutions because lines can be blended: equivalent-width methods will overestimate the abundance, and synthetic spectral-fitting techniques might produce inaccurate results if the nearby lines have poor-quality atomic data.

A strategy that minimizes some of these difficulties is to follow a purely line-by-line differential approach. For instance, this could be done by calibrating the absorption lines' $\log(gf)$ value to better reproduce each line profile in a reference star with very well-known atmospheric parameters (typically, the Sun) and then using this calibrated atomic data to derive atmospheric parameters. In the case of equivalent widths, a different but equivalent approach is to measure the abundance of all the lines in a reference star, subtract the result for all the lines measured in the target star, and use these differential abundances (instead of absolute abundances) to reach ionization balance and excitation equilibrium.

Nevertheless, given the goal of this study, I preferred to avoid the calibration of $\log(gf)$ values and to use the same atomic data for all the different radiative transfer codes. This avoids introducing another degree of freedom that might make the comparison more difficult, although it makes the line selection process particularly important.

I used the NARVAL solar spectrum with the highest signal-to-noise ratio in the Gaia Benchmark Stars library for the line selection process described in the following subsections. The spectrum was convolved to a resolution of 47\,000, corrected from its radial velocity and normalized following the same procedure as any other spectra in this study.

\subsubsection{Matching absorption lines to atomic data}

The first required step in the line selection process is to identify which lines from the GES atomic line list are the main contributors to the observed absorption lines in the solar spectrum. For this, first I only considered lines from the GES line list that have a theoretical depth greater than 0.01 and a reduced equivalent width\footnote{$\log_{10}\left(\frac{EW}{\lambda}\right)$ where EW is the equivalent width and $\lambda$ is the wavelength position} greater than -7 for solar atmospheric parameters. Then, I used iSpec to fit Gaussian profiles for all the lines in this subset and discarded those whose fit failed or that had a depth greater than 1 or lower than 0.01. There may be more than one close atomic line (closer than 0.001 nm) blended in a single unique observed absorption line. In these cases I discarded all the atomic lines except the one with the greatest theoretical equivalent width (i.e. the one that has the highest probability of being the main contributor to the observed line). From this process, a total of 2\,496 atomic lines were selected.

\subsubsection{Deriving solar abundances}
\label{s:deriving_abundances}

The second step is to determine abundances with all the codes and methods for each of the selected lines in the solar spectrum by fixing the following reference solar parameters (based on the Gaia Benchmark Stars recommended values):

\begin{itemize}
    \item effective temperature (T$_{\text{eff}}$): 5771~K,
    \item surface gravity ($\log(g)$): 4.44~dex,
    \item metallicity ([M/H]): 0.00~dex,
    \item microturbulence velocity (V$_{\text{mic}}$): 1.07 km/s,
    \item macroturbulence velocity (V$_{\text{mac}}$): 4.21 km/s,
    \item projected rotational velocity  ($v\sin(i)$): 1.60 km/s,
    \item limb-darkening coefficient: 0.6.
\end{itemize}

I used solar abundances from \cite{2007SSRv..130..105G} to be consistent with the MARCS model atmosphere. I implemented the macroturbulence broadening using the radial-tangential formalism as described in \cite{2014dapb.book.....N} (adapted from SME), and applied the projected rotational velocity plus the limb-darkening coefficient following equation (17.12) from \cite{2008oasp.book.....G} (adapted from SYNSPEC, \cite{2011ascl.soft09022H}). All these effects are directly implemented into iSpec and are applied independently of the selected radiative transfer code.

\subsubsection{Equivalent width}
\label{s:ew_abundances}

Using MOOG and WIDTH9 radiative transfer codes, I determined the abundances for all the selected lines using the equivalent width (EW) derived from the previously fitted Gaussian profiles. In addition, in order to assess the quality of each line I also derived abundances when the metallicity is artificially increased by 0.10~dex (in order to assess the impact of errors in the metallicity) and abundances when the EW is drawn from a random distribution using the fitted EW as the mean and its error (computed following \cite{2006AN....327..862V}) as the sigma (in order to test the signal-to-noise ratio influence).

\subsubsection{Synthetic spectral-fitting technique when interpolating model atmospheres}
\label{s:ssf_abundances}

In this method, the codes SPECTRUM, Turbospectrum, SME, MOOG and SYNTHE are used. For each line, I first computed a small synthetic spectrum that includes the target line and used it to adjust the line mask (the spectral region used by the minimization algorithm). This way, if nearby lines are present in the synthetic spectra but not in the observed one, their impact can be reduced by excluding them from the mask. I cross-correlated the same synthetic spectrum with the observed one to detect and correct small line shifts. I then determined the abundances by letting only the corresponding element be a free parameter. In order to be able to later assess the quality of each line (i.e. filtering lines with differences larger than certain limits), I also derived abundances for that line when the metallicity is artificially increased by 0.10~dex, when a new realization of the spectrum is created (fluxes are drawn from a Poisson distribution using the fluxes as mean values and errors as the sigma) and when the atomic line list contains only the target atomic data and no other blended lines.

\subsubsection{Synthetic spectral-fitting technique when interpolating pre-computed spectra}
\label{s:ssf_grid_abundances}

In the previous two methods, iSpec interpolated model atmospheres using the MARCS grid and provided the model to the corresponding radiative transfer code. In this method, the atmosphere is not interpolated, but a grid of synthetic spectra was pre-computed using SPECTRUM and matching the exact atmospheric parameters that the MARCS grid provides but with two different alpha abundance variations ($[\alpha/\textrm{Fe}]=\pm0.40$~dex, where the alpha elements correspond to neon, magnesium, silicon, sulphur, argon, calcium and titanium) and four microturbulences (0.00, 1.00, 2.00 and 4.00 km s-1 ). Thus, the dimensions of the grid are effective temperature, surface gravity, metallicity, alpha enhancement and microturbulence. The grid is computed with a very high resolution (R $>$ 300\,000). This allows iSpec to interpolate a very high-resolution spectrum that can then be degraded to the target resolution, and effects such as macroturbulence, rotation and limb-darkening can be applied.

For each line, I performed the same radial velocity correction as detailed in the previous section and I derived the abundance for each line. In this case, I let the metallicity parameter be free because the grid does not have a dimension for every possible chemical element. This strategy is also followed by other authors and surveys (e.g. APOGEE) when using pre-computed grids. To be able to assess the quality of the line, an abundance is also derived for a new realization of the spectrum (as explained in the previous section).

\subsubsection{Selecting lines}
\label{s:selecting_lines}

I created two line selections for each code: one optimized to be used for determining atmospheric parameters, and a second less strict one that can be used for a line-by-line determination of individual chemical abundances. To make these line selections, I defined the following simple criterion to evaluate if an absorption line should be selected: a line can be considered good when I am able to derive an accurate solar abundance. In practice, this means that the derived abundance with respect to the solar abundance of reference (understood as $[X/H]$) should be close to zero within a certain margin.

After several tests, I found that the optimal margin for the determination of atmospheric parameters with equivalent-width methods is $\pm$0.10~dex (a sufficient number of neutral and ionized iron lines need to pass this filter), while for the synthetic spectral-fitting technique the margin can be $\pm$0.05~dex. For the former, only iron lines are considered, while for the latter, I considered only lines not affected by hyperfine structure splitting and that belong to iron peak elements (iron, chromium, nickel) and alpha elements (silicon, calcium, titanium).

It is worth noting that these selected lines will not be blindly used with all the target spectra. Before executing the determination of atmospheric parameters, I fitted Gaussian profiles to all the selected lines using the target spectrum and I discarded those that do not contain valid fluxes, are affected by telluric lines, have a bad line mask, have a Gaussian profile fit that failed or too large a root mean square (i.e. rms error >= 1.00), have a reduced equivalent width greater than -4.2 or lower than -6 (to avoid saturated or too weak lines), or have an excitation potential too extreme (e.g. greater than 6 eV, where there are almost no lines and an outlier can deeply affect trend computations). Hence, the line selection will be further fine-tuned and adapted to each target spectrum to be analysed.

Once the atmospheric parameters of a star are found, the determination of individual chemical abundances can be calculated differentially line-by-line, and it is not necessary to be so strict with the selection criteria, especially if we want to include elements that have only a few difficult lines. Thus, if an element has more than 10 lines, the strict margin is applied ($\pm$0.10 and $\pm$0.05~dex for equivalent-width and synthesis, respectively) but if not, a more generous margin of $\pm$0.50~dex is enforced (a value also determined from experimental tests). With this strategy, we can maximize the number of elements for which we can derive abundances without affecting the quality of the elements that already have many lines.

Some more quality controls are applied to the line selection for atmospheric parameters and the line selection for individual chemical abundances. Absorption lines close to the known strong lines H-$\alpha$ (652-660 nm), H-$\beta$ (483.5-489.5 nm) and the Mg triplet (514-521 nm) are discarded. Weak lines are impacted to a greater extent by continuum placement: to reduce this effect I discarded absorption lines with depths lower than 0.05. The determination of line shifts can be inaccurate for very weak and noisy lines or for extremely blended lines. Based on some manual tests, the best indicator to identify these cases is the error on the radial velocity that comes from the cross-correlation process, which should be lower than 100 km s-1 (from visual inspection, errors larger than this are correlated with problematic lines, while slightly lower errors can be caused by problematic lines or by overestimated errors). Abundance errors are also good indicators for identifying good fits. I discarded any line with an error greater than 0.25~dex. Abundances derived from new realizations of the equivalent-width or the spectrum fluxes should not be more than 0.10~dex away from the main derived abundance; otherwise, the line is too sensitive to the noise. Finally, in the case of the synthetic spectral-fitting technique using atmospheric model interpolation, I discarded lines with abundances more different than 0.10~dex when the metallicity was artificially set 0.10~dex higher, and lines for which the abundance was 0.50~dex different when using atomic line lists without blended lines.

\subsection{Atmospheric parameters and abundances}
\label{s:ap_determination}

The determination of parameters can take place in one or two full iterations, depending on how we want to normalize the spectra (each full iteration includes a normalization and the determination of parameters executed by the minimization algorithm, which goes through multiple iterations exploring the parameter space until convergence). The first full iteration will normalize the spectra (before deriving any parameter) by applying a median and a maximum filter with different window steps (0.05 and 1.0 nm, respectively) and fitting the continuum with a B-spline of 2 degrees every 5 nm (ignoring strong lines automatically detected by iSpec). It is worth noting that the same normalization process was used in the line selection described in Section~\ref{s:line_selection}. The second full iteration, if enabled, will synthesize a spectrum (i.e. template) with the atmospheric parameters found in the first full iteration. The observed spectrum is divided by the template, and I apply median and Gaussian filters with different window steps (0.05 and 10 nm, respectively) to find the continuum. The advantage of using a synthetic spectrum as a template is that areas with strong lines and many blended lines (e.g. the blue part of the visual range for cooler stars tends to be very crowded and blended) will be better normalized, and differences between spectra from the same star but with different noise levels will be normalized more similarly; hence the determination of parameters in that second full iteration may improve with respect to the first full iteration. The risks are that the first full iteration may have led to inaccurate parameters (i.e. the template will be synthesized with bad parameters), and that lines existing in the synthetic spectrum but not in the observed one can create normalization artefacts.

To accelerate the convergence process during the determination of atmospheric parameters, it is optimal to start with initial parameters as close as possible to the type of star we are analysing. I pre-computed with each code a very limited grid of synthetic spectra that covers only four temperatures (3500, 4500, 5500 and 6500~K), two surface gravities (1.5 and 4.5~dex) and three metallicities (-2.0, -1.0 and 0.0~dex). The normalized observed spectrum is compared with all the spectra in the grid, and the parameters of the one with the lowest $\chi^2$ are selected as initial values. This process allows me to quickly distinguish between metal-poor/rich dwarfs and giants and start the minimization algorithm with values closer to the final solution, thereby speeding up the convergence.

For the equivalent-width method, iSpec lets the effective temperature, surface gravity and/or microturbulence velocity be set as free parameters, and a maximum of 20 iterations are allowed. The synthetic spectral-fitting technique uses the effective temperature, surface gravity, metallicity, alpha enhancement, microturbulence velocity and resolution as free parameters, with a maximum of six iterations (several tests showed that these are reasonable maximums to obtain accurate results in an optimal computation time; see \citealt{2014A&A...569A.111B}). The resolution, macroturbulence and projected rotational velocity are degenerate parameters, which are very difficult to disentangle by relying only on spectroscopy. After several tests, the most accurate results were obtained by fixing the rotation to 1.6 kms-1 and letting the macroturbulence follow an empirical relation established by GES (although setting this parameter to zero leads to similar results). This empirical relation was built by GES considering the effective temperatures, surface gravities and metallicities from their data set.

The determination of individual chemical abundances follows the same structure as described in Section~\ref{s:deriving_abundances}, where individual line shifts are detected by cross-correlating the spectrum region of the target line with a synthetic template, and then abundances are derived using the corresponding method and some additional controls are executed as explained in Sections~\ref{s:ew_abundances}, \ref{s:ssf_abundances} and \ref{s:ssf_grid_abundances}. I discarded lines for which it was not possible to derive the main abundance or any of the quality-control abundances, and lines with extreme abundances that fall outside the metallicity range considered in the atmospheric models grid (i.e. abundances with respect to the Sun greater/lower than 1.0/-5.0~dex). From the remaining data set, I discarded lines that do not pass the quality controls following the same criteria as explained in Section~\ref{s:selecting_lines}, except for elements that have only one line (I relaxed the criteria to maximize the number of measured elements).

To partially compensate for modelling errors, it is useful to perform a differential abundance analysis. For certain studies it could be convenient to use more than one reference star, which are at different evolutionary stages \citep{2018A&A...618A..65B, 2016sf2a.conf..333B}, but for this work I used the Sun as the only reference and included seven solar spectra from the Gaia Benchmark Stars library, which were analysed using the same process as for the rest of the spectra (i.e. pre-processing, normalization, determination of atmospheric parameters and abundances). For most of the selected lines, I obtained seven different measurements (one per spectrum) and I used them to compute an averaged abundance and a dispersion (to be used as the error). To ensure good reference values, I filtered lines that were not measured in more than three solar spectra (some lines can fail because of quality issues in the observed spectrum or missing fluxes). For the rest of stars, the final differential abundances were derived by subtracting the reference abundance line by line, while errors were added quadratically.

\subsection{The non-observed data set experiment}
\label{s:non_observed_dataset_experiment}

As described in Section~\ref{s:data}, the main analysis in this work uses the high-resolution spectra from the Gaia FGKM Benchmark Star public library. Observed data may be affected by many different variables that depend on the instrument used, the night conditions, the treatment of the raw data, etc. In addition, there is no model that can perfectly reproduce all the physical processes that take place in a star (e.g. theoretical assumptions are made to make the problem tractable with our current computer resources and time constraints). To remove the possibility that any of these variables is playing a role in the main analysis done in this work, I created a purely theoretical data set by synthesizing spectra using the Gaia FGKM Benchmark Star reference values, the signal-to-noise ratios from the public library, and all the synthesis codes used in this study plus interpolation from a pre-computed grid method. This produced a data set of 672 normalized synthetic spectra, which was analysed following the same procedure as for the observed data set.

\subsection{The one variable at a time experiment}
\label{s:one_variable_at_a_time_experiment}

To understand what each code does and how they differ in detail, there are mainly two broad strategies: (1) read the documentation, ask the author(s) and invest a large amount of time interpreting the thousands of lines present in each source code (if available); (2) design an experiment in which one output variable is measured while all the input variables remain constant except one. These are not exclusive strategies, and following both of them would be instructive, but given the complexity of the codes (and that fact that they are written in different programming languages, and one of them is not public) and the limited amount of resources, I mainly followed the second one.

In the case of the equivalent-width method, I considered the lines in common in the two codes (MOOG EW and WIDTH) plus the equivalent widths measured in the solar spectrum used in Section~\ref{s:line_selection}, and I measured the median abundance. In the case of the spectral-fitting technique, I did not use any observed data and I measured the synthetic flux depth around 556.45 nm (a region with practically no blend, and thus close to the continuum) and the flux depth at the line peaks for the common selection. In both cases, I used the solar parameters detailed in Section~\ref{s:deriving_abundances}, all the default values for the atomic line list and atmospheric model, and the results from MOOG (EW) and SPECTRUM, respectively, as reference points.

In the experiment, I tracked the changes to the reference equivalent-width abundance, continuum and absorption line peak depths while (1) changing (one at a time) the effective temperature, surface gravity, metallicity, alpha enhancement, microturbulence, number of layers in the atmospheric model; (2) multiplying by a factor (between 0.25 and 1.75) the values of column mass (rhox), temperature, gas pressure (pgass), electron density (xne), Rosseland mean absorption coefficient (abross), radiation pressure (accrad), microturbulence velocity (vturb), optical depth (logtau5) and electro pressure (pelectron) on each atmospheric model layer; (3) multiplying by a factor (between 0.25 and 1.75) the values of the oscillator strength (loggf), radiative damping parameter (rad), Stark damping parameter (stark), van der Waals damping parameter (waals). The factor multiplication was applied as a logarithmic addition for parameters expressed in logarithmic terms such as logtau5, loggf, rad, stark and waals.

\section{Results}
\label{s:results}

\subsection{Line selection}
\label{s:line_selection_results}

\begin{figure}
 \includegraphics[width=\columnwidth]{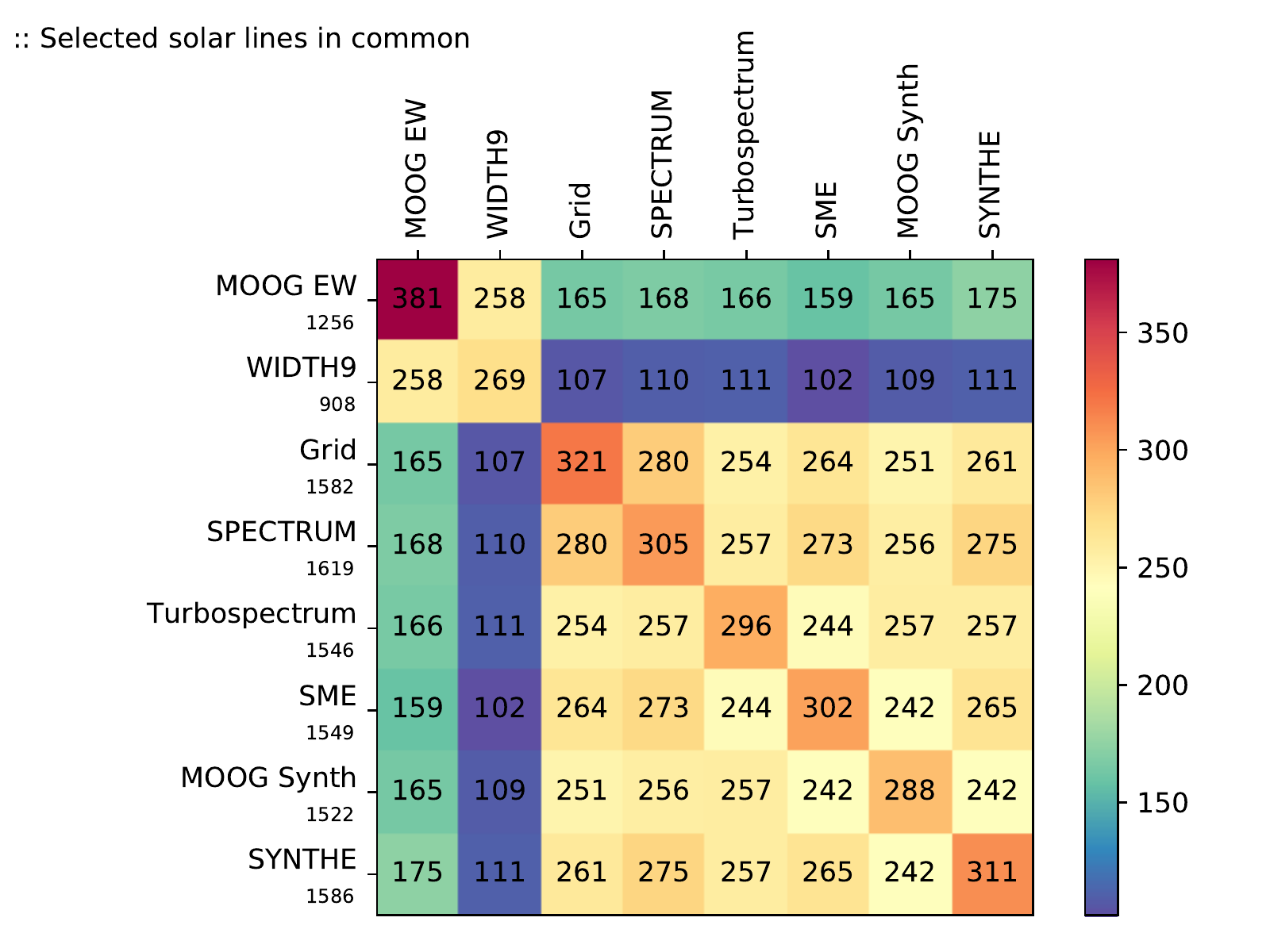}
 \caption{Absorption lines in common for which a $\pm0.05$~dex abundance ($\pm0.10$ for MOOG EW and WIDTH9) was derived with a particular code when analysing the NARVAL solar spectrum with the highest signal-to-noise ratio. The number below each code name correspond to the sum of all the values in the row minus the lines that correspond to the same code. The label Grid corresponds to the results obtained when interpolating from a pre-computed grid of synthetic spectra.}
 \label{fig:accuracy_solar_lines}
\end{figure}

The total numbers of selected lines for each code (as described in Section~\ref{s:line_selection}) are shown in the diagonal of the table drawn in Fig.~\ref{fig:accuracy_solar_lines}. The rest of the table shows the number of lines in common between each pair of codes. MOOG EW (i.e. MOOG using equivalent width) is the code with which the greatest number of accurate solar abundances were obtained, followed by Grid (synthetic spectral-fitting using a grid of pre-computed spectra with SPECTRUM) and SYNTHE. Recall that lines from the equivalent-width codes were selected using a looser criterion than the rest (as described in Section~\ref{s:selecting_lines}) and these numbers are not directly comparable.

\begin{figure*}
 \includegraphics[width=\linewidth]{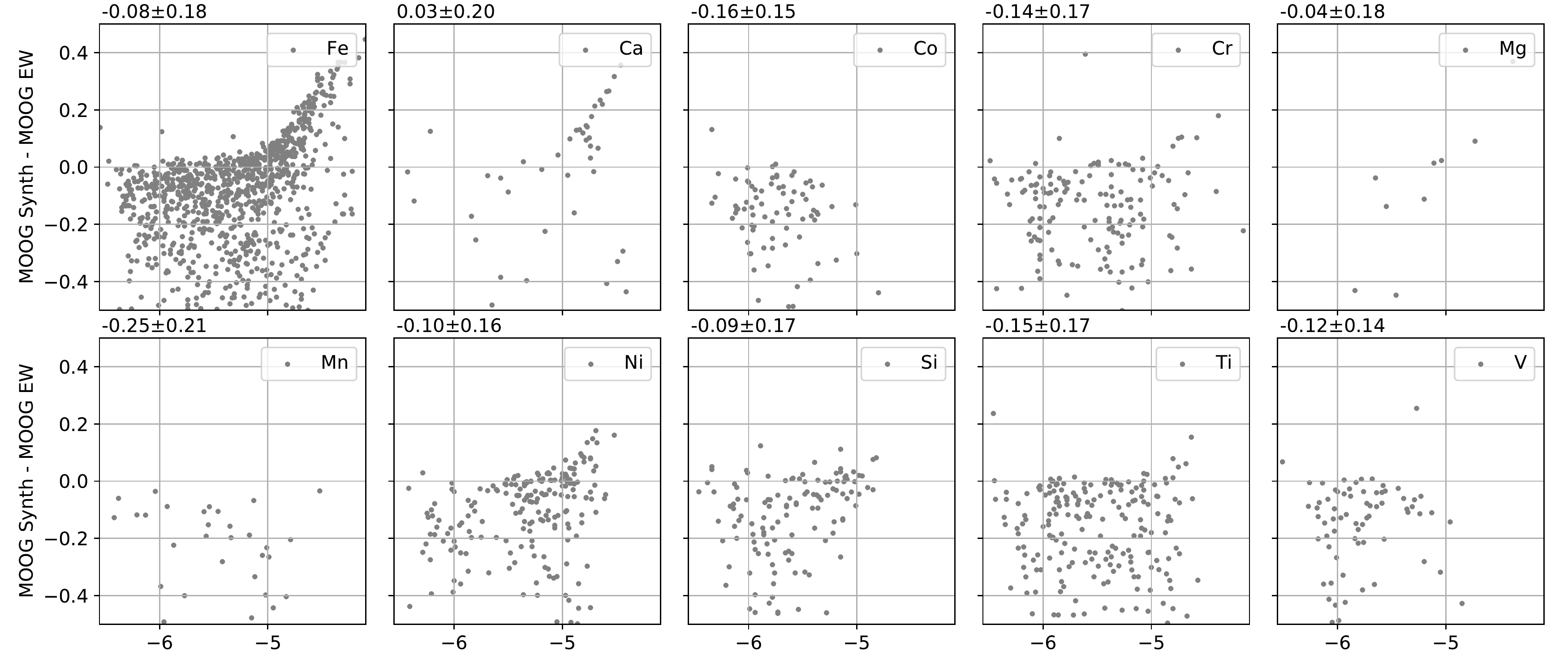}
 \includegraphics[width=\linewidth]{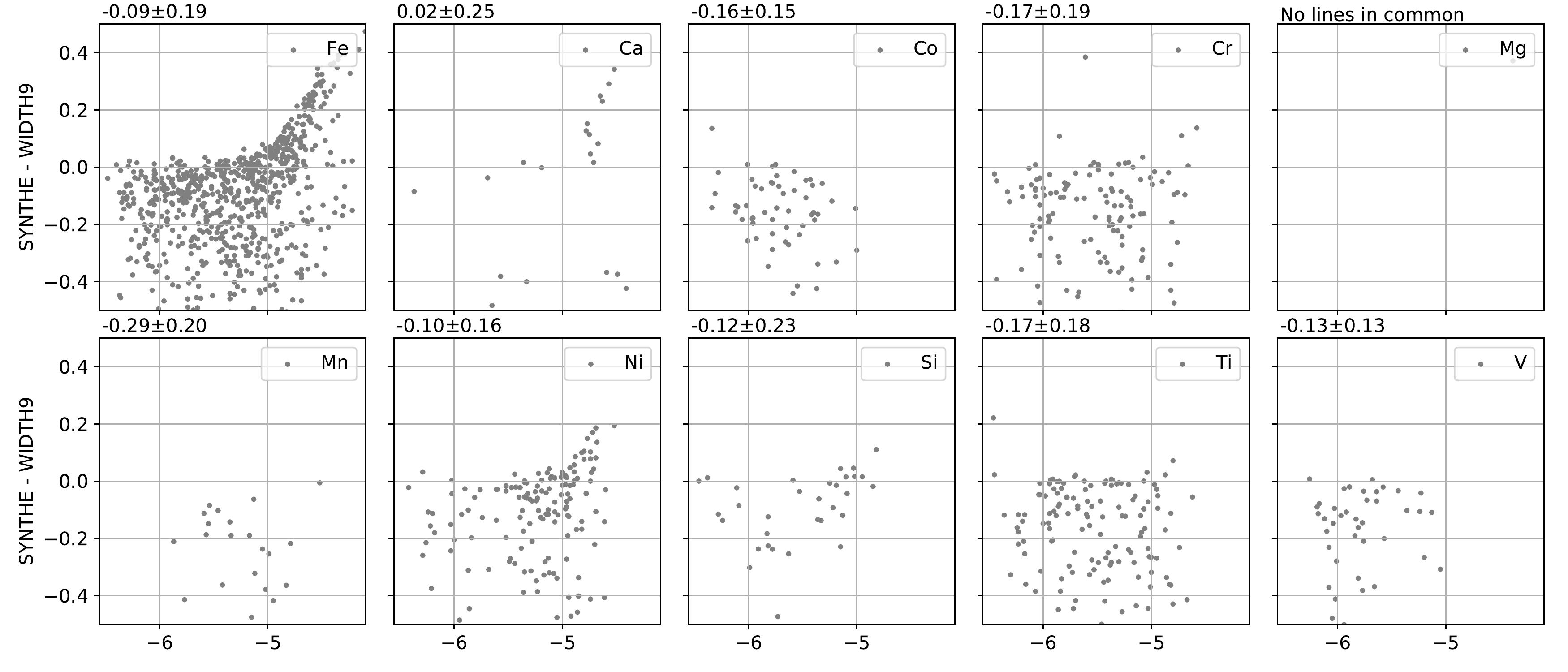}
 \caption{Solar abundance difference between equivalent-width and synthesis codes as a function of reduced equivalent width for various elements. The median and absolute median deviation are indicated in the upper left of each subplot.}
 \label{fig:precision_solar_lines_ew_vs_synth}
\end{figure*}

Merely by looking at the colour-coding of the results we can see two separate islands: equivalent-width codes (MOOG EW and WIDTH9) and synthesis codes (Grid, SPECTRUM, Turbospectrum, SME, MOOG Synth AND SYNTHE). These two groups have a greater number of lines in common within themselves but not so many across each other. These differences can be intrinsic to how these methods work: one only considers the area of an absorption line, while the other takes into account the full shape of the line profile including blends (see also section~5.1 in \citealt{2017MNRAS.470.4363C}). When comparing line-by-line abundances for each element from the two methods, it can be seen that the equivalent-width method provides larger abundances than synthesis, because the latter can reproduce and account for blends (if the synthesis is forced to ignore blends, the agreement with equivalent-width results increases, as shown in section~5.1 in \citealt{2017MNRAS.470.4363C}). At the same time, saturated lines (those with a greater reduced equivalent width) depart from a Gaussian profile (used to determine the equivalent width), and the equivalent-width method derives smaller abundances. The example shown in Fig.~\ref{fig:precision_solar_lines_ew_vs_synth} compares codes that use the same radiative transfer core code; thus, these differences do not arise from major differences in their implementation but from the intrinsic differences between the two methods.

In a previous discarded analysis (not included in this work), when interpolating from a grid of spectra that was computed with only two microturbulences (0.00 and 4.00 km s$^{-1}$ ), a lower number of lines in common with pure synthesis was found. Hence, increasing the number of data points to cover four microturbulences (0.00, 1.00, 2.00, and 4.00 km s$^{-1}$ ) led to a higher agreement between Grid and the rest of the synthesis codes.

\begin{figure*}
 \includegraphics[width=\linewidth]{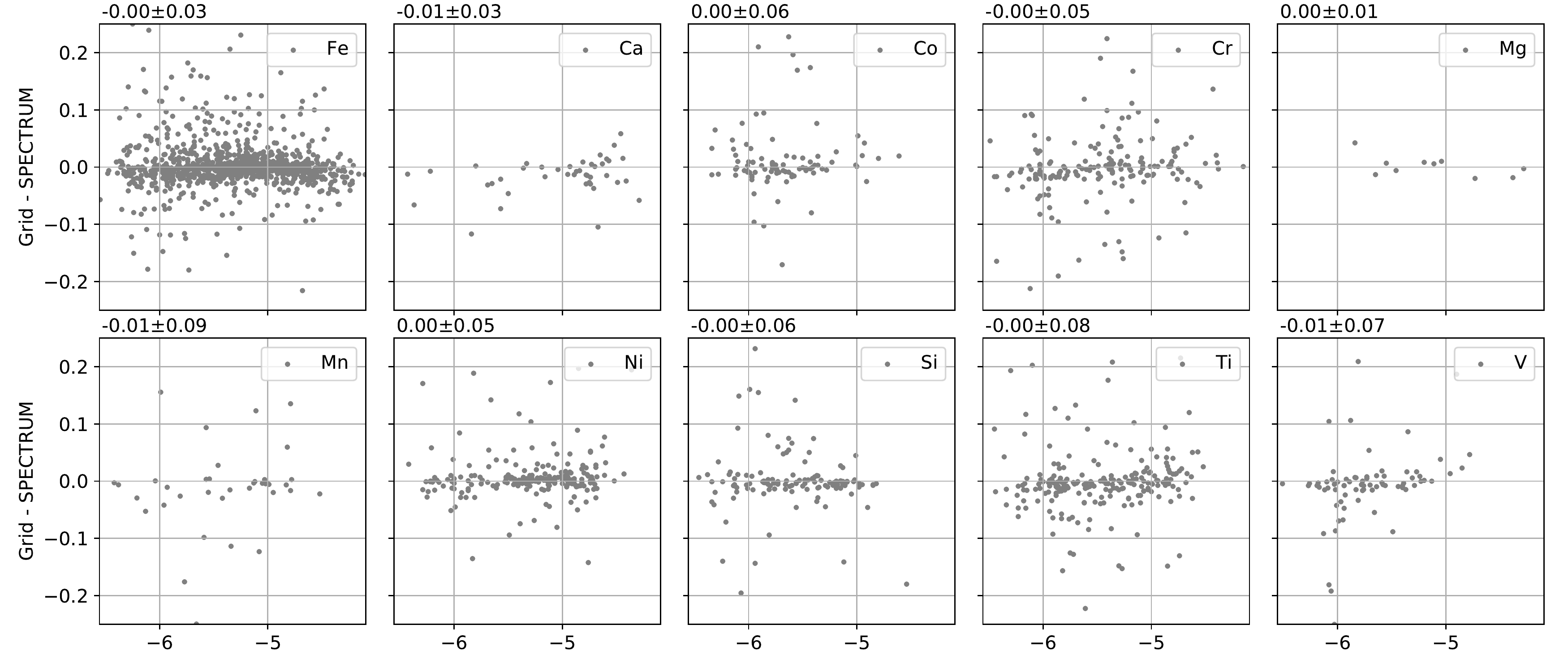}
 \caption{Solar abundance difference between synthesis and interpolation from pre-computed grid of spectra (i.e., Grid) as a function of reduced equivalent width for different elements. Median and absolute median deviation are indicated on the upper left of each subplot.}
 \label{fig:precision_solar_lines_grid_vs_synth}
\end{figure*}

\begin{figure*}
 \includegraphics[width=\linewidth]{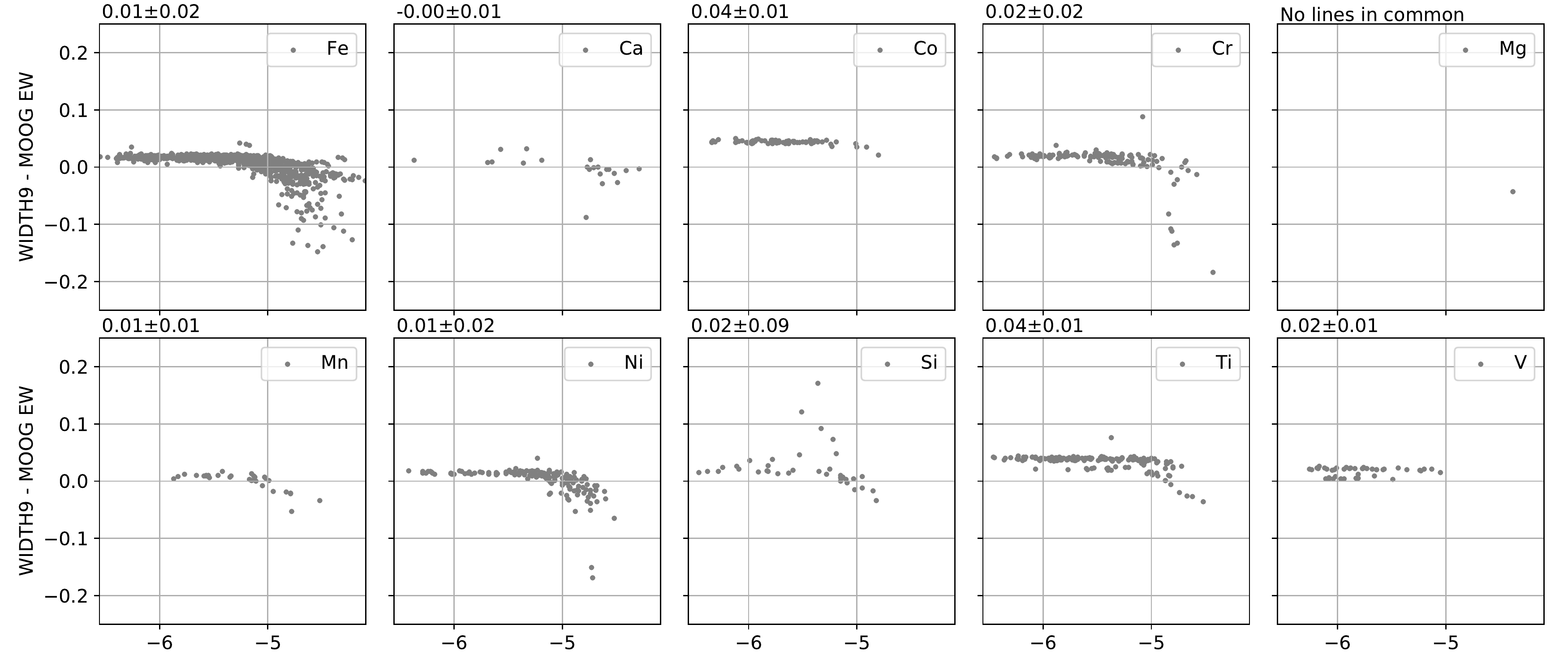}
 \caption{Solar abundance difference between equivalent width codes as a function of reduced equivalent width for different elements. Median and absolute median deviation are indicated on the upper left of each subplot.}
 \label{fig:precision_solar_lines_ew}
\end{figure*}

\begin{figure*}
 \includegraphics[width=\linewidth]{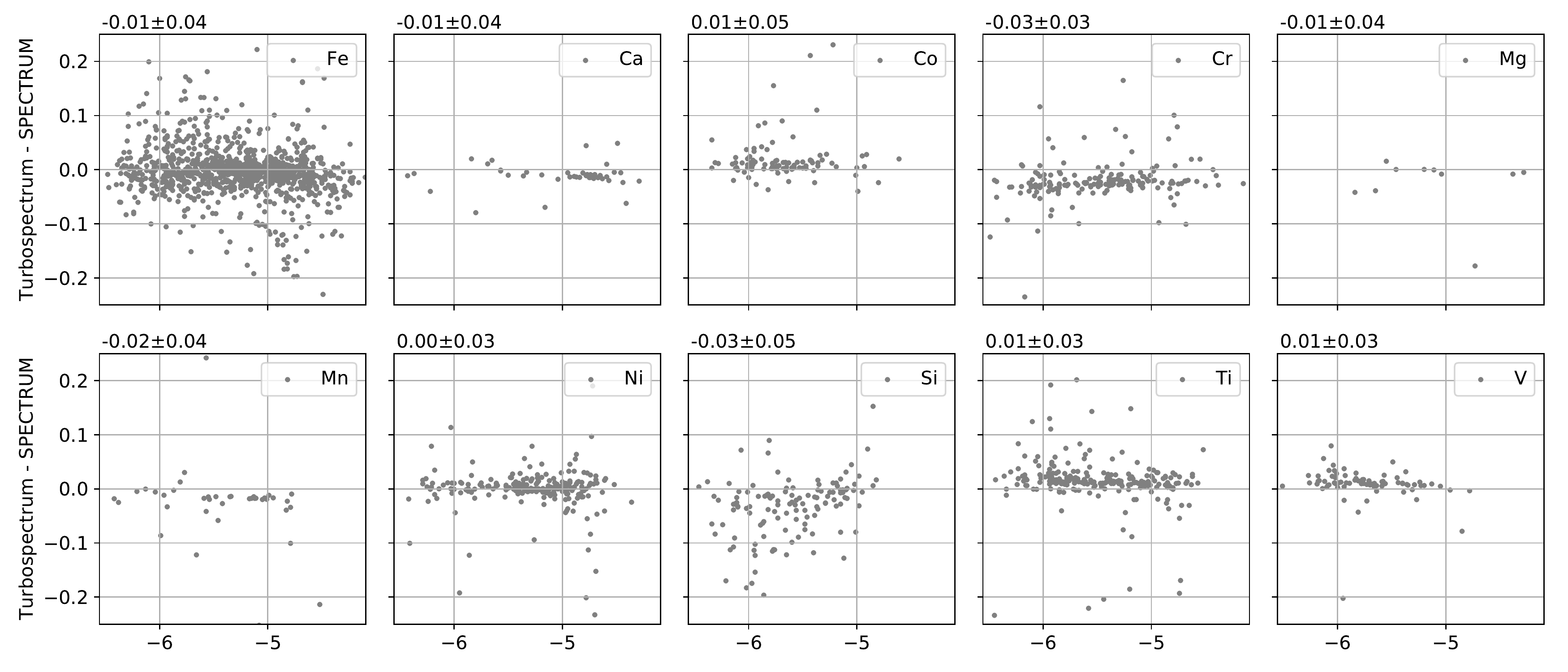}
 \includegraphics[width=\linewidth]{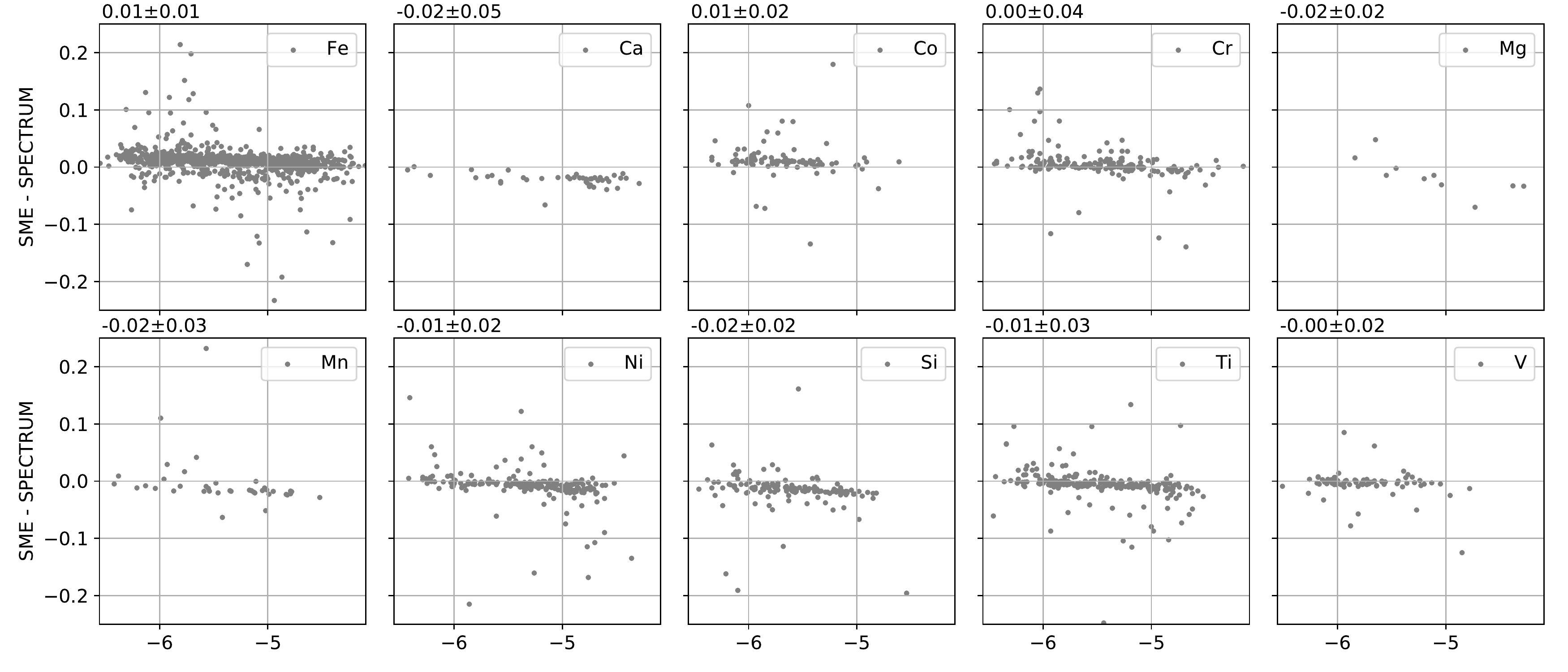}
 \caption{Solar abundance difference between synthesis codes as a function of reduced equivalent width for different elements. Median and absolute median deviation are indicated on the upper left of each subplot.}
 \label{fig:precision_solar_lines_synth_1}
\end{figure*}

\begin{figure*}
 \includegraphics[width=\linewidth]{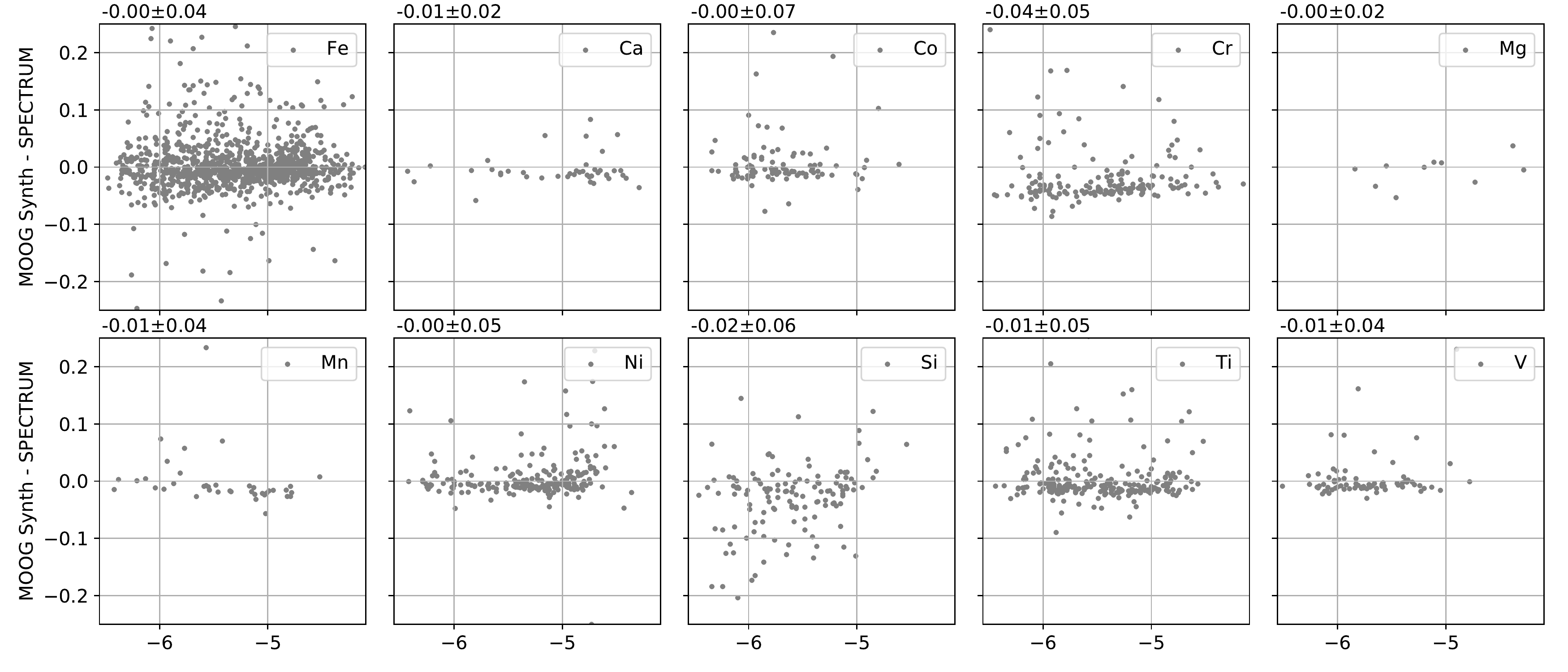}
 \includegraphics[width=\linewidth]{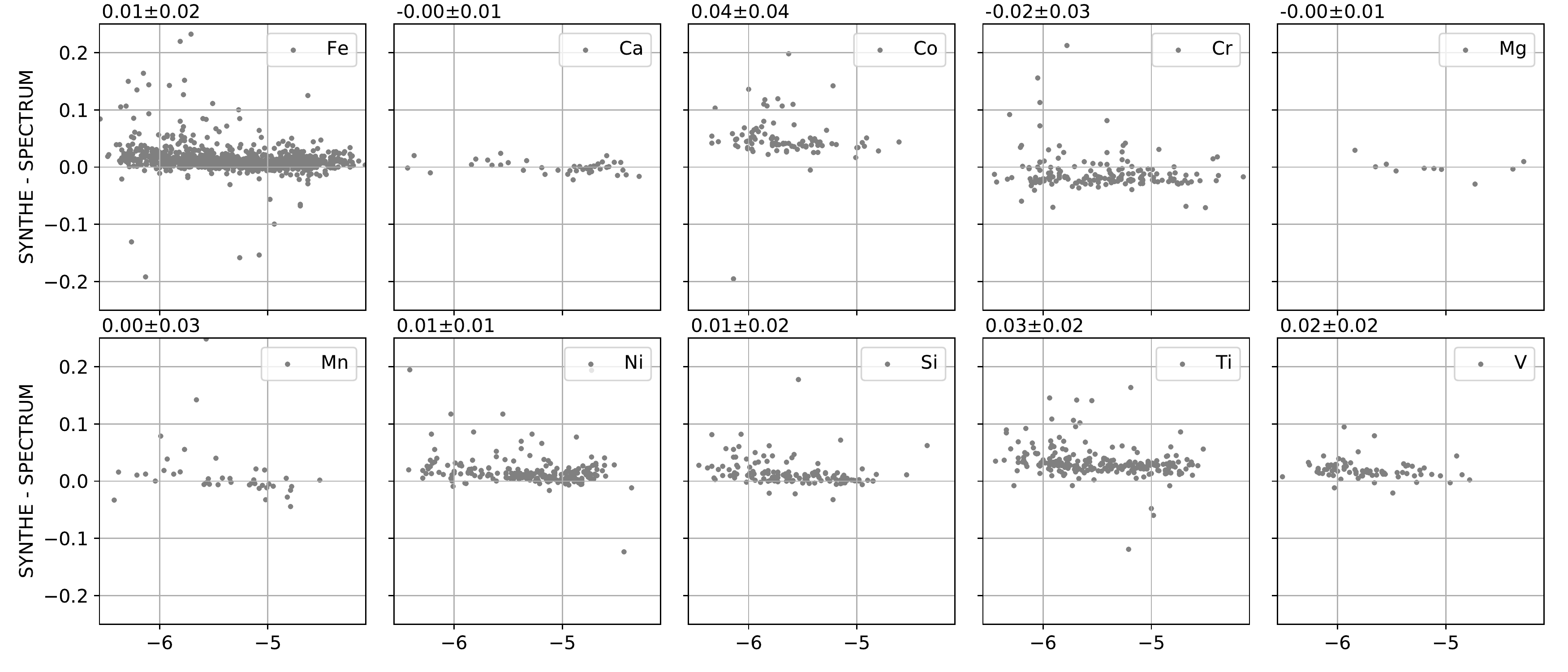}
 \caption{As Fig.~\ref{fig:precision_solar_lines_synth_1}}
 \label{fig:precision_solar_lines_synth_2}
\end{figure*}

A line-by-line comparison between Grid and SPECTRUM (the code used to pre-compute the grid) shows no systematics for any particular element (see Fig.~\ref{fig:precision_solar_lines_grid_vs_synth}). When comparing line-by-line equivalent-width code results, some systematics are observed for certain elements, and major disagreements appear with larger reduced equivalent widths (see Fig.~\ref{fig:precision_solar_lines_ew}). Similar systematics are also observed for synthesis codes, but the size of the reduced equivalent width does not seem to have a major impact on the results between different codes (see Figs~\ref{fig:precision_solar_lines_synth_1} and \ref{fig:precision_solar_lines_synth_2}). This shows the importance of executing line-by-line differential analysis to minimize different systematics between codes.

In total, considering the abundances within $\pm0.05$~dex with respect to the solar abundance, there are only 45 absorption lines in common between all the codes (of which 26 correspond to neutral iron and one to ionized iron). It would not be possible to determine atmospheric parameters with this limited number of lines. Instead, given the different nature of the equivalent-width method and the synthetic spectral technique, I created one line selection for each approach (hereafter, the common line selection). The common line selection is composed of 258 lines (where 146 correspond to neutral iron and 11 to ionized iron) for equivalent-width methods (i.e. MOOG EW and WIDTH9), and of 205 lines for Grid plus the rest of the synthesis codes. The numbers are higher for the former because a less strict limit was required for this method (i.e. the limit was set to $\pm0.10$ instead of $0.05$, as explained in Section~\ref{s:selecting_lines}; otherwise, not enough ionized iron lines would be left).

\begin{figure}
 \includegraphics[width=\columnwidth]{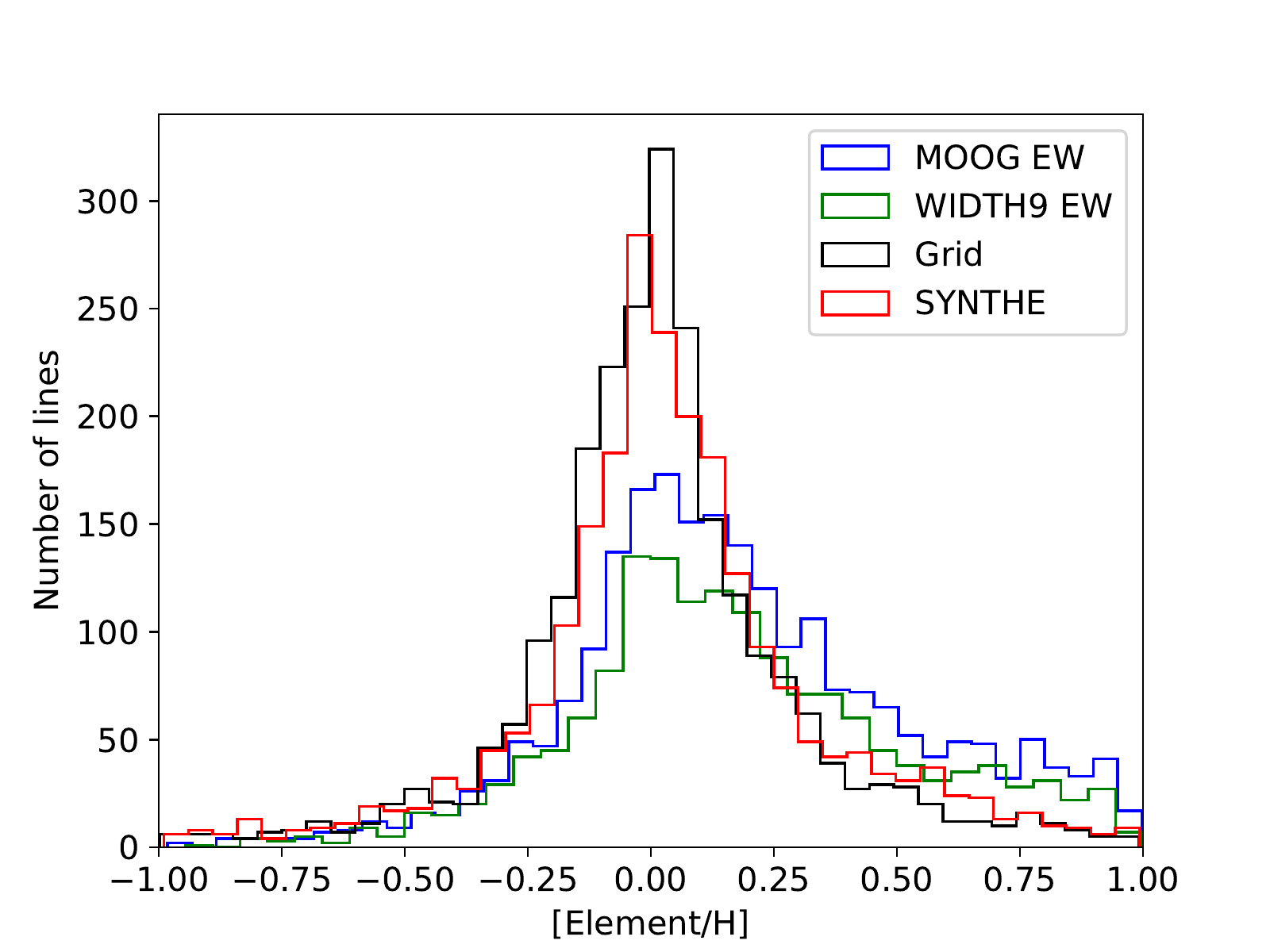}
 \caption{Original distribution of derived abundances for the solar spectrum before any filtering was applied.}
 \label{fig:distribution_solar_lines}
\end{figure}

Regarding lines selected for chemical abundance determination where the constraints were more relaxed, as described in Section~\ref{s:selecting_lines}, an average of $\sim$1\,200 lines were selected for all the codes, with the exception of WIDTH9 and Grid, for which $\sim$900 and $\sim$1\,400 lines were selected. This is coherent with the original distribution of derived abundances shown in Fig.~\ref{fig:distribution_solar_lines}, where WIDTH9 underperforms compared with MOOG EW. Equivalent-width codes show a larger variance with a skewed distribution favouring larger abundances, and Grid has the largest number of lines around zero.

\subsection{Impact on atmospheric parameters}
\label{sec:impact_on_ap}

\subsubsection{Full Gaia Benchmark Stars data set}
\label{sec:impact_on_ap_full}

\begin{figure*}
 \includegraphics[width=\columnwidth]{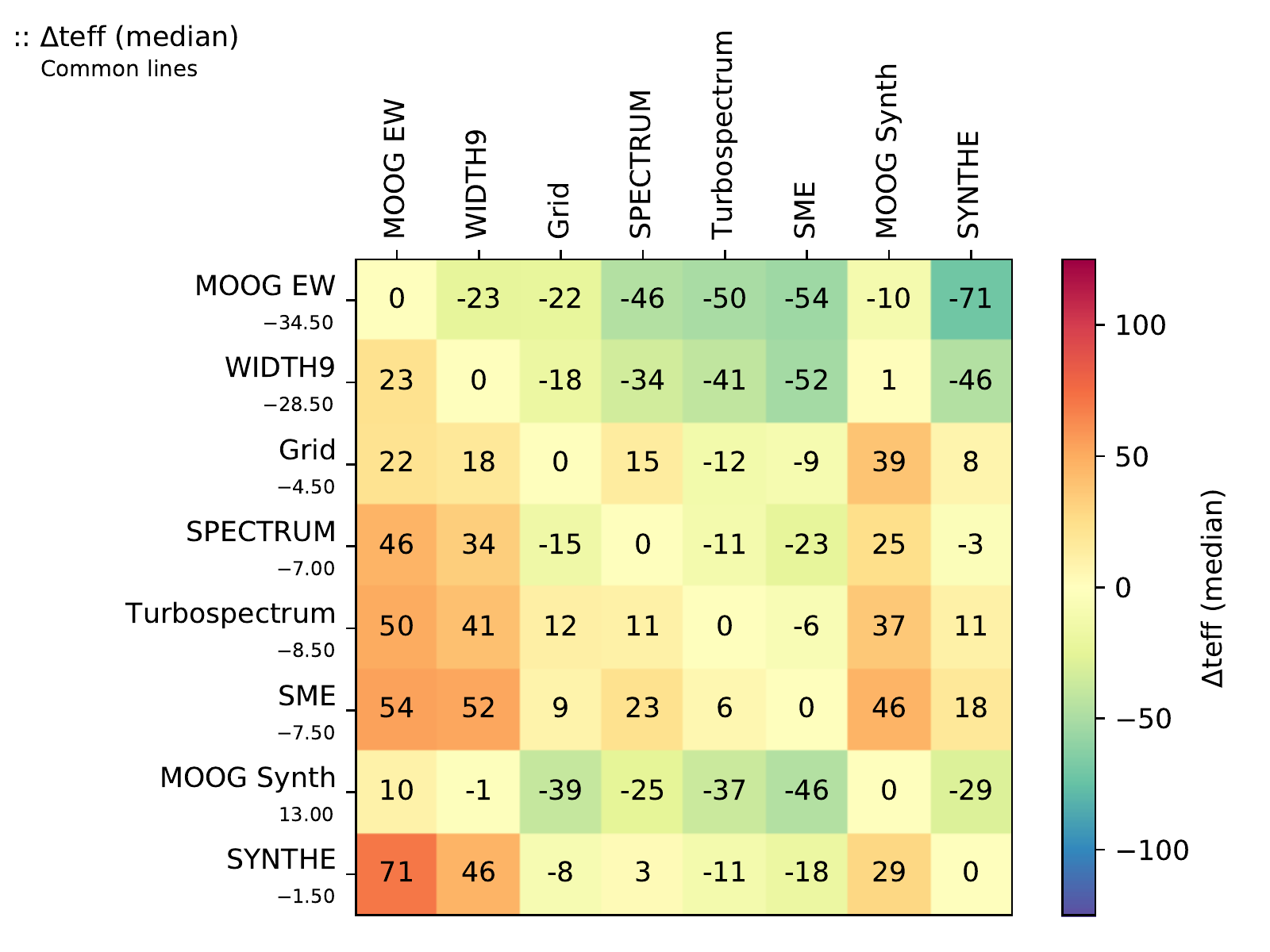}
 \includegraphics[width=\columnwidth]{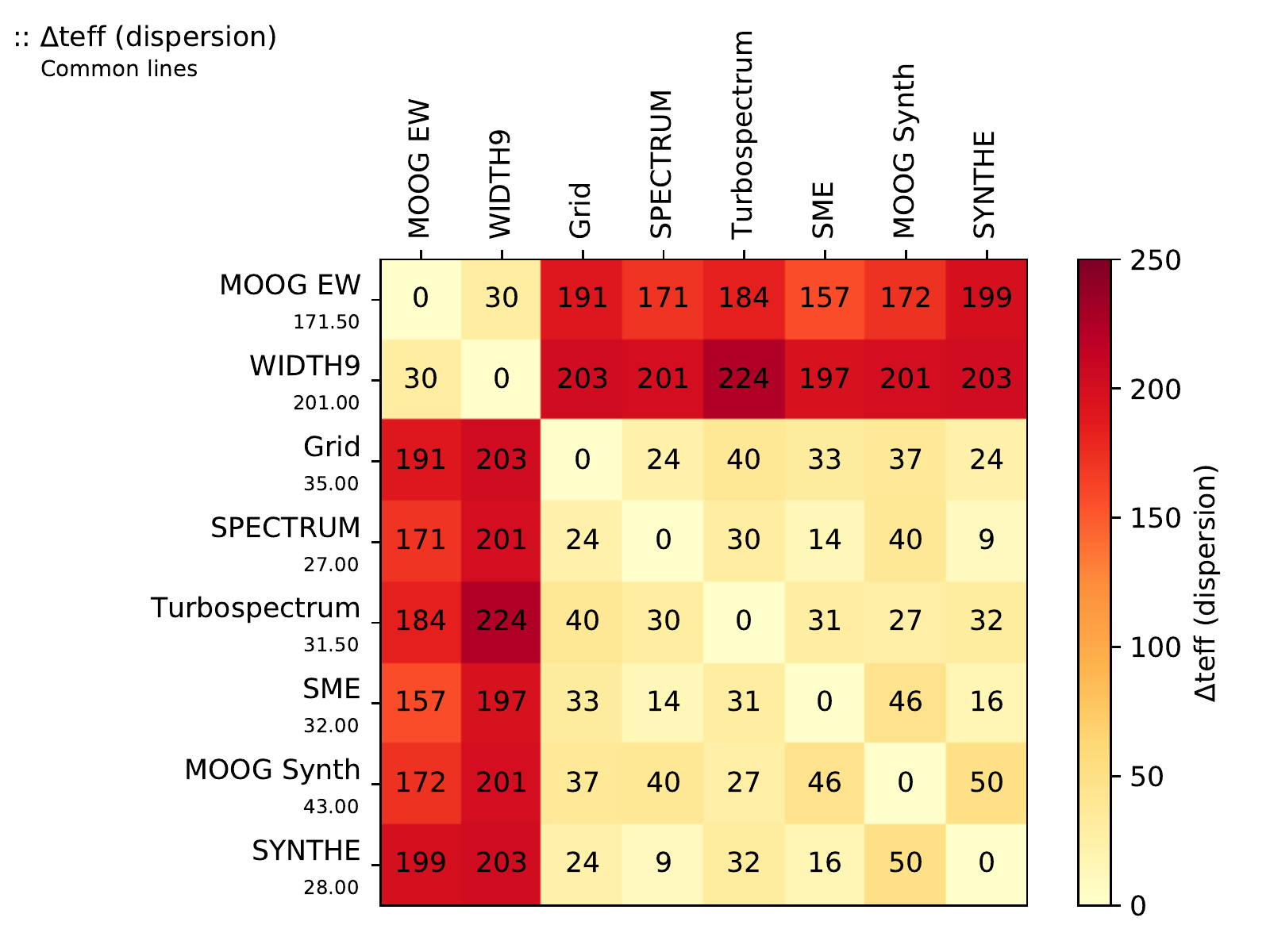}
 \includegraphics[width=\columnwidth]{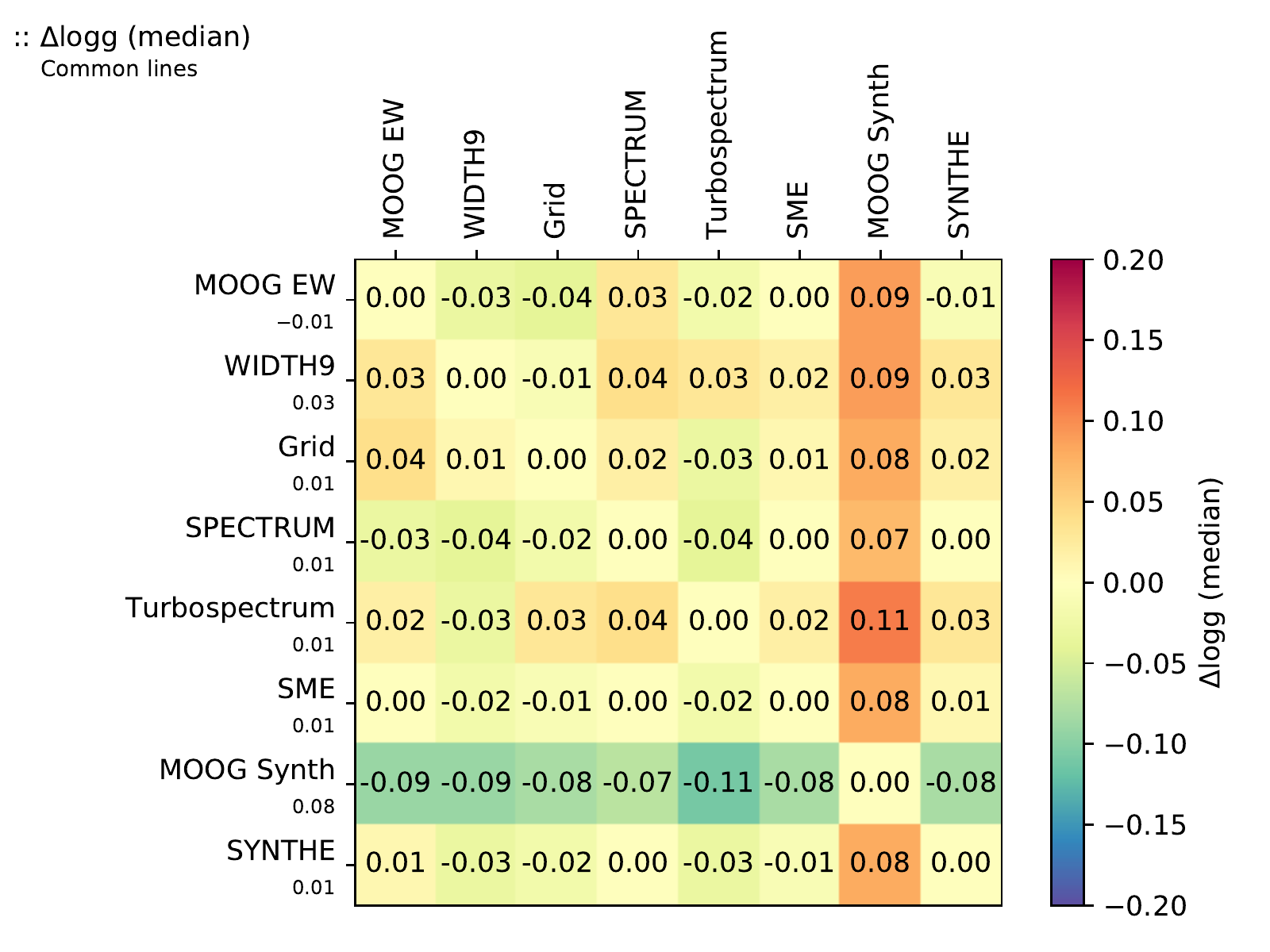}
 \includegraphics[width=\columnwidth]{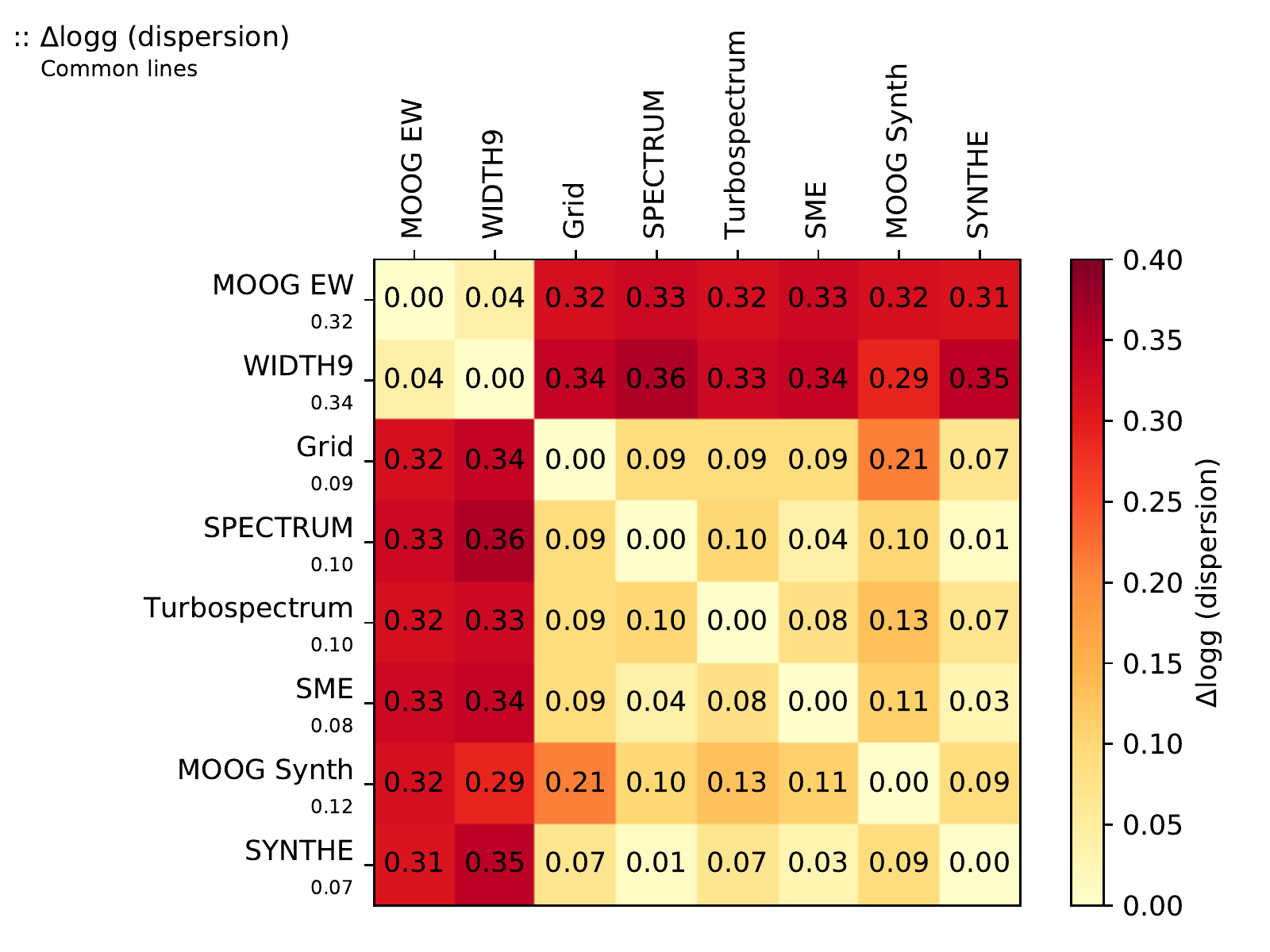}
 \includegraphics[width=\columnwidth]{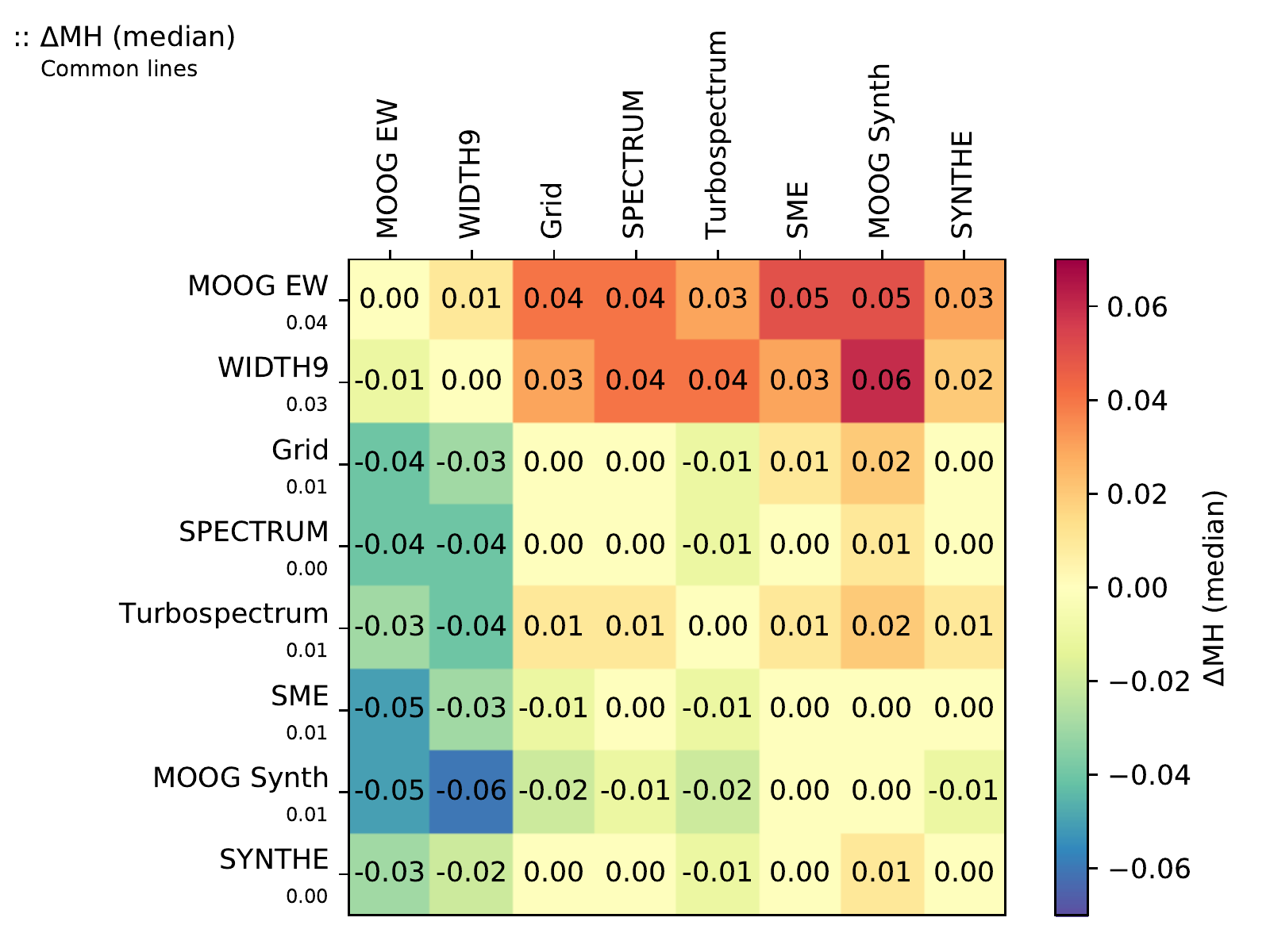}
 \includegraphics[width=\columnwidth]{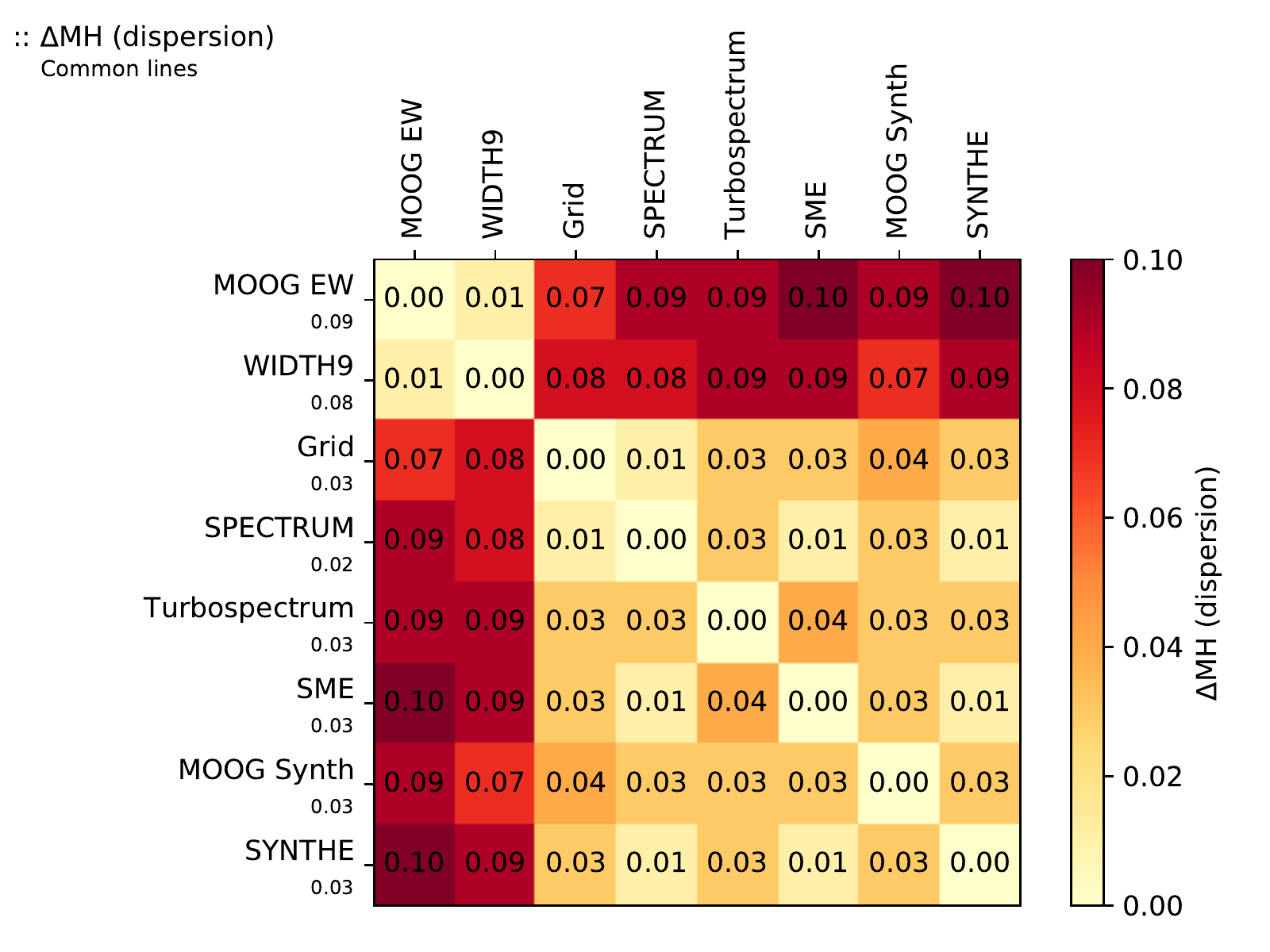}
 \caption{Median and robust standard deviation of the difference in effective temperature, surface gravity or metallicity between different radiative transfer codes when analysing the Gaia Benchmark Stars and using the common line selection (subtraction sense: column minus row).}
 \label{fig:precision_teff_logg_MH}
\end{figure*}

Using the common line selection, I compared the derived atmospheric parameters for the Gaia Benchmark Stars by computing the mean difference between each radiative transfer code and calculating the robust standard deviation\footnote{Function mad\_std from the astropy.stats package \citep{2018arXiv180102634T, 2013A&A...558A..33A}: $\sigma \approx \frac{\textrm{MAD}}{\Phi^{-1}(3/4)} \approx 1.4826 \ \textrm{MAD}$ where $\Phi^{-1}(P)$ is the normal inverse cumulative distribution function evaluated at probability $P = 3/4$} (i.e. dispersion) of these differences (Fig.~\ref{fig:precision_teff_logg_MH}). Ideally, we would like both quantities to be as close as possible to zero, meaning that the precision is high between different pairs of radiative transfer codes. The dispersion found in the three atmospheric parameters again shows two islands that separate equivalent-width methods from synthesis methods with precisions higher within each group but not lower between the two groups.

\begin{figure*}
 \includegraphics[width=\columnwidth]{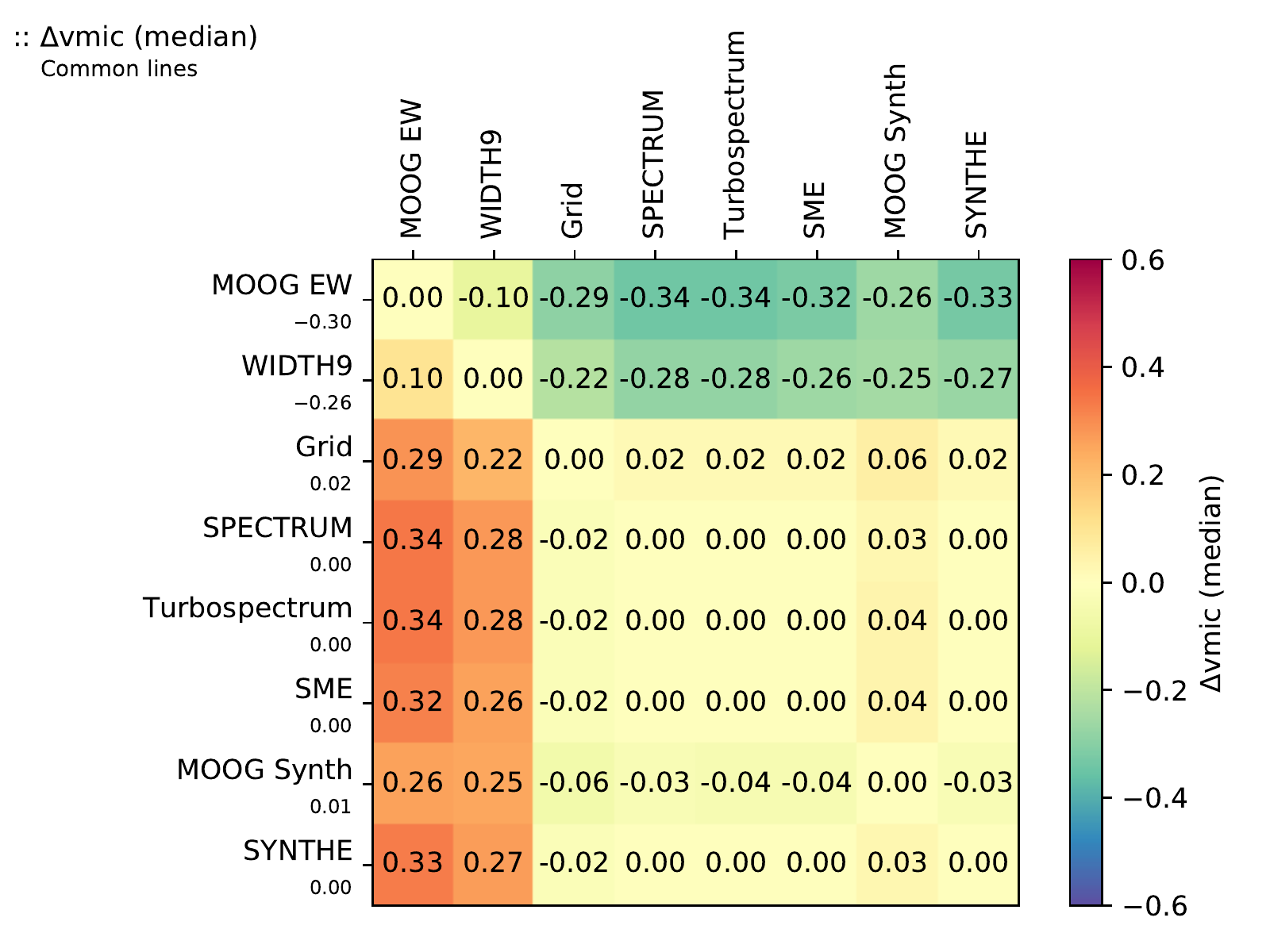}
 \includegraphics[width=\columnwidth]{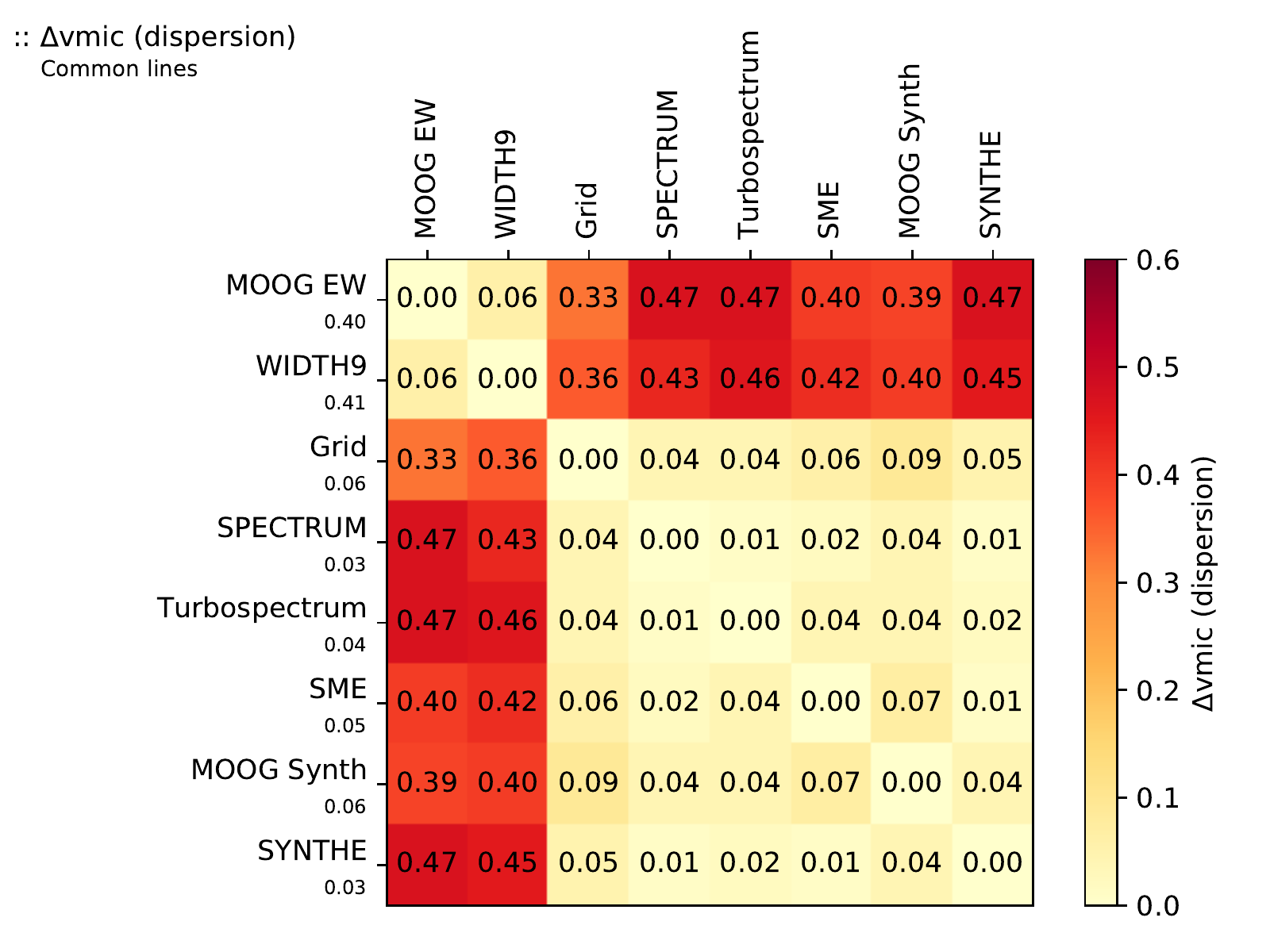}
 \caption{Median and robust standard deviation of the difference in microturbulence velocity between different radiative transfer codes when analysing the Gaia Benchmark Stars and using one common line selection for equivalent width methods plus another common one for synthetic spectral-fitting technique. 
 }
 \label{fig:precision_vmic}
\end{figure*}

In terms of median differences, a clear bias is observed for the surface gravity and a less significant systematic in effective temperature, where the equivalent-width methods provide consistently lower and higher values, respectively. This effect may be driven by the differences in the microturbulence velocity, as shown in Fig.~\ref{fig:precision_vmic}. The microturbulence parameter represents ensemble velocity fields that are not available in 1D model atmospheres (in 3D models, the microturbulence parameter is not necessary), and these velocity fields have broadening effects (depth-independent) on the line opacity (the parameter serves to desaturate the line). The differences shown for the microtubulence velocity, especially between the equivalent-width method and the synthetic spectral-fitting technique, could be caused by a compensatory effect on differences in the derived effective temperatures and surface gravities or by real differences between the methods and codes. The latter is explored with the experiment described in Section~\ref{s:one_variable_at_a_time_experiment}, the results of which are presented in Section~\ref{s:one_variable_at_a_time_results}.

\begin{figure*}
 \includegraphics[width=\columnwidth]{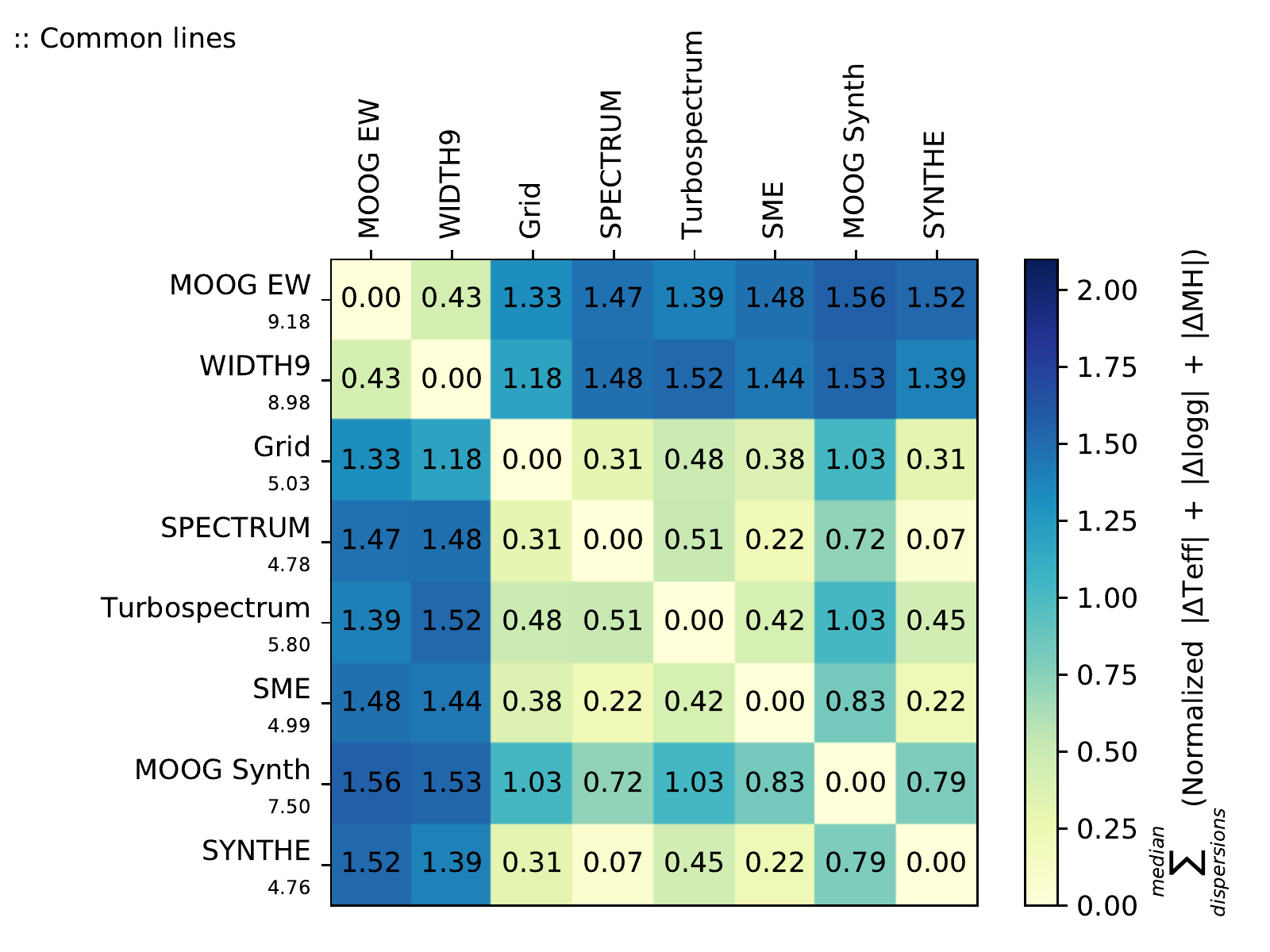}
 \includegraphics[width=\columnwidth]{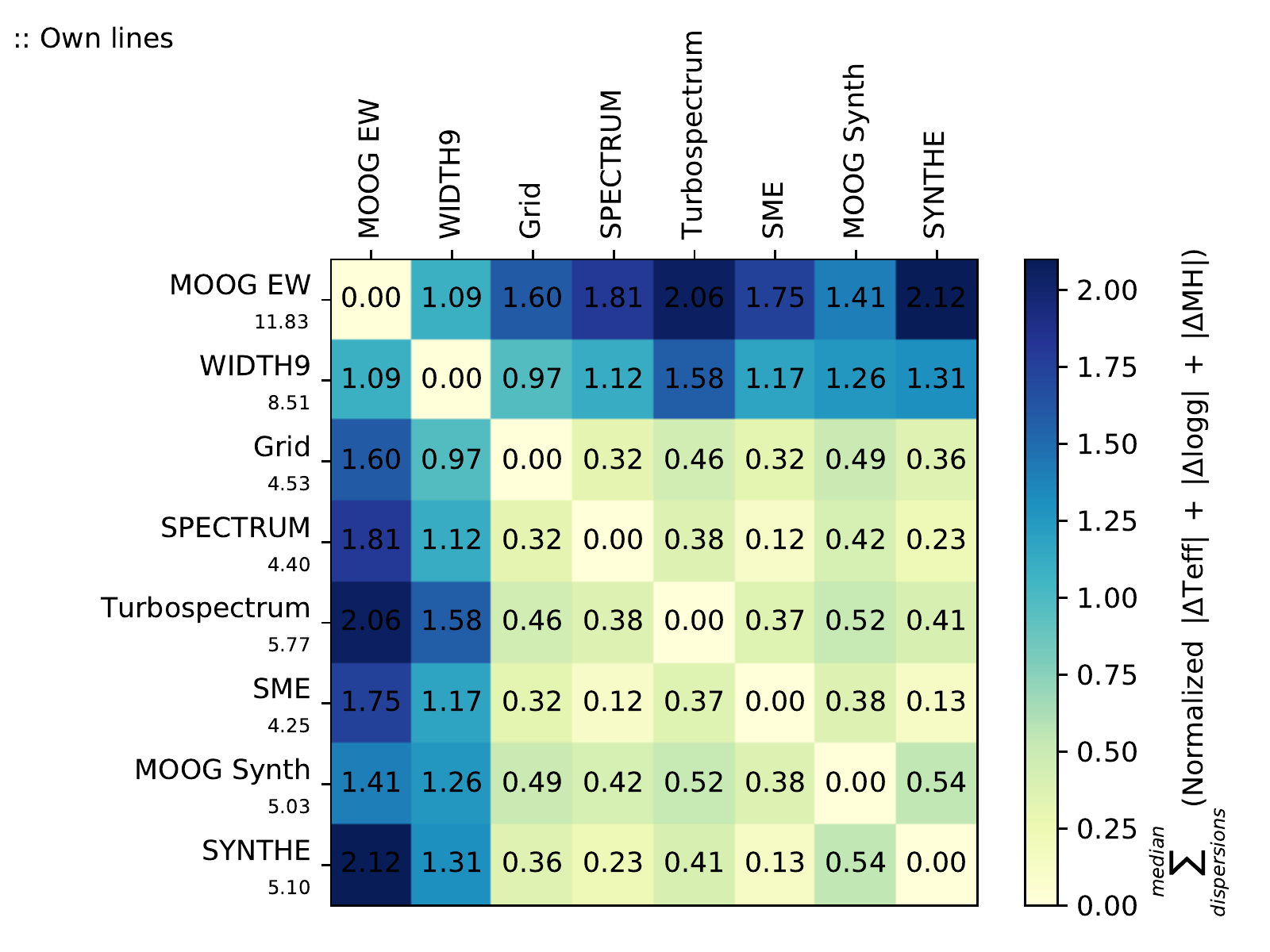}
 \caption{Sum of the normalized absolute median differences and normalized robust standard deviation for effective temperature, surface gravity and metallicity when analysing the Gaia Benchmark Stars. Lower numbers indicate the codes lead to more similar results (higher precision).}
 \label{fig:best_precision}
\end{figure*}

To be able to visually compare all the parameters for all radiative transfer codes at the same time, I normalized\footnote{All the values were scaled to unit norm (vector length) using the sklearn.preprocessing.normalize function \citep{sklearn_api, scikit-learn}.} all the values from Fig.~\ref{fig:precision_teff_logg_MH} and added them together as shown in Fig.~\ref{fig:best_precision} (left plot). In addition, I repeated the same operation for all the results obtained when the best line selection (hereafter the 'own lines' selection) for each code is used (right plot). Using the best line selection improves the statistics by increasing the number of lines, but it introduces more inhomogeneities into the analysis. The former effect dominated for synthesis codes because the agreement among them slightly increased (mainly for MOOG SYNTH), while the latter was more significant for equivalent-width methods, where WIDTH9 results separated from MOOG and they got slightly more similar to the synthesis results.

\begin{figure*}
 \includegraphics[width=7.4cm]{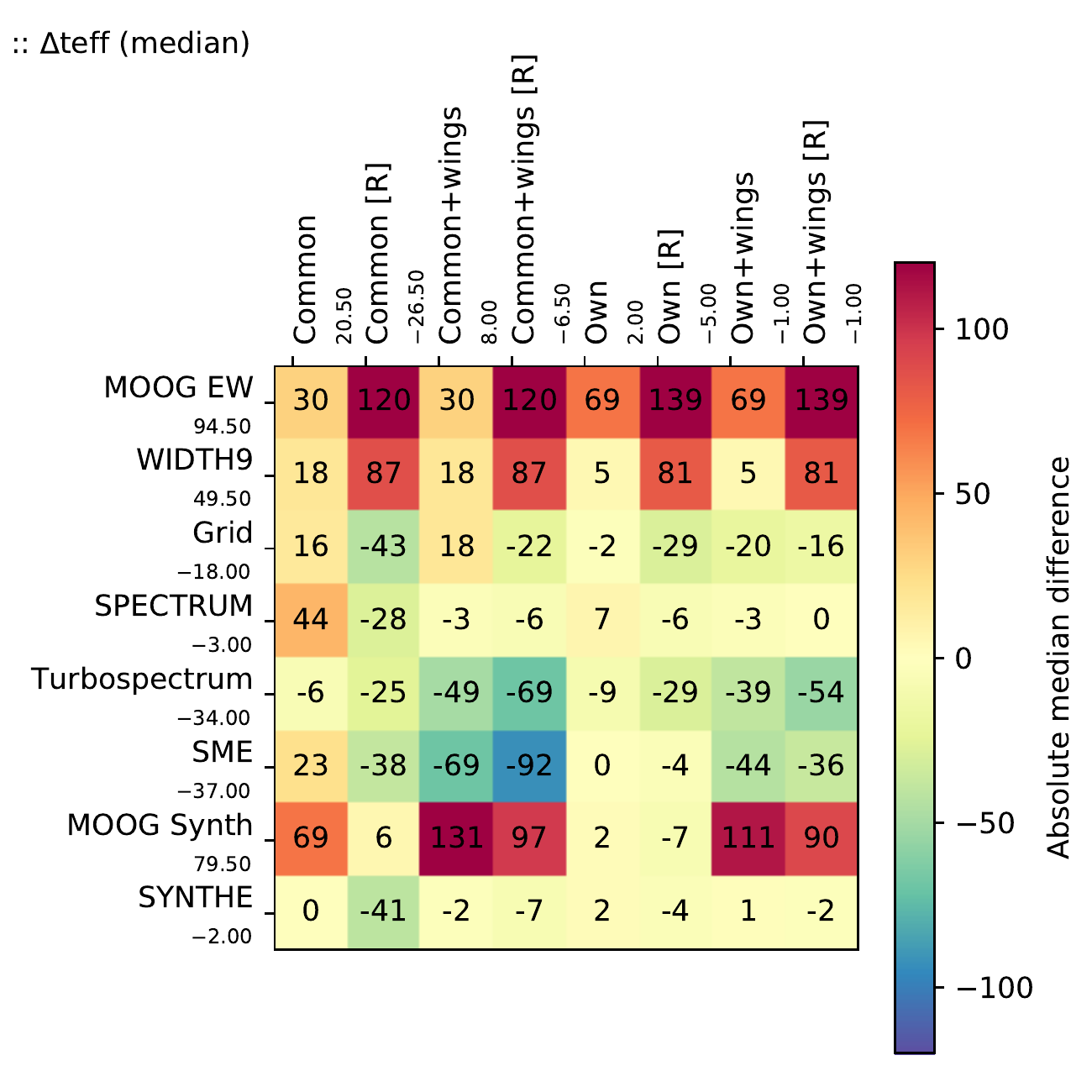}
 \includegraphics[width=7.4cm]{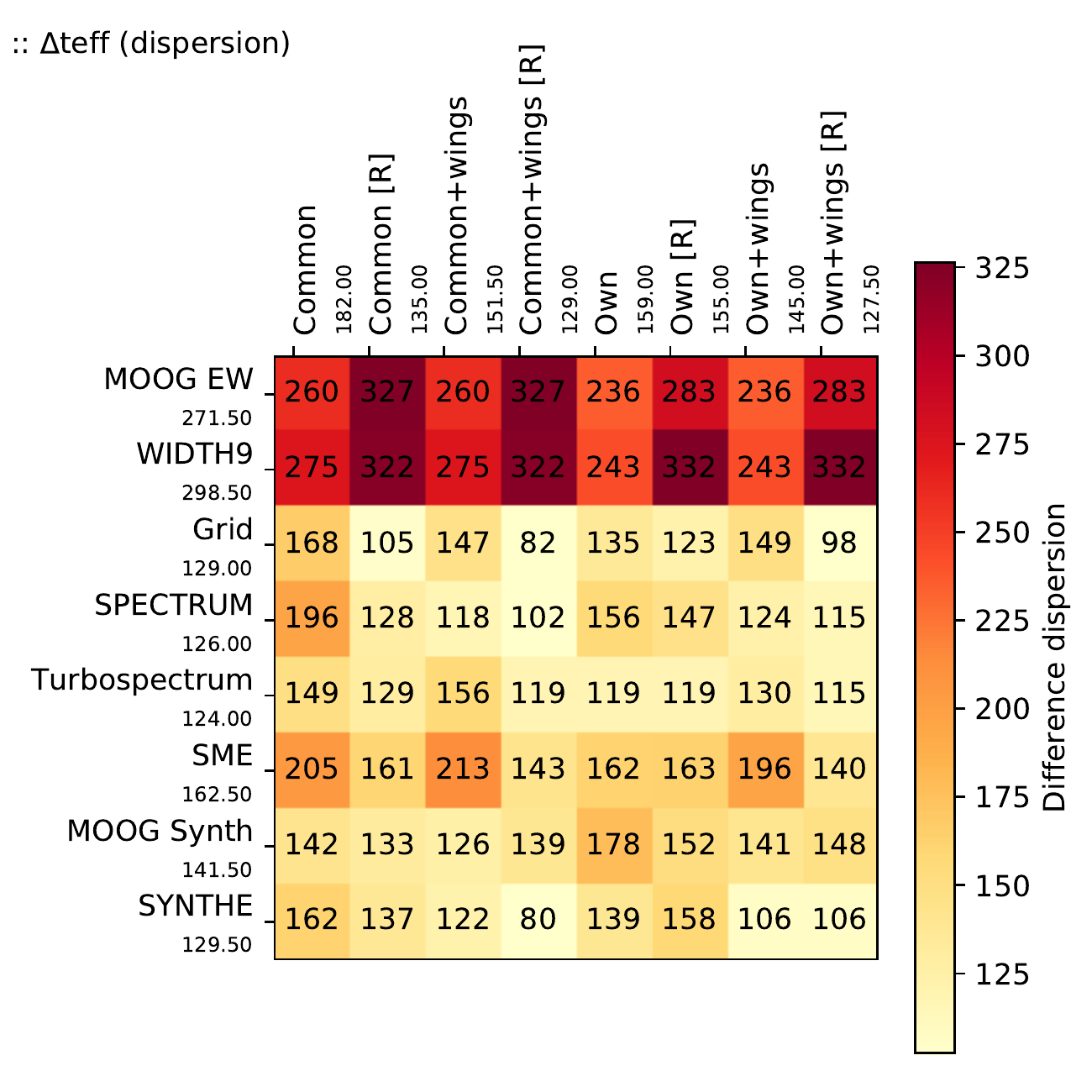}
 \includegraphics[width=7.4cm]{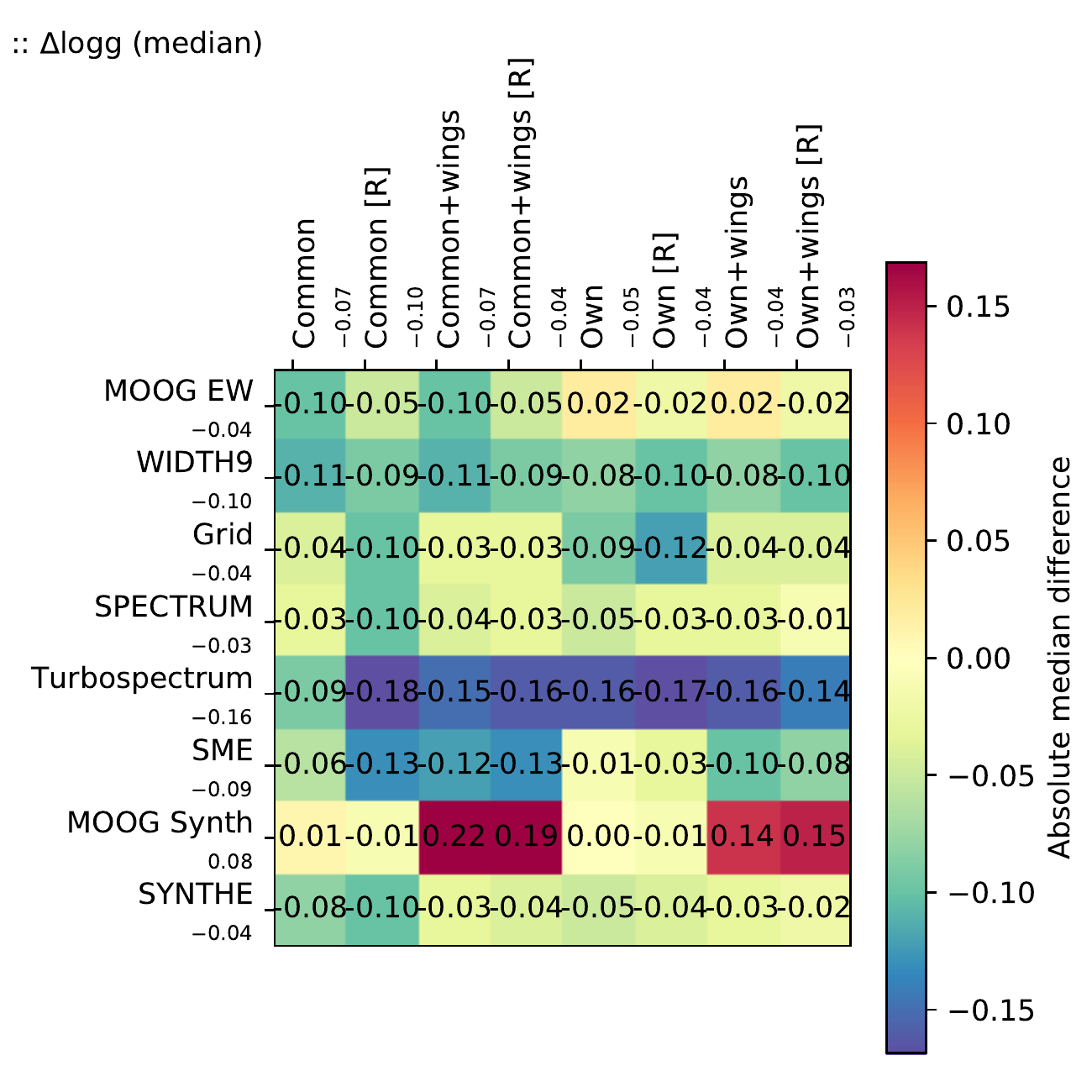}
 \includegraphics[width=7.4cm]{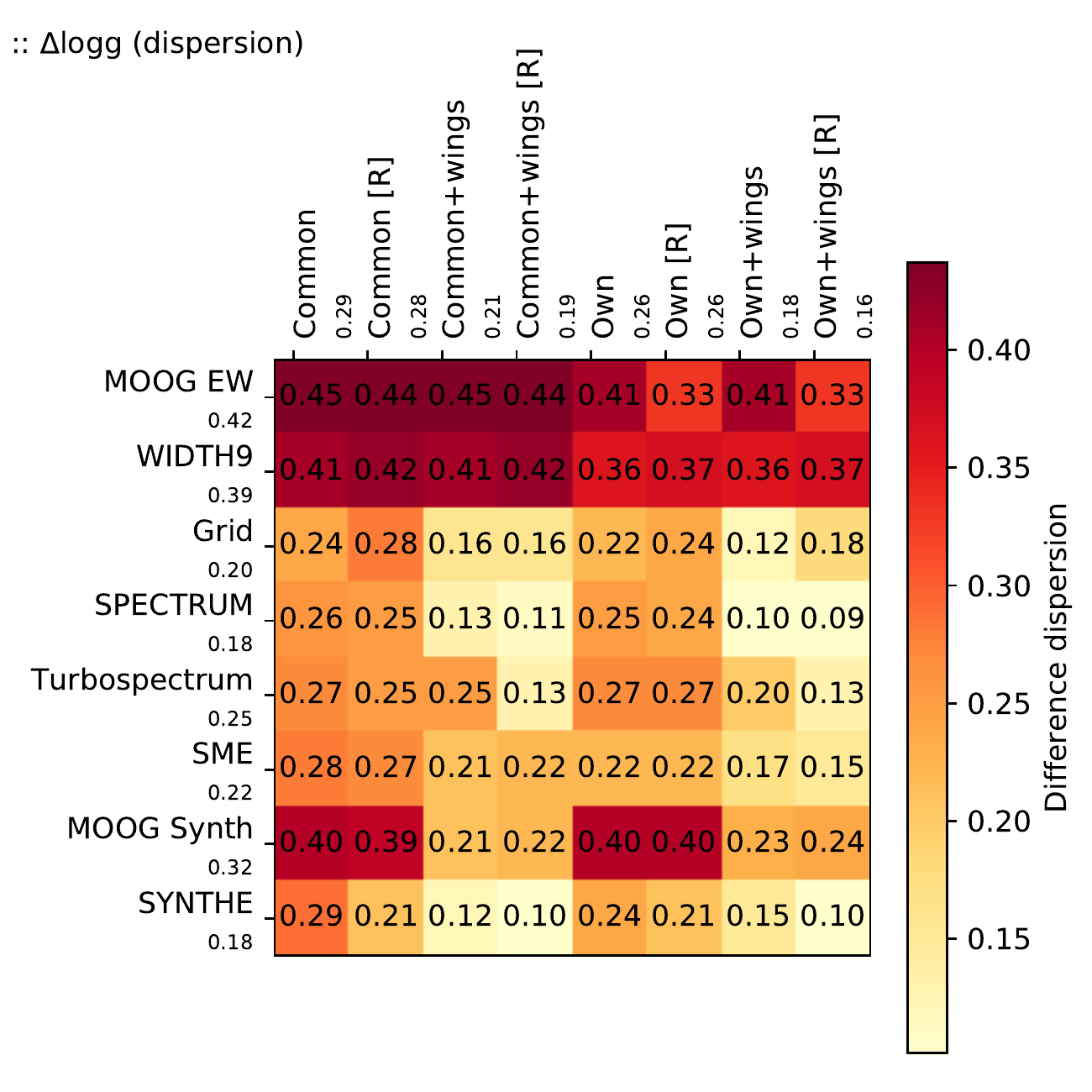}
 \includegraphics[width=7.4cm]{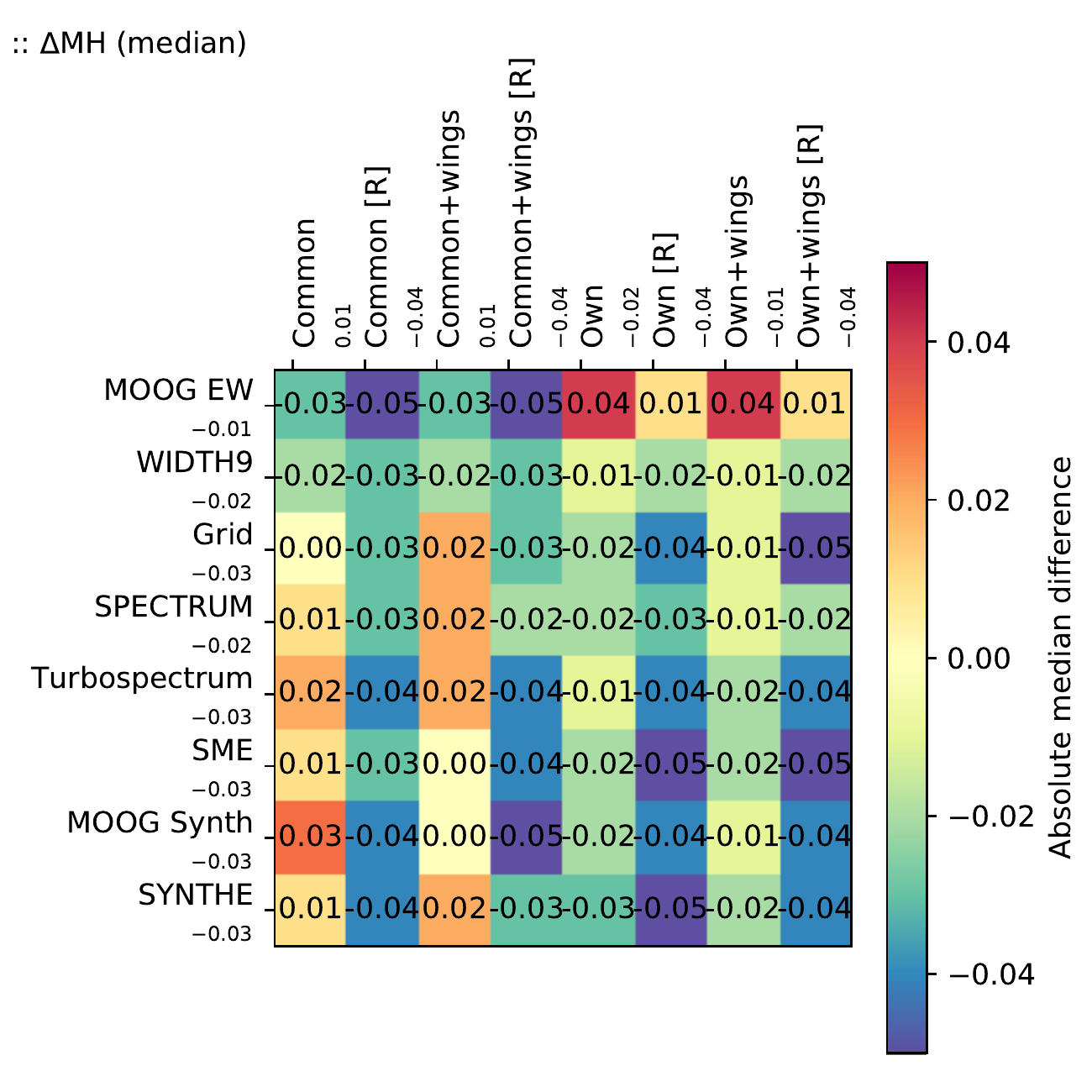}
 \includegraphics[width=7.4cm]{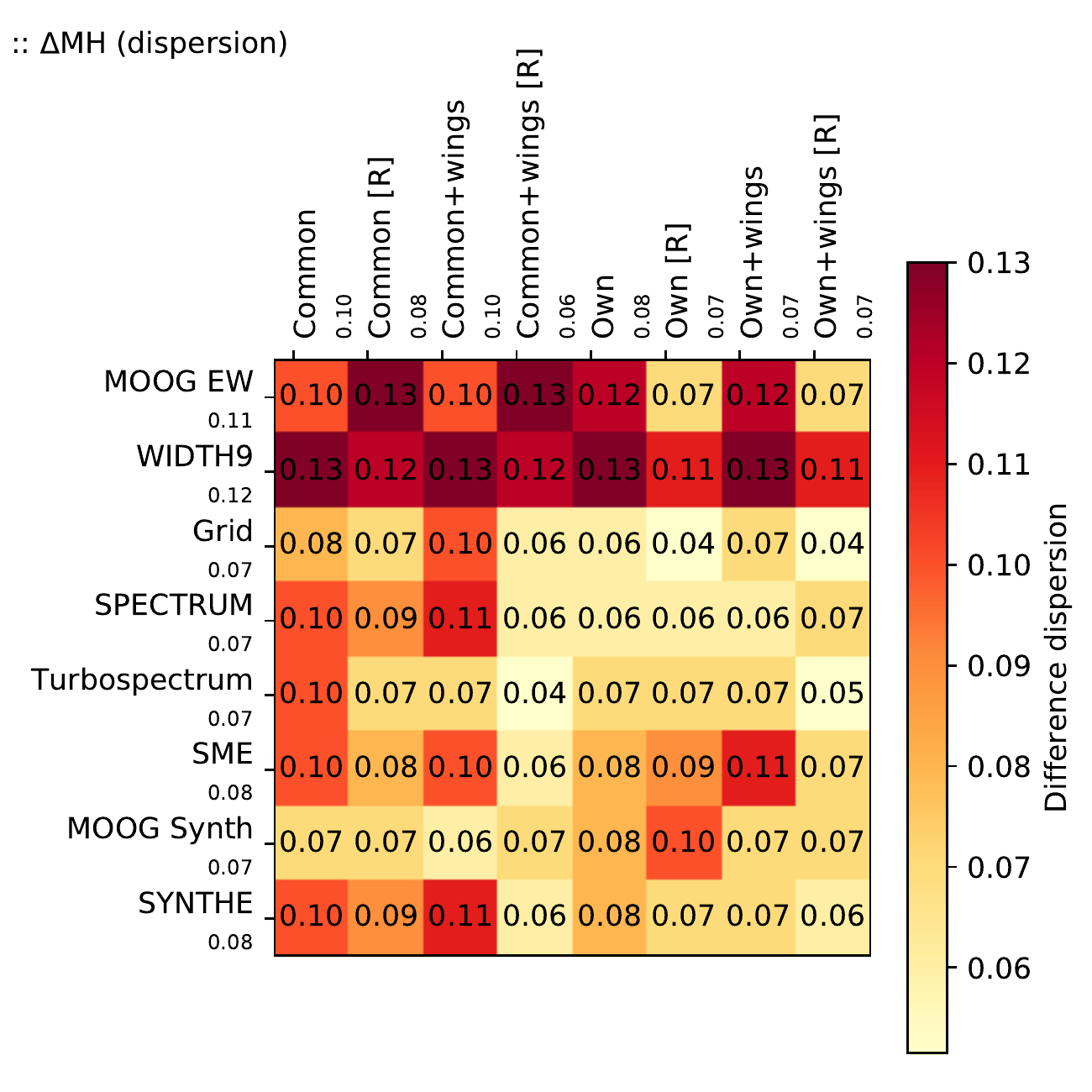}
 \caption{Median and robust standard deviation of the difference in effective temperature with respect to the reference values when analysing the Gaia Benchmark Stars and using different radiative transfer codes with several set-ups: using lines in common within equivalent-width and synthesis methods (labelled as 'Common'); using the best lines for each code (i.e. their own lines, which are not necessarily good for other codes, labelled as 'Own'); using the wings of H-$\alpha$/$\beta$ and the Mg triplet (labelled as 'Wings'); and repeating the normalization (labelled as ''[R]'') but using a synthetic spectrum that matches the atmospheric parameters found in the first iteration.}
 \label{fig:accuracy_teff_logg_MH}
\end{figure*}

Assessing the precision between pairs of codes allows us to verify what codes lead to the most similar results, but does not verify which code and/or set-up obtains the results closest to the expected reference parameters (e.g. a pair of codes may be very imprecise because only one of them is very accurate). The accuracy of the results (the difference with respect to the Gaia Benchmark Stars reference values) is shown in Fig.~\ref{fig:accuracy_teff_logg_MH}. I ran the analysis with eight different set-ups using the selection of common lines or the best lines for each code plus enabling/disabling the following options.

\begin{itemize}
    \item In addition to the selected lines, consider the wings of H-$\alpha$/$\beta$ and the Mg triplet. Enabling this option is not possible for the equivalent-width methods, for which the results in Fig.~\ref{fig:accuracy_teff_logg_MH} are just duplicated.
    \item Run a second full iteration as described in Section~\ref{s:ap_determination}, where the normalization is repeated but using a synthetic spectrum as a template (which matches the atmospheric parameters found in a first full iteration) and re-determine the atmospheric parameters.
\end{itemize}

\begin{figure}
 \includegraphics[width=\columnwidth]{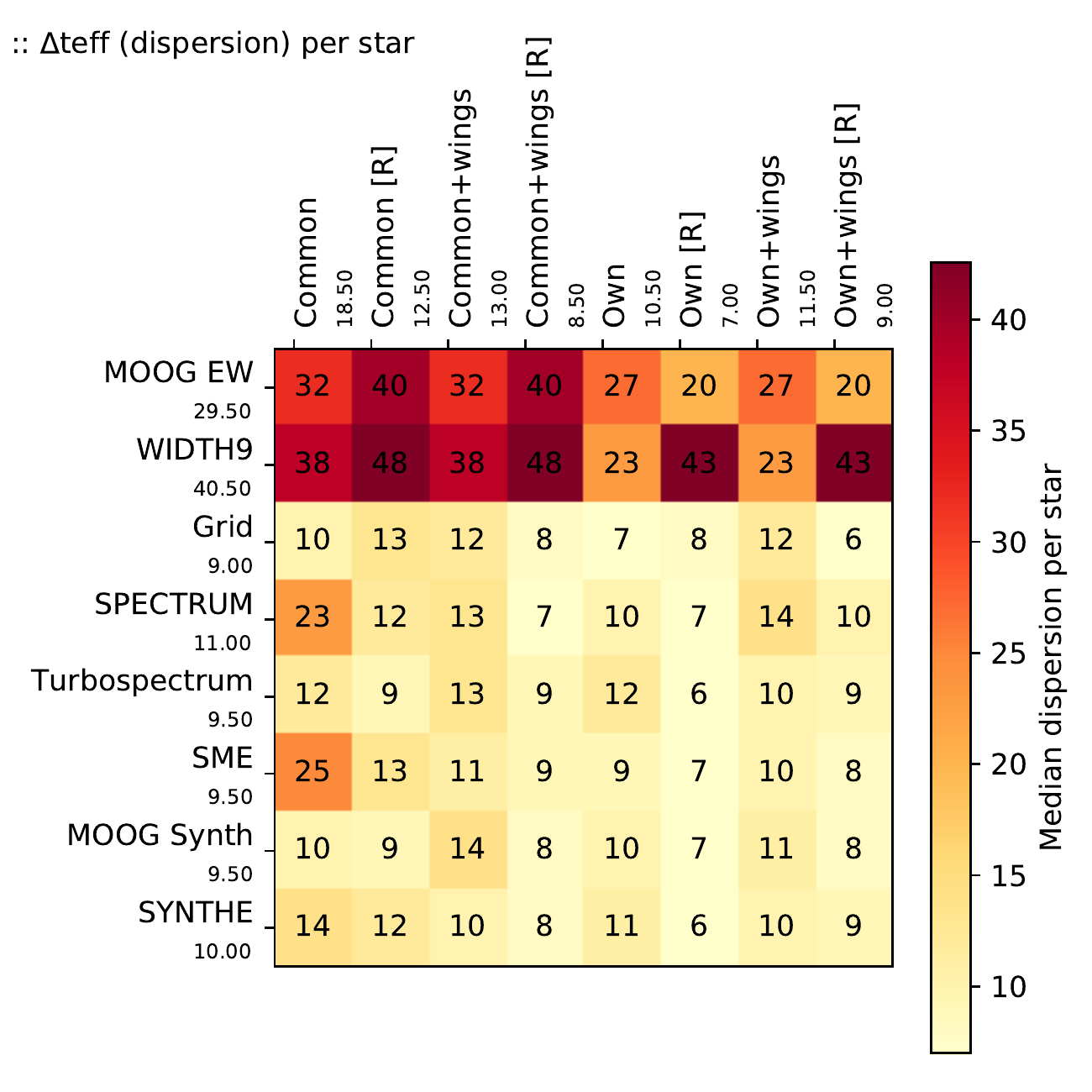}
 \includegraphics[width=\columnwidth]{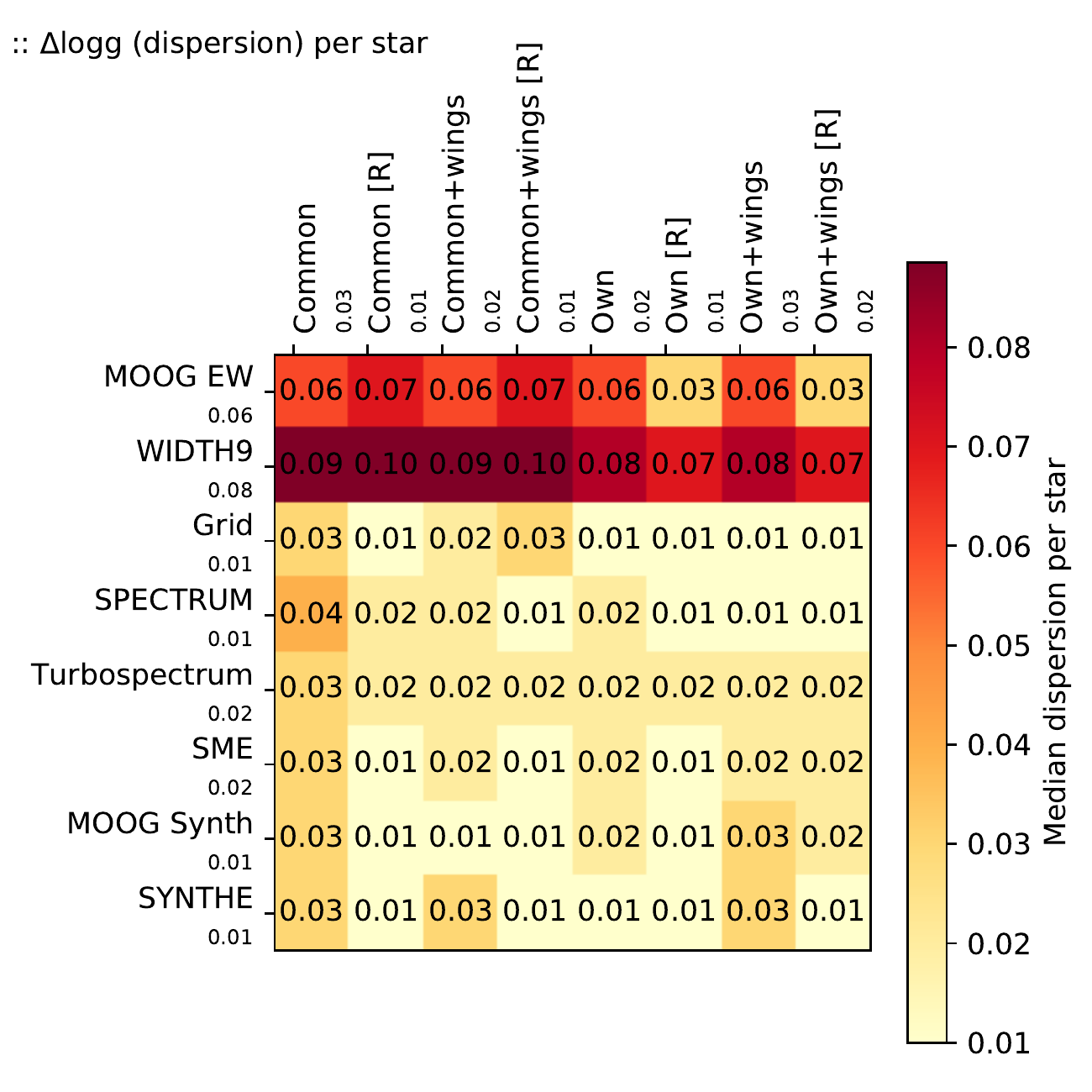}
 \caption{Median robust standard deviation of the effective temperature and surface gravity per star (when multiple spectra were available) when analysing the Gaia Benchmark Stars and using different radiative transfer codes with several set-ups: using lines in common within equivalent-width and synthesis methods; using the best lines for each code (i.e. their own lines, which are not necessarily good for other codes); using the wings of H-$\alpha$/$\beta$ and the Mg triplet; and repeating the normalization (labelled as ''[R]'') but using a synthetic spectrum that matches the atmospheric parameters found in the first iteration.}
 \label{fig:precision_teff_logg_per_star}
\end{figure}

The equivalent-width method presents a higher dispersion for all the atmospheric parameters. It also has the lowest level of agreement when analysing several spectra corresponding to the same star, as shown in Fig.~\ref{fig:precision_teff_logg_per_star}. Metallicity is not included in that figure because the results are very similar across codes: the median robust standard deviation per star is about 0.03~dex for equivalent-width methods and 0.01~dex for synthesis methods. The synthetic spectral-fitting technique performs better in this test mainly because the Gaia Benchmark Stars include a wide range of FGKM stars and the equivalent-width method is not the best option for all of them (see Section~\ref{sec:impact_on_ap_limited}). For instance, the accuracy of the equivalent-width method degrades more strongly with cooler stars owing to blends and with metal-poor stars owing to the lack of iron lines.

\begin{figure}
 \includegraphics[width=\linewidth]{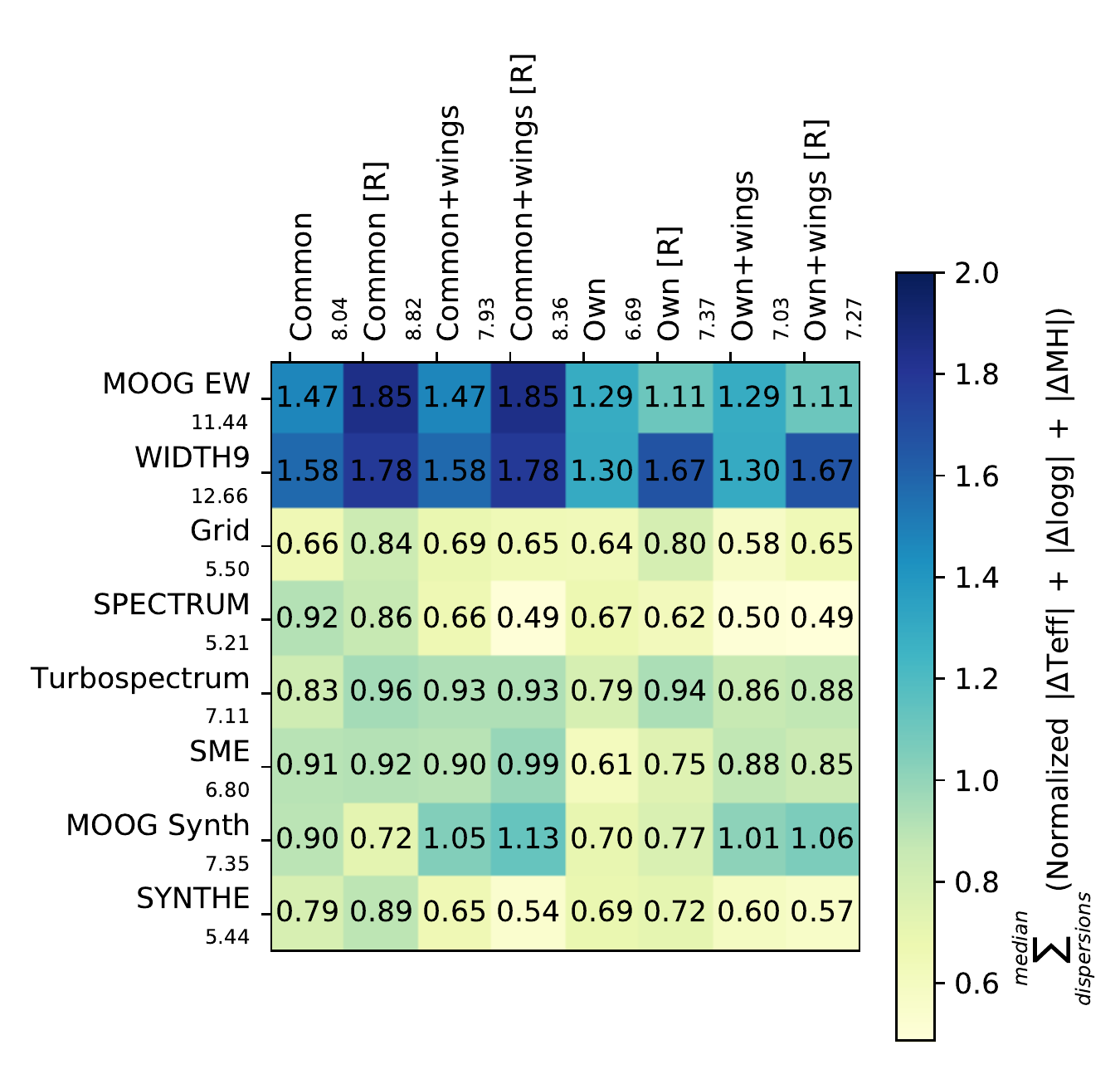}
 \caption{Sum of the normalized absolute median differences and normalized robust standard deviation and normalized median robust standard deviation for effective temperature, surface gravity and metallicity when analysing the Gaia Benchmark Stars. Lower numbers indicate results closer to reference values and lower dispersion (higher accuracy).}
 \label{fig:best_accuracy}
\end{figure}

In order to visually compare all the parameters for all radiative transfer codes and set-ups at the same time, I normalized and added all the values from Figs~\ref{fig:accuracy_teff_logg_MH} and \ref{fig:precision_teff_logg_per_star}, as shown in Fig.~\ref{fig:best_accuracy}. As a general rule, using the best line selection instead of the common line selection leads to a better accuracy for all the codes, thanks to the increase in the statistics without sacrificing quality.

The codes MOOG EW and WIDTH lead to similar results for the equivalent-width method, with MOOG EW being the best of the two when using its own line selection and executing a second full iteration normalizing with a synthetic spectrum matching the atmospheric parameters found in the first full iteration. This second full iteration does not have the same positive effect for all the codes. Its major contribution is improving the dispersion per star for most synthesis codes, as shown in Fig.~\ref{fig:precision_teff_logg_per_star}, but sometimes it slightly worsens the overall results. The effect of this second full iteration could be due to the template-based normalization or to the execution of an extra batch of iterations until convergence for the determination of atmospheric parameters is reached. I executed a validation test with Grid using only one full iteration with the best line selection but allowing the process that determines the atmospheric parameter to run for a greater maximum number of iterations (12 instead of 6), and the results did not change significantly (rms decreased by less than 0.01 for 11 spectra, and the rest remained at roughly the same level). This is a strong indication that the effects of adding the second full iteration are caused mainly by the template-based normalization.

\begin{figure}
 \includegraphics[width=\columnwidth]{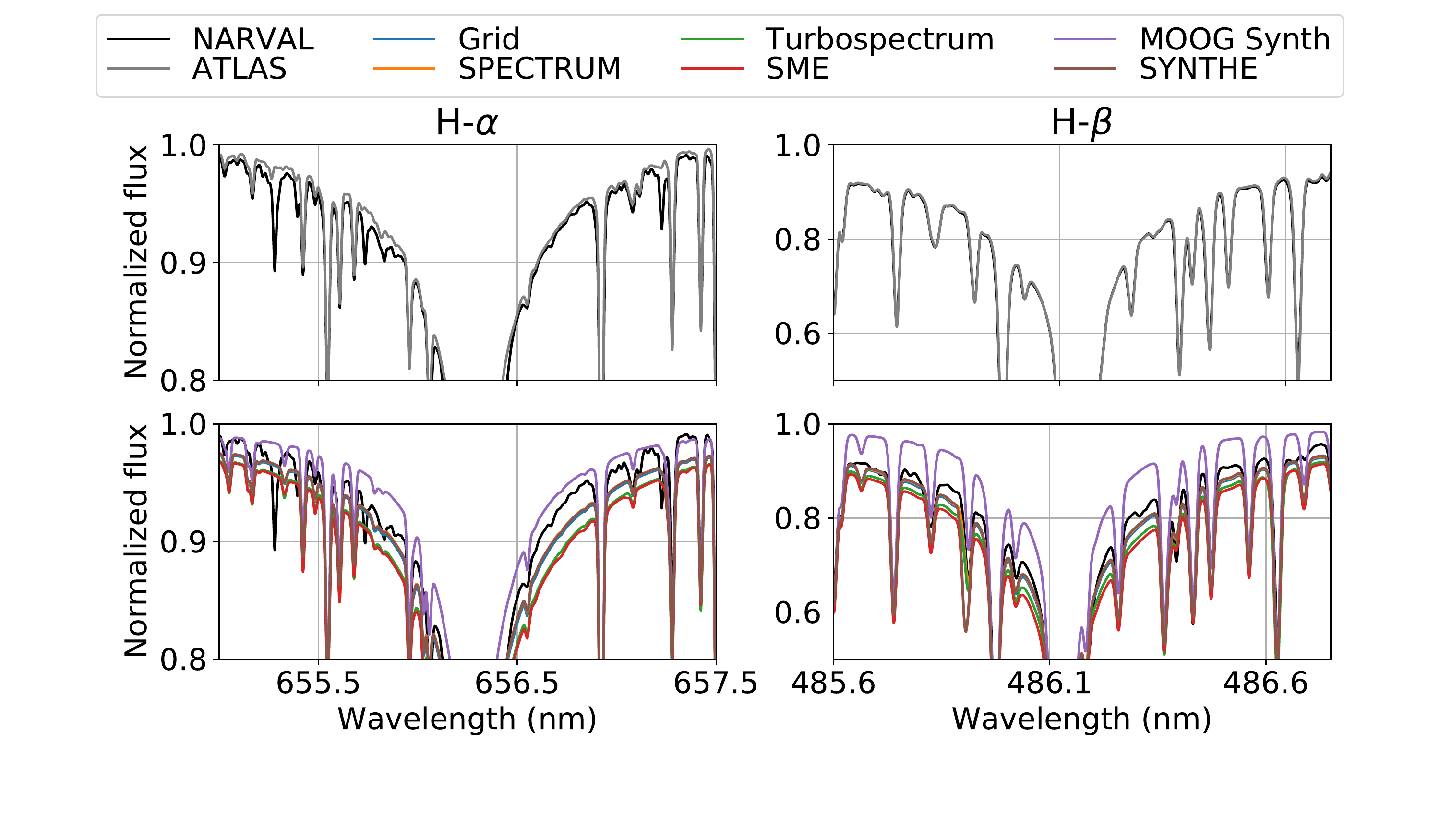}
 \caption{Hydrogen line regions from a NARVAL-observed solar spectrum and the solar ATLAS (top), plus the same NARVAL-observed solar spectrum and the corresponding synthetic spectra generated with each code (bottom).}
 \label{fig:hydrogen_lines}
\end{figure}

For the synthetic spectral-fitting methods, adding the wings of H-$\alpha$/$\beta$ and the Mg triplet generally improves the results for Grid, SPECTRUM and SYNTHE, while it worsens the results for Turbospectrum, SME and MOOG Synth. To rule out that the normalization process is not favouring some of the codes, a comparison of an observed NARVAL solar spectrum and an observed solar ATLAS \citep{2000vnia.book.....H} is shown in the top subplots of Fig.~\ref{fig:hydrogen_lines}, where the agreement is outstanding. Synthetic solar spectra for each code are also shown in the bottom subplots of the same figure. SME and Turbospectrum are close together with lower normalized fluxes, while MOOG is on the other extreme with higher normalized fluxes. Grid, SPECTRUM and SYNTHE have a large region of overlap with each other and they are generally closer to the observed spectrum. These discrepancies reflect differences in how the broadening of the hydrogen lines is computed by each code.

\begin{figure*}
 \includegraphics[width=5.8cm]{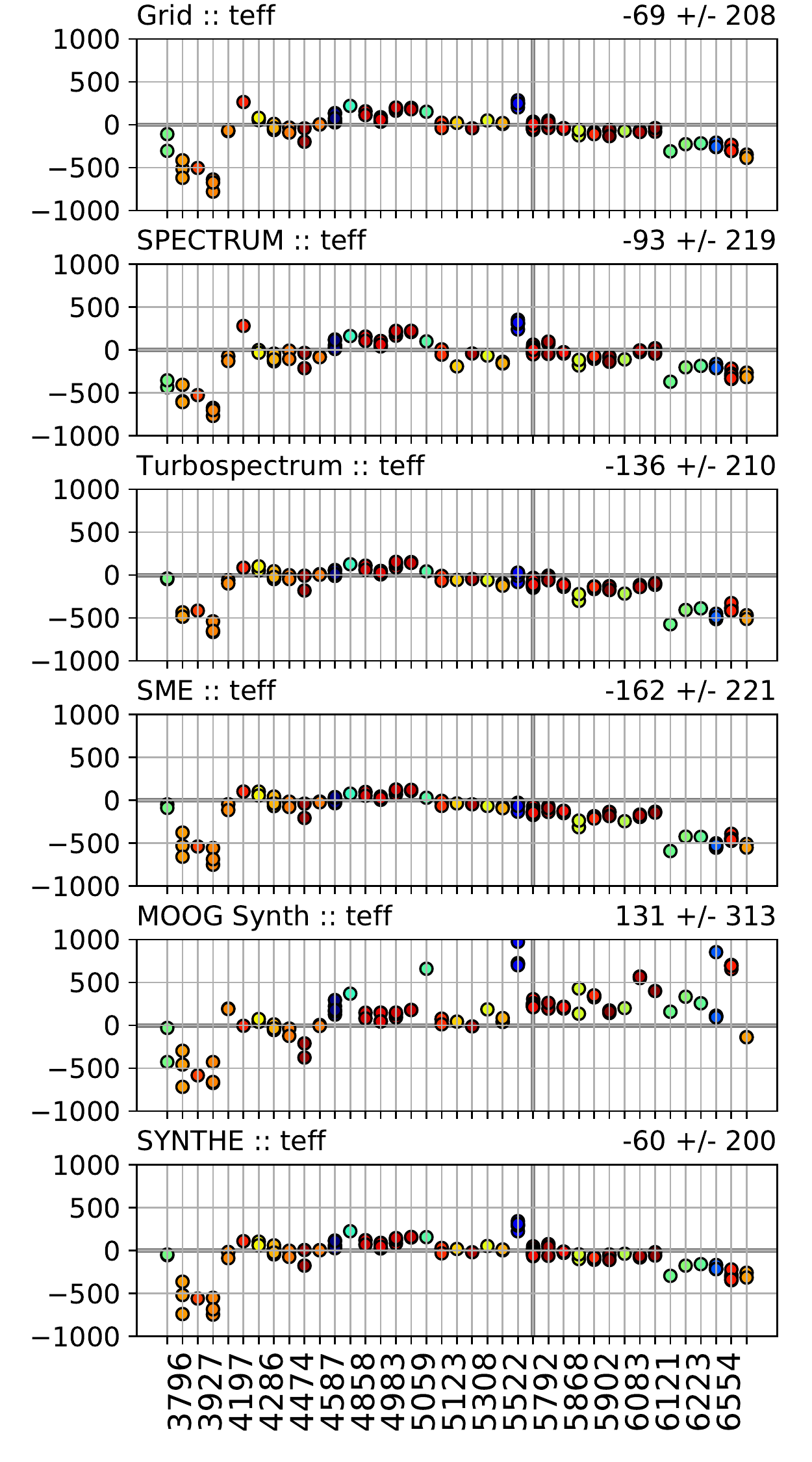}
 \includegraphics[width=5.8cm]{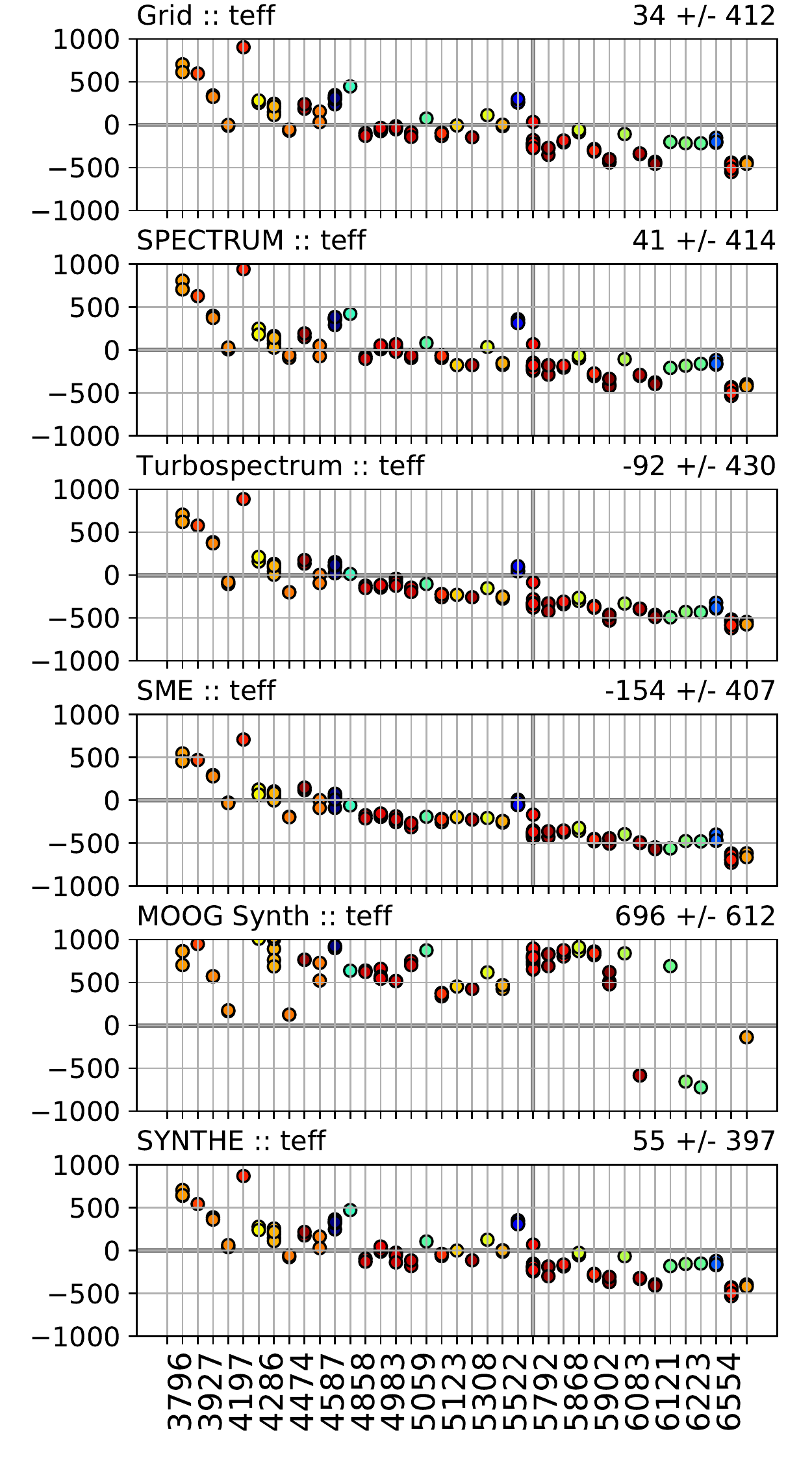}
 \includegraphics[width=5.8cm]{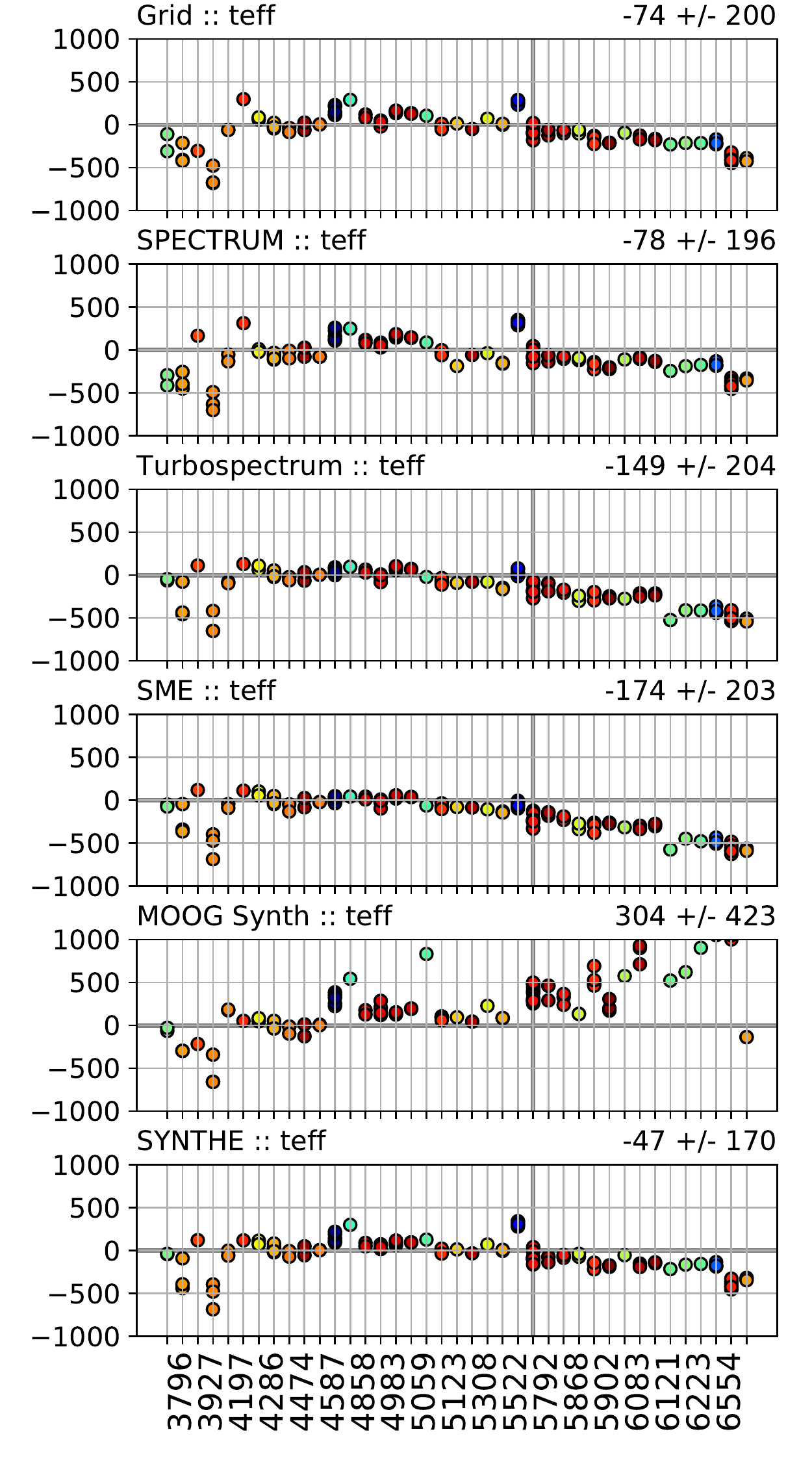}
 \caption{Difference between the derived effective temperature and its reference value for each spectrum analysed with the synthetic spectral-fitting codes when using only the wings of H-$\alpha$ (left subplots), H-$\beta$ (middle subplots) and both of them together (right subplots). The vertical thick grey line denotes the Sun. The colour-coding represents the metallicity of each star (see right subplots in Fig.~\ref{fig:ew_accuracy_own_lines} for colour-code interpretation). All the subplots are sorted taking into account the reference effective temperature. The median and absolute median deviation are indicated in the upper right of each subplot.}
 \label{fig:accuracy_teff_hydrogen}
\end{figure*}

\begin{table}
    \begin{center}
        \tabcolsep=0.08cm
        \begin{tabular}{l|c|c|c|c|c|c}

    &   \scriptsize Grid    &   \scriptsize SPECTRUM    &   \scriptsize Turbospec.   &   \scriptsize SME &   \scriptsize MOOG    &   \scriptsize SYNTHE   \\
\hline
\scriptsize H-$\alpha$    &   -30$\pm$27  &   -6$\pm$30  &   -105$\pm$29  &   -138$\pm$26  &   239$\pm$22  &  -21$\pm$31     \\
\hline
\scriptsize H-$\beta$    &   -93$\pm$45  &   -79$\pm$46  &   -185$\pm$46  &   -237$\pm$49  &   367$\pm$85  &  -87$\pm$46     \\
\hline                                          
\scriptsize H-$\alpha$+H-$\beta$    &   -217$\pm$69  &   -185$\pm$69  &   -322$\pm$65    &   -381$\pm$59  &   795$\pm$128  &   -188$\pm$71  \\
        \end{tabular}
    \end{center}
    \caption{Median and absolute median deviation for differences between effective temperatures derived using the wings of hydrogen lines and the solar reference value.}
    \label{tab:solar_teff_from_hydrogen}
\end{table}

Determining the effective temperatures using the wings of the hydrogen lines is a recognized strategy \citep{2014dapb.book.....N, 2011A&A...531A..83C}, and is understood to be very difficult \citep{2002A&A...385..951B}. In this context, \cite{2018arXiv181112274G} showed how normalization plays a major role and identified a systematic of 28~K for the Sun when using H-$\alpha$. Given the differences found in this work for the considered radiative transfer codes, I executed an extra analysis in which I determined the effective temperature for all the Gaia Benchmark Stars using the wings of H-$\alpha$ and H-$\beta$ separately and together, while the rest of parameters were fixed to their reference values. The results are shown in Fig.~\ref{fig:accuracy_teff_hydrogen}. Hβ is the worse modelled line of the two, as shown in the middle subplots; however, when combined with H-$\alpha$ (right subplots) the results improve or remain similar for all of the codes except MOOG. Turbospectrum, SME and MOOG show the largest systematics, and this may be the reason why adding these regions to the different analyses presented in this work does not improve the overall results. If I limit the validation to the solar spectra, the closest effective temperature to the reference value is obtained when using only H-$\alpha$ with Grid, SPECTRUM and SYNTHE, as shown in Table~\ref{tab:solar_teff_from_hydrogen}.

\begin{figure*}
 \includegraphics[width=16cm]{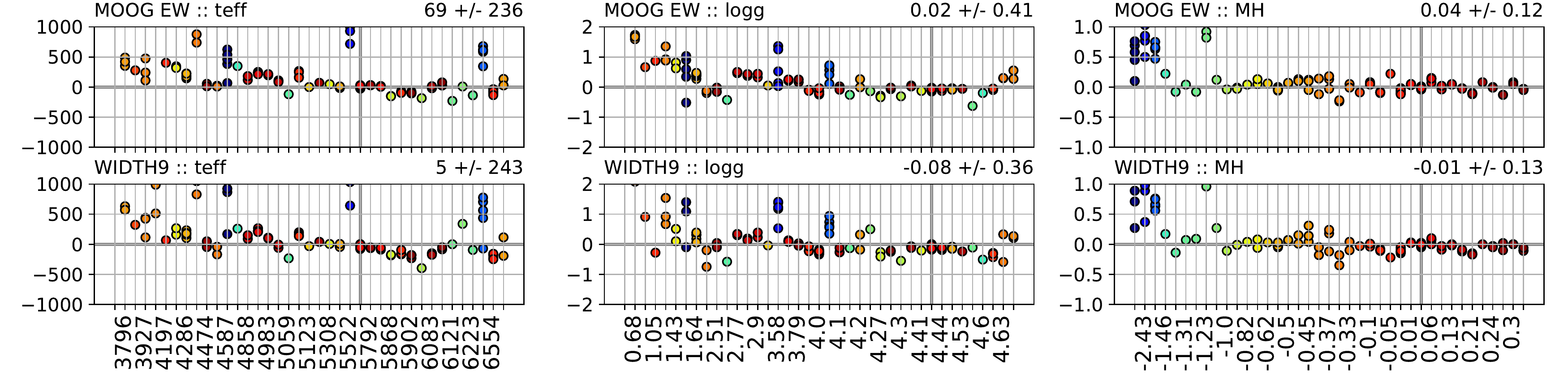}
 \caption{Difference between the derived parameter and its reference value for each spectra analyzed with the equivalent method width when using their best line selection (i.e., own lines). The vertical thick gray line denotes the Sun. The color coding represent the metallicity of each star. All the subplots are sorted taking into account the reference value of the corresponding atmospheric parameter. Median and absolute median deviation are indicated on the upper right of each subplot.}
 \label{fig:ew_accuracy_own_lines}
\end{figure*}

\begin{figure*}
 \includegraphics[width=\linewidth]{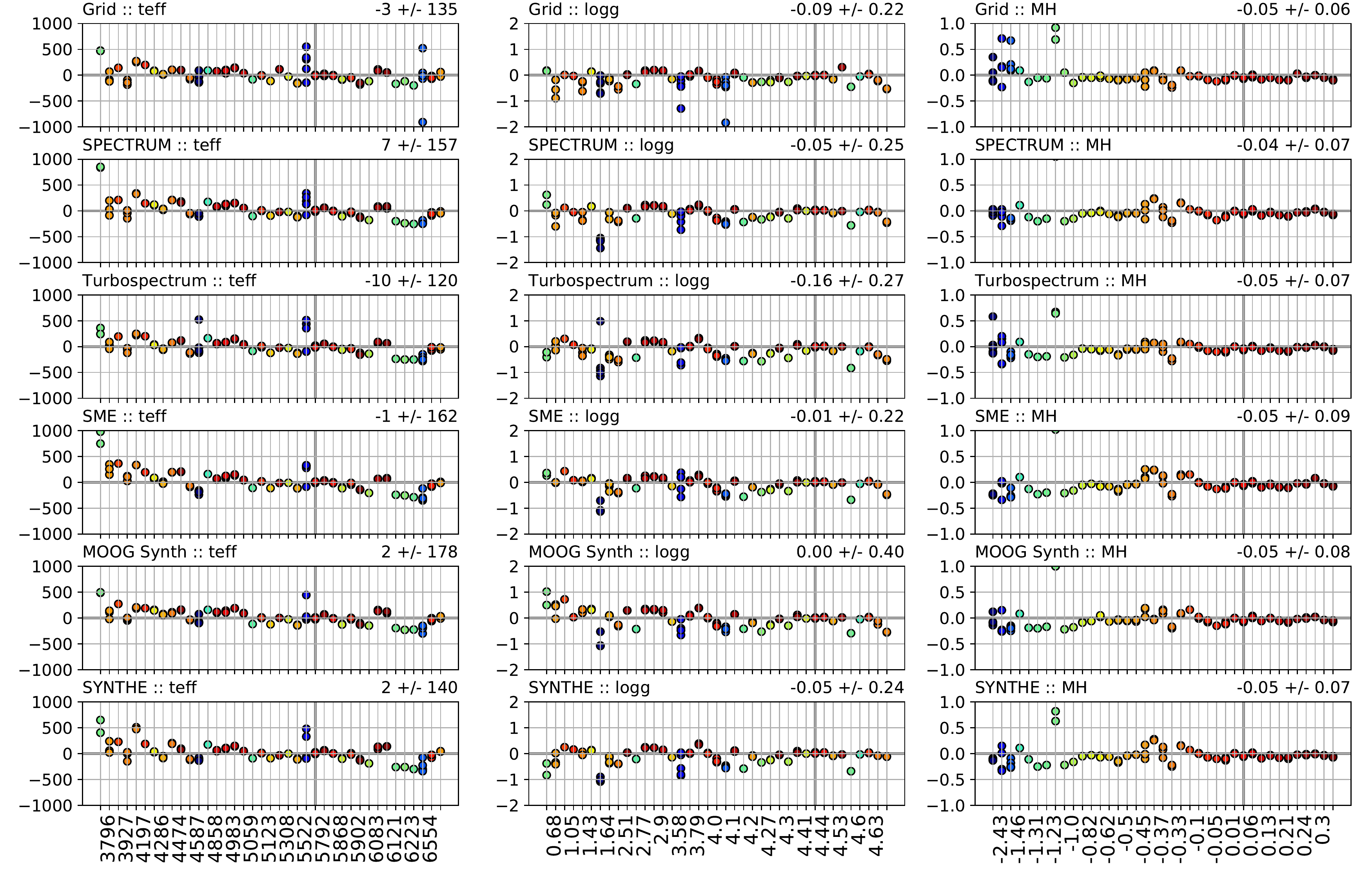}
 \caption{As Fig.~\ref{fig:ew_accuracy_own_lines}, but for spectra analyzed with the synthetic spectral-fitting technique.}
 \label{fig:synth_accuracy_own_lines}
\end{figure*}

\begin{figure}
 \includegraphics[width=4.6cm]{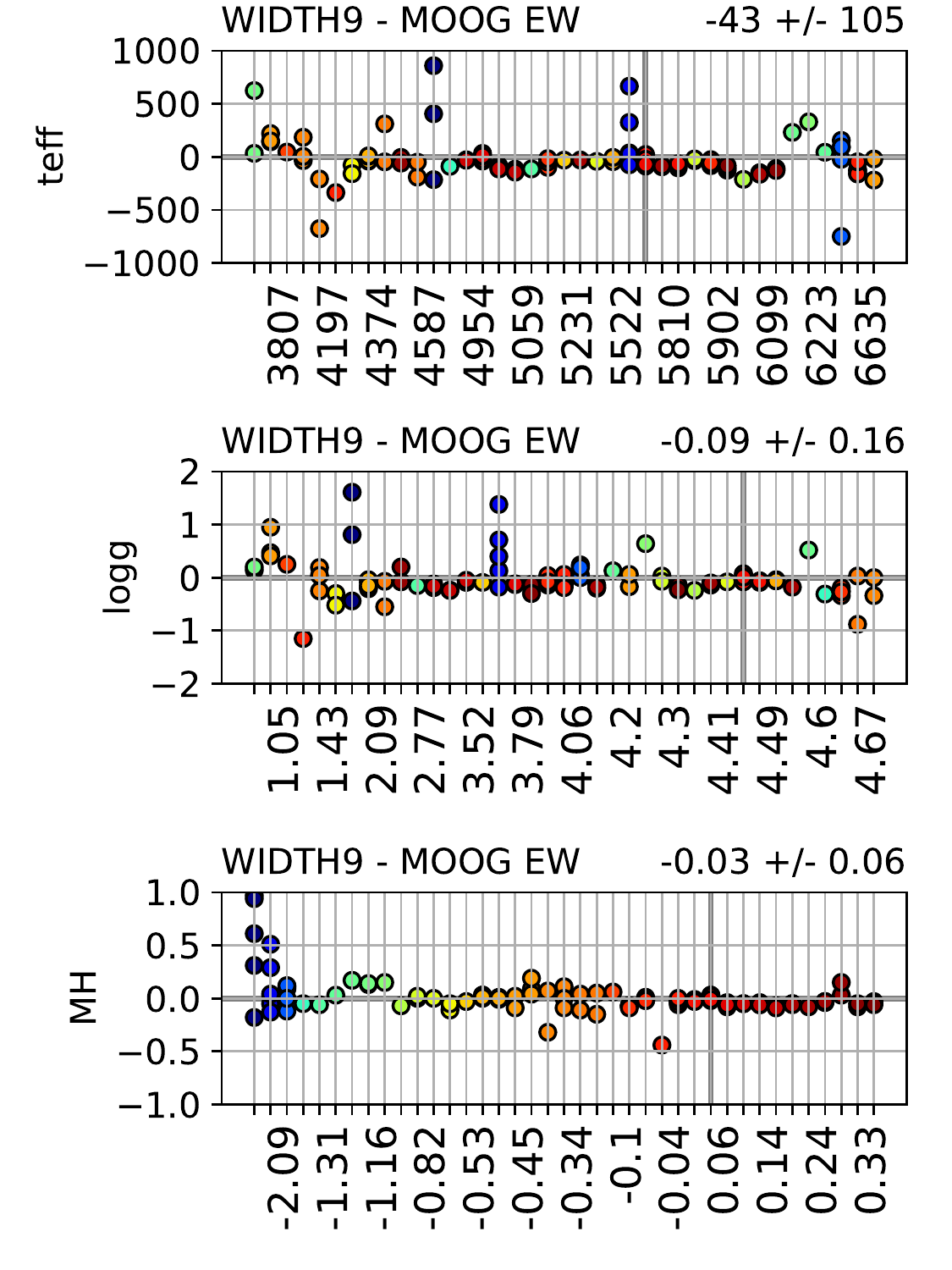}
 \includegraphics[width=3.6cm, trim={2.5cm 0 0 0}, clip]{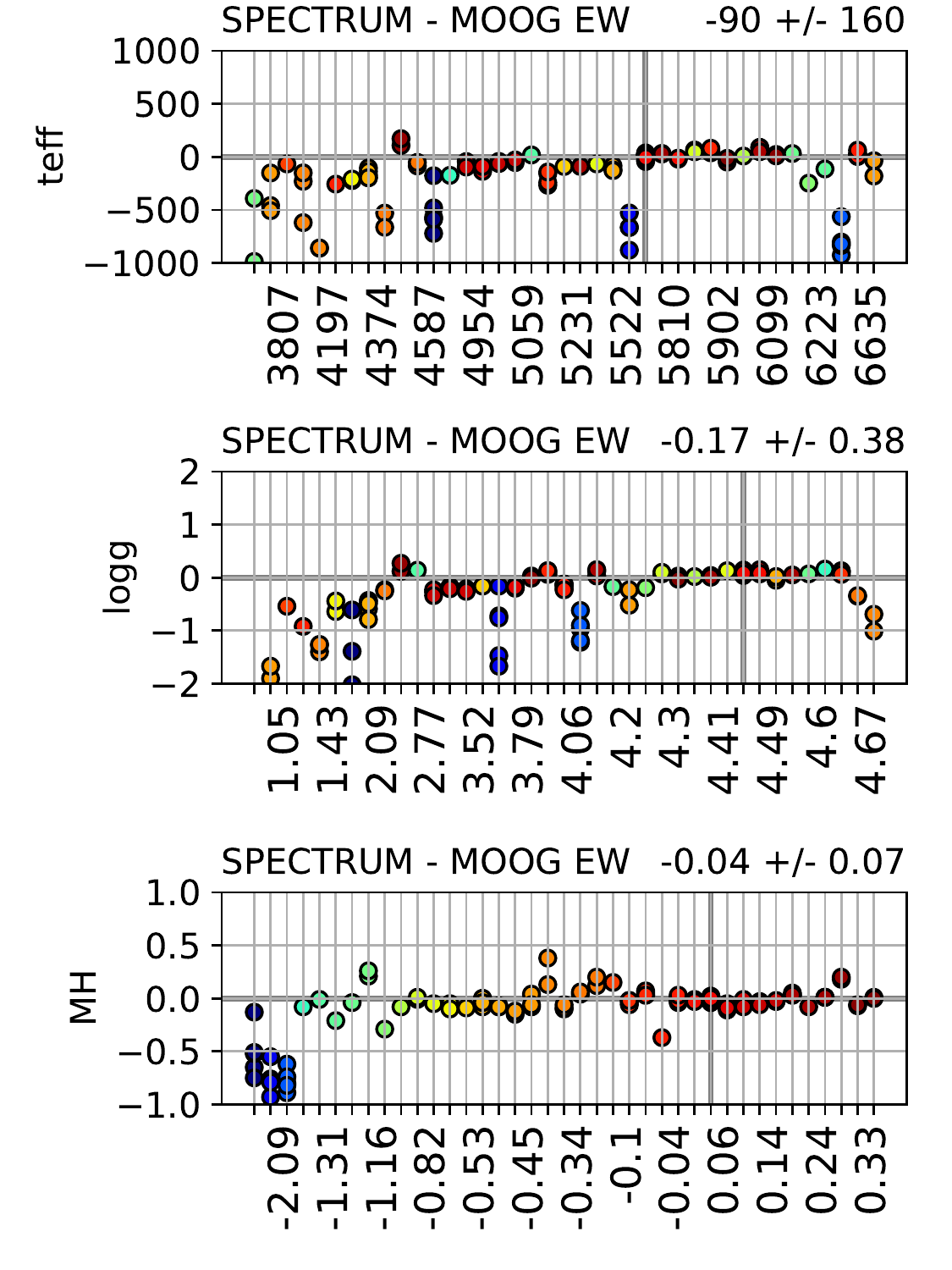}
 \caption{Differences between the two equivalent widths codes (left subplots), and one code from each method (right subplots) when using their best line selection (i.e. own lines). The vertical thick grey line denotes the Sun. The colour-coding represents the metallicity of each star. All the subplots are sorted taking into account the reference value of the corresponding atmospheric parameter. The median and absolute median deviation are indicated in the upper right of each subplot.}
 \label{fig:ew_synth_precision_own_lines}
\end{figure}

\begin{figure*}
 \includegraphics[width=\linewidth]{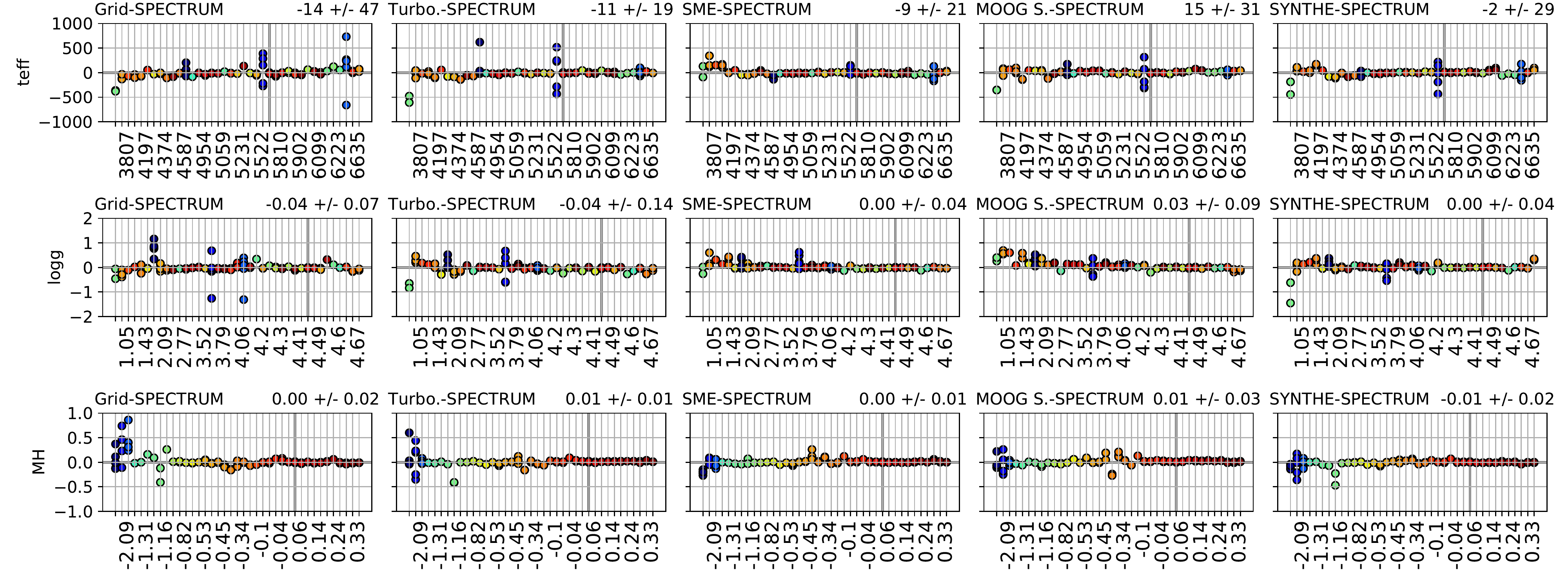}
 \caption{As Fig.~\ref{fig:ew_synth_precision_own_lines}, but only for the synthetic spectral-fitting technique.}
 \label{fig:synth_precision_own_lines}
\end{figure*}

The best global results (i.e. considering all the methods/codes) are obtained using each code's own line selection. I used these results to assess the accuracy and precision as a function of effective temperature, surface gravity and metallicity. Figs~\ref{fig:ew_accuracy_own_lines} and \ref{fig:synth_accuracy_own_lines} illustrate accuracies by comparing the results to the reference values, and they show that the biggest disagreements tend to happen with cold and/or metal-poor stars, which is especially significant for the equivalent-width method, as explained in the next section. In terms of agreement between the two equivalent-width codes (see the left subplots in Fig.~\ref{fig:ew_synth_precision_own_lines}), discrepancies seem to be mainly influenced by the stellar metallicity, although higher discrepancies also appear for lower and higher effective temperatures and surface gravities. Metallicity also affects the level of agreement between synthetic spectral-fitting codes (see Fig.~\ref{fig:synth_precision_own_lines}), and certain code pairs also show higher disagreements for giant stars and the coolest star. When one code from each method is compared, as shown in the right subplots in Fig.~\ref{fig:ew_synth_precision_own_lines}, the highest discrepancies are found for stars with low temperatures, gravities or metallicities.

\subsubsection{Limited Gaia Benchmark Stars data set}
\label{sec:impact_on_ap_limited}

\begin{figure*}
 \includegraphics[width=\columnwidth]{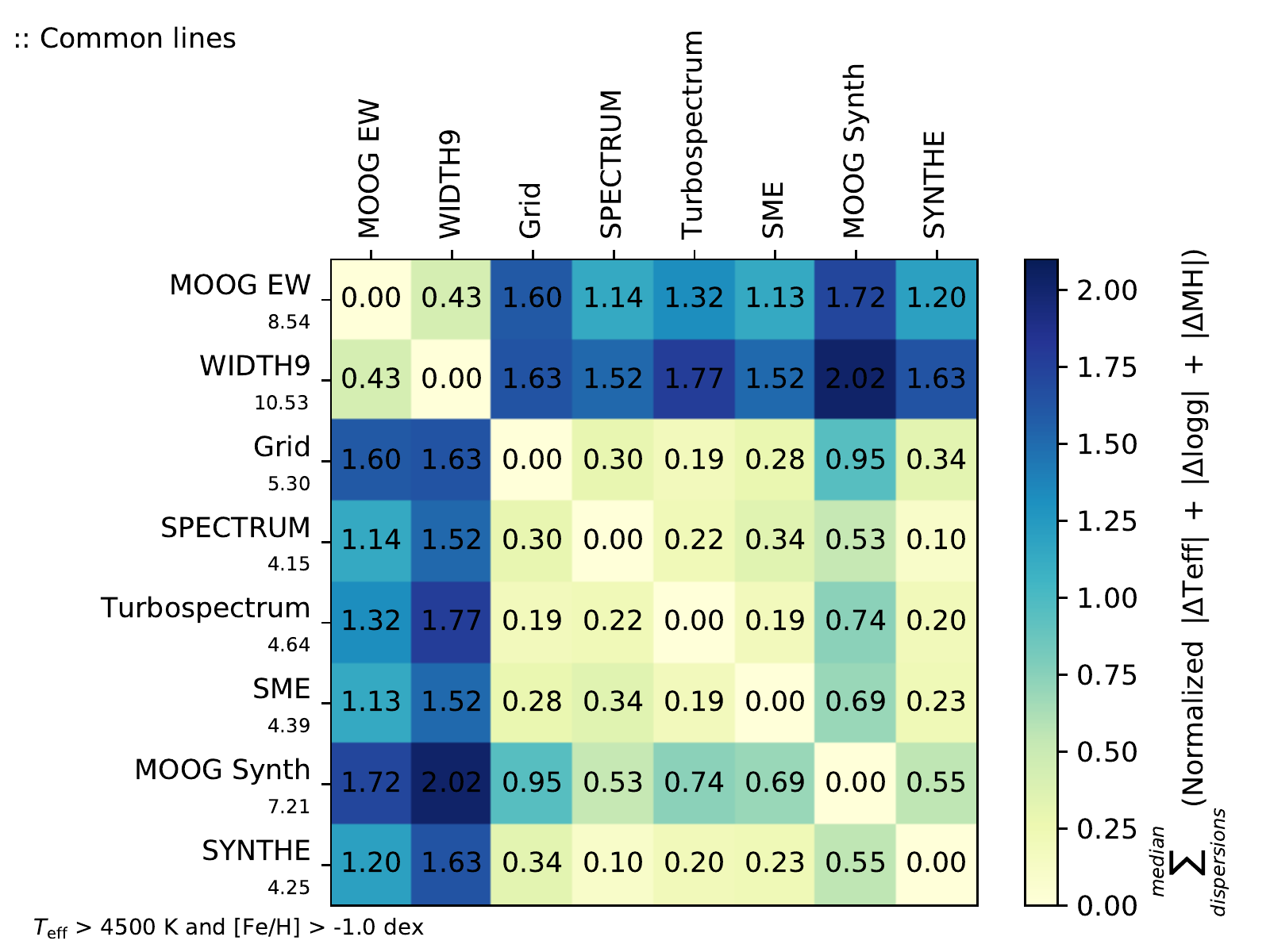}
 \includegraphics[width=\columnwidth]{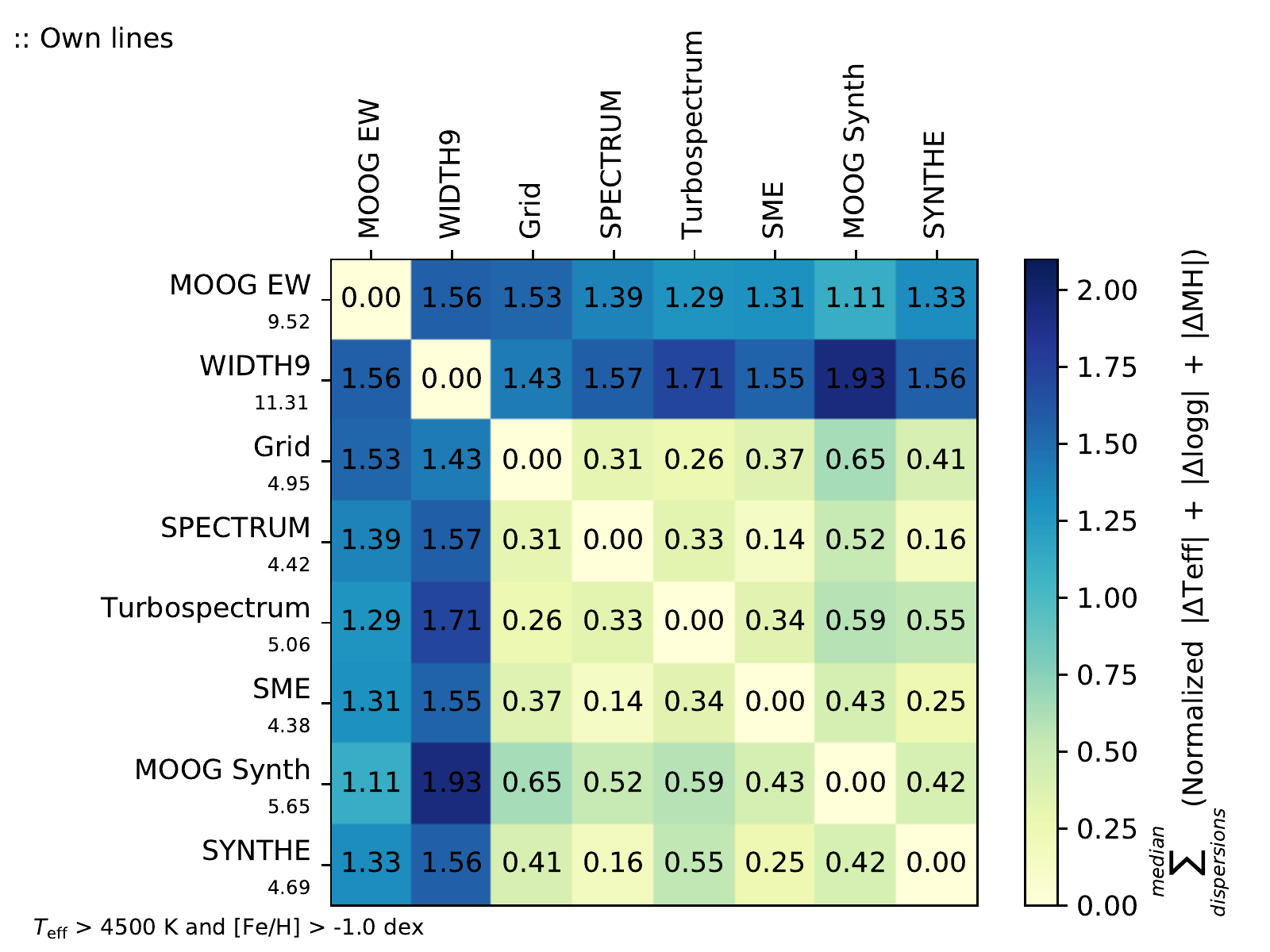}
 \caption{As Fig.~\ref{fig:best_precision}, but considering only Gaia Benchmark Stars with effective temperatures greater than 4\,500~K and metallicities greater than -1.0~dex.}
 \label{fig:best_precision_ew_range_limited}
\end{figure*}

The accuracy of the equivalent-width method can be affected by the presence of strongly blended lines and by the lack of enough observed iron lines, while the synthetic spectral-fitting technique is more lenient. Hence, the cooler and/or more metal-poor Gaia Benchmark Stars are challenging targets for the equivalent-width method, and to account for this I repeated the previous assessment but selecting only the derived atmospheric parameters for the Gaia Benchmark Stars that have a reference effective temperature greater than 4\,500~K and a metallicity higher than -1.0~dex. In Fig.~\ref{fig:best_precision_ew_range_limited}, I show the added normalized median differences and robust standard deviation between codes (i.e. precision) using the common and their own line selections. Interestingly, the similarities between codes are in line with what was observed in Fig.~\ref{fig:best_precision}, but in this case it is MOOG EW that comes closer to the synthesis codes when using their own line selection instead of WIDTH9.

\begin{figure}
 \includegraphics[width=\linewidth]{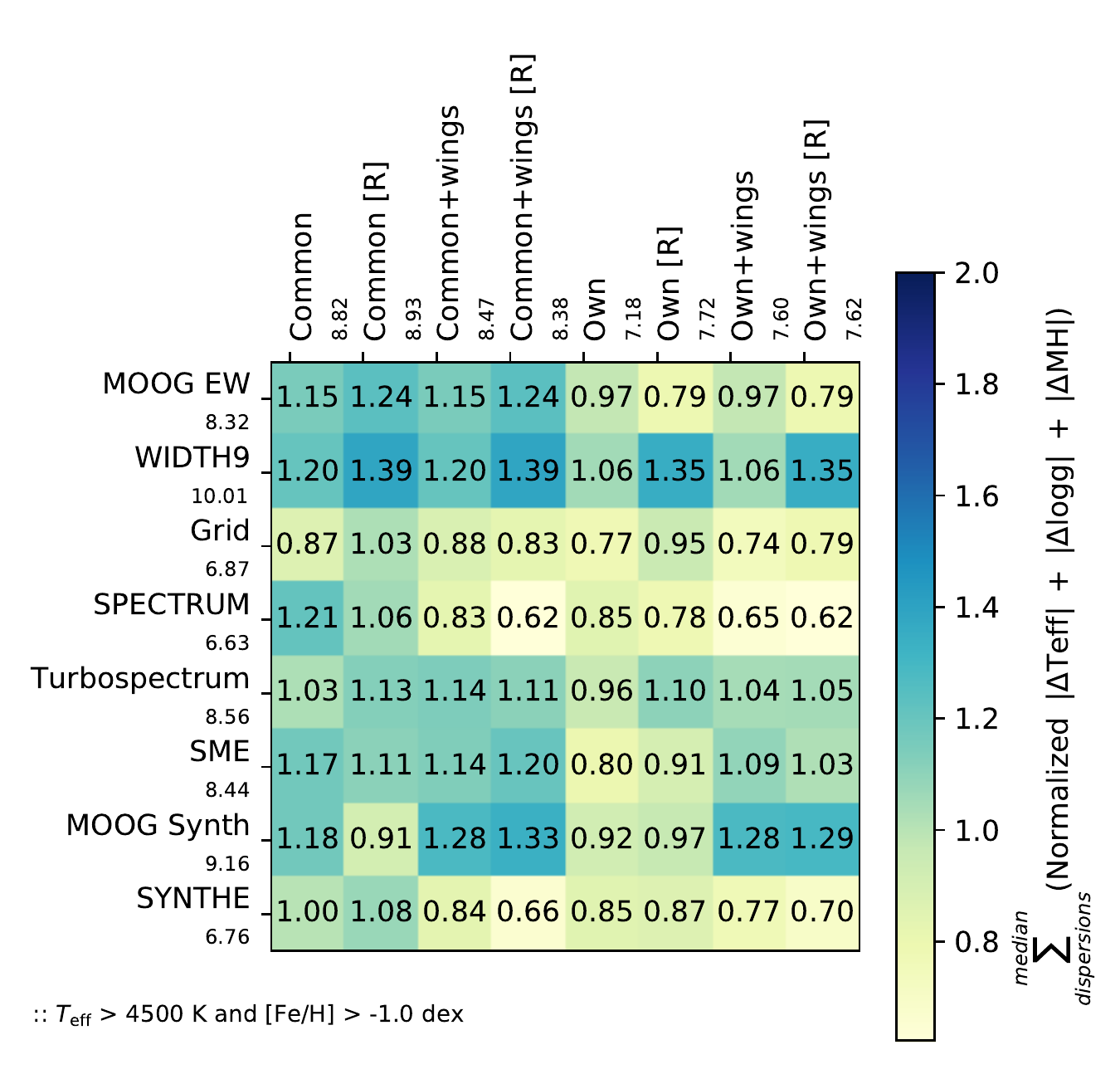}
 \caption{As Fig.~\ref{fig:best_accuracy}, but considering only Gaia Benchmark Stars with effective temperatures greater than 4\,500~K and metallicities greater than -1.0~dex.}
 \label{fig:best_accuracy_ew_range_limited}
\end{figure}

Regarding the accuracy of the codes depending on their set-up, Fig.~\ref{fig:best_accuracy_ew_range_limited} shows a similar pattern to Fig.~\ref{fig:best_accuracy}, except that the equivalent-width results are significantly improved and become more accurate when this limited subset of the Gaia Benchmark Stars is considered, with MOOG EW the best code for equivalent width.

\subsection{Impact on chemical abundances}
\label{s:abundances}

\subsubsection{Full Gaia Benchmark Stars data set}
\label{s:abundances_full}

\begin{figure*}
 \includegraphics[width=\columnwidth]{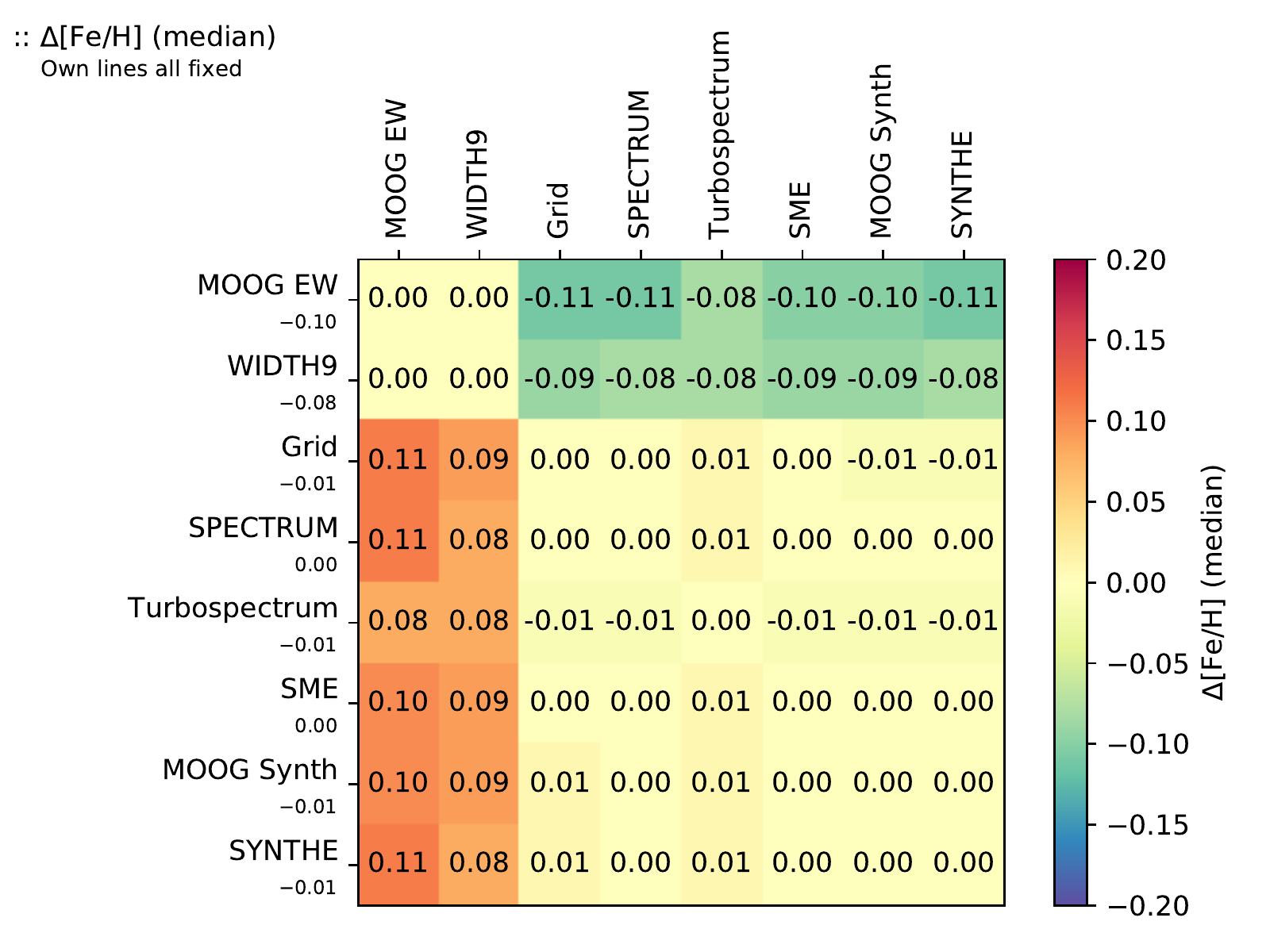}
 \includegraphics[width=\columnwidth]{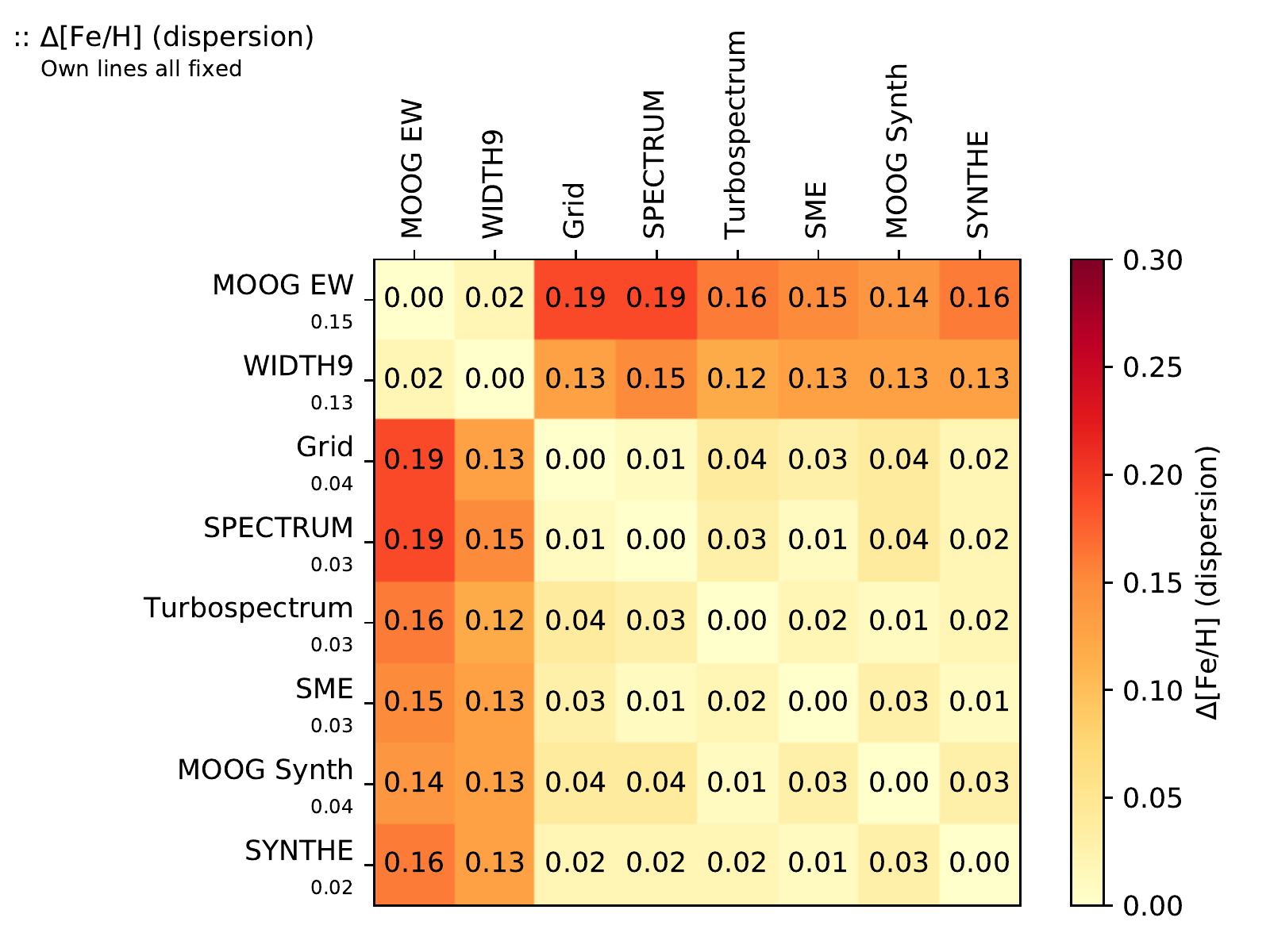}
 \caption{Median and robust standard deviation of the difference in iron abundance between different radiative transfer codes when analysing the Gaia Benchmark Stars and fixing all the atmospheric parameters to the reference ones and using the best line selection for each code.}
 \label{fig:precision_Fe}
\end{figure*}

Individual chemical abundances were derived by fixing the atmospheric parameters to the reference values and using the best line selection for each method. The iron abundances tend to be used as a proxy for metallicity, and indeed Fig.~\ref{fig:precision_Fe} shows very similar patterns to Fig.~\ref{fig:precision_teff_logg_MH} (bottom plot), where equivalent-width codes and synthesis codes form separate islands. However, the dispersion is worse in Fig.~\ref{fig:precision_Fe} for the equivalent-width codes. Imposing the same atmospheric parameters on all the spectra probably worsens the results because, as shown in Section~\ref{sec:impact_on_ap}, the microturbulence velocity does not have exactly the same effect for all the methods. The rest of the parameters may play a role too, and the abundance determination may compensate for the discrepancies from other parameters when enforcing a certain temperature or gravity that does not match what our analysis would have found with our models, codes and set-up.

\begin{figure}
 \includegraphics[width=\columnwidth]{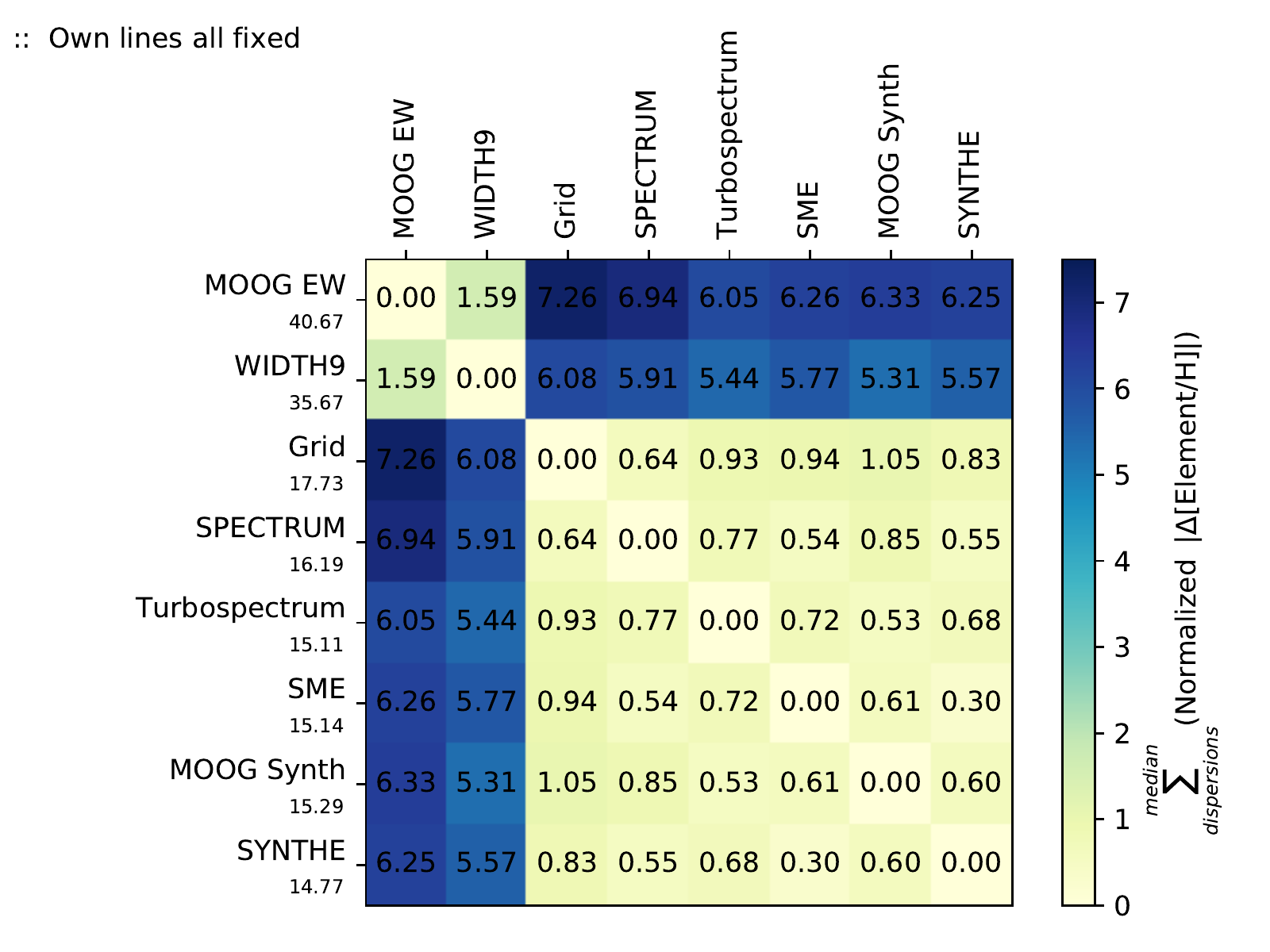}
 \caption{Sum of the normalized absolute median differences and robust standard deviation for iron, calcium, cobalt, chromium, magnesium, manganese, nickel, silicon, titanium and vanadium abundances when analysing the Gaia Benchmark Stars. Lower numbers indicate the codes lead to more similar results (higher precision).}
 \label{fig:best_abundance_precision}
\end{figure}

In order to compare visually the precision for all the analysed elements simultaneously among all the codes, I added the normalized differences and dispersion and represent them in Fig.~\ref{fig:best_abundance_precision}. The results indicate that the pattern observed for iron abundances can be generalized for the rest of the elements.

\begin{figure*}
 \includegraphics[width=\columnwidth]{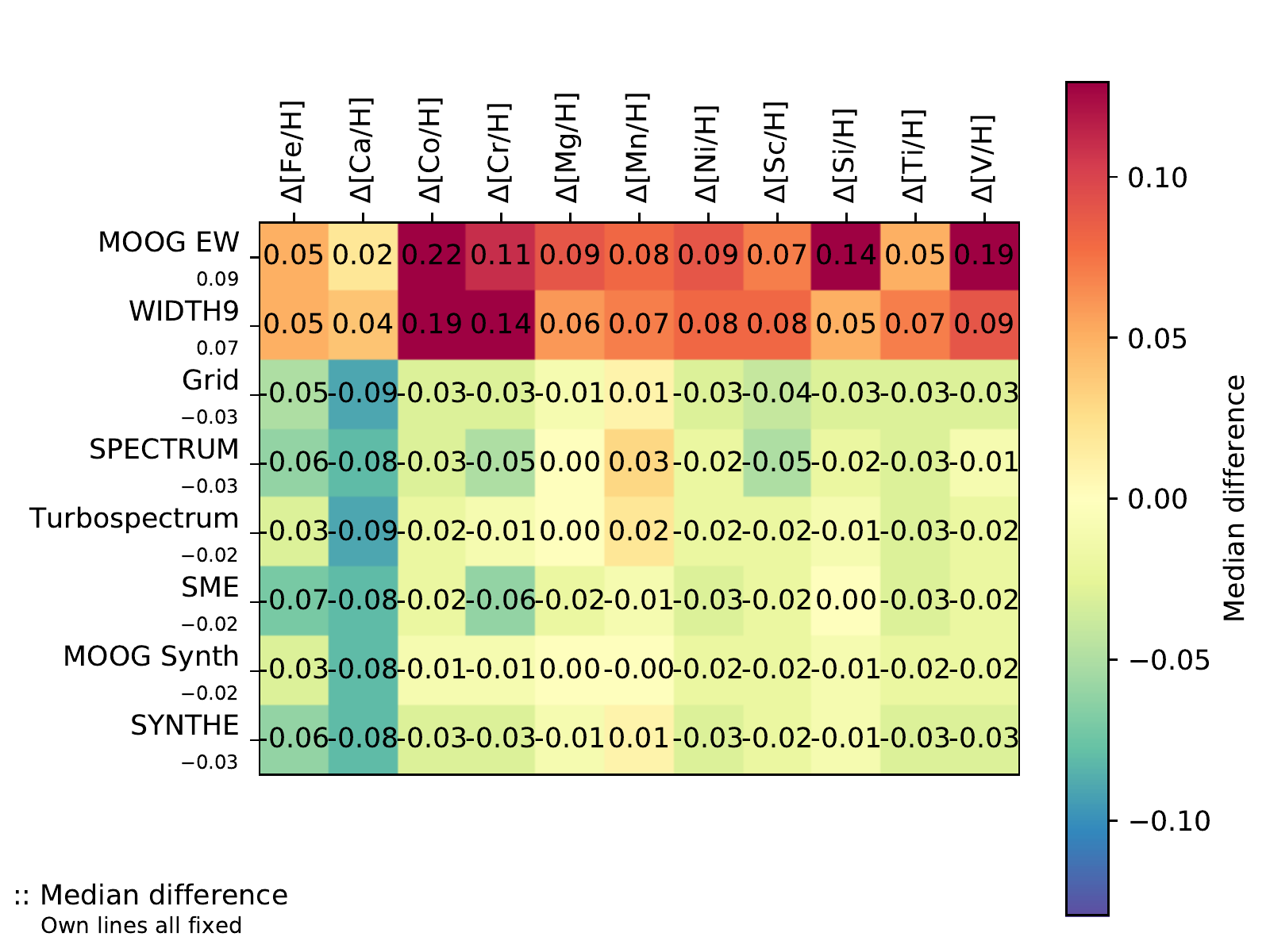}
 \includegraphics[width=\columnwidth]{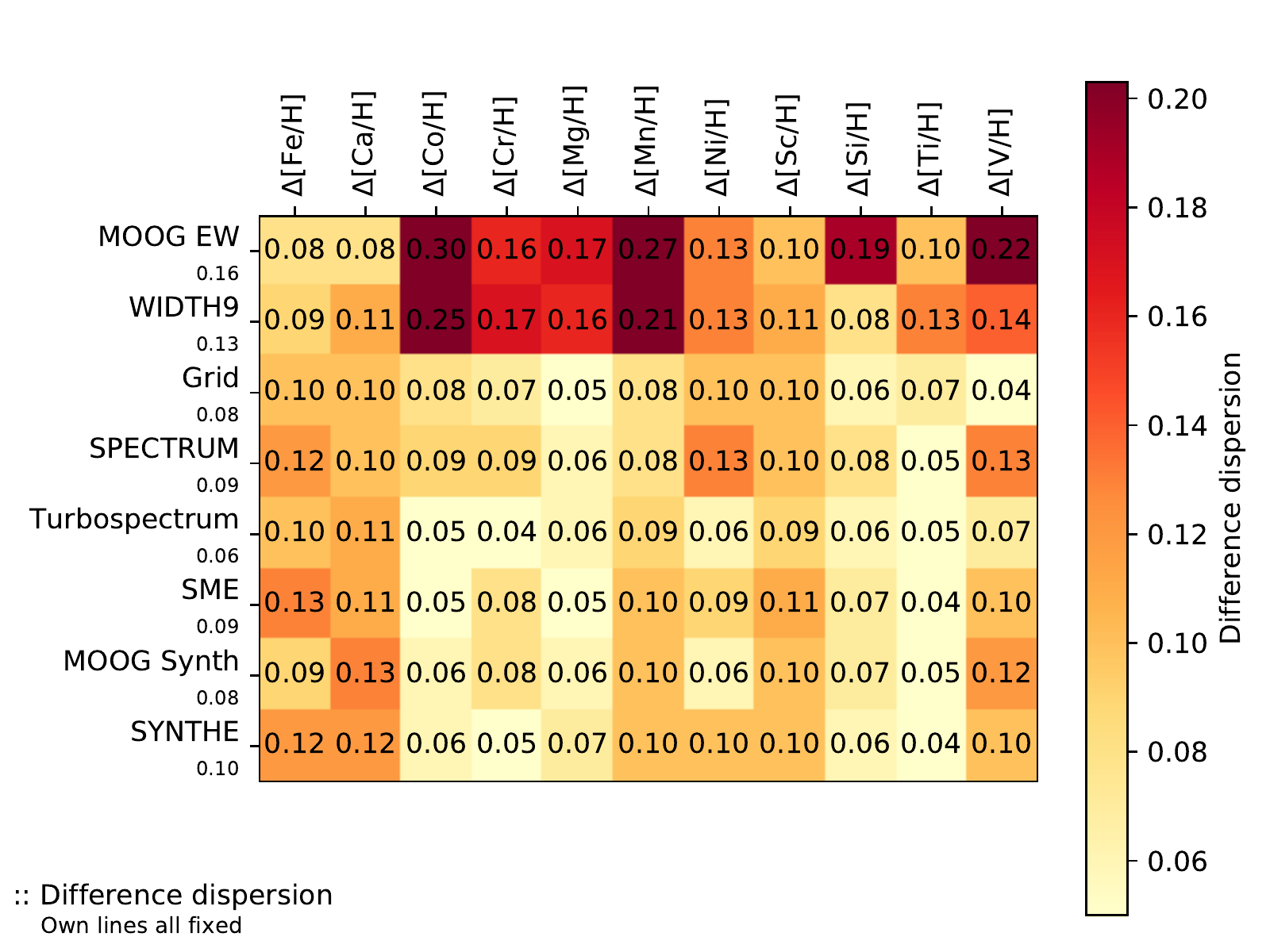}
 \caption{Median and robust standard deviation of the difference in individual abundances between the reference values and different radiative transfer codes when fixing all the atmospheric parameters to the reference ones when analysing the Gaia Benchmark Stars. The best line selection for each code was used. The median dispersion per star is equal or below 0.02~dex for all the cases.}
 \label{fig:accuracy_abundances}
\end{figure*}

Regarding the accuracy of all the derived individual chemical abundances with fixed atmospheric parameters (Fig.~\ref{fig:accuracy_abundances}), equivalent-width methods outperform synthesis methods for calcium abundances but not for the rest of the elements. It is worth remembering that the reference chemical abundances were determined by combining spectroscopic results obtained by different groups using their own techniques (abundances cannot be obtained independently from spectroscopy, in contrast to the effective temperature and surface gravity). Thus, depending on how many different methods were used, how many stars and elements each group analysed, how results were combined, and how outliers were treated or removed \citep{2015A&A...583A..94A}, the reference values may be biased towards one method or code, depending on the element. For instance, when considering calcium abundances derived with Turbospectrum but using only lines in common with the groups ULB and GAU from \citealt{2015A&A...582A..81J} (which used synthesis with Turbospectrum) then the median difference and robust standard deviation go down to $-0.03\pm0.08$ and $-0.01\pm0.06$, respectively, from the original $-0.09\pm0.11$~dex. Similarly, there are several elements for which equivalent-width codes strongly underperform. For instance, if I consider cobalt results with MOOG EW and I use the same spectra and absorption lines in common with the groups EPI, POR and UCM from \citealt{2015A&A...582A..81J} (which used equivalent-width with MOOG) then the median difference and robust standard deviation go to $0.00\pm0.05$ and $0.00\pm0.03$ and $0.00\pm0.02$, respectively, from the original $0.22\pm0.30$~dex.

I also showed in Section~\ref{sec:impact_on_ap} that the microturbulence velocity does not have exactly the same effect for all the methods, thus imposing the same reference value affects each code differently. To assess this effect, I used Grid and SPECTRUM (i.e. interpolating from a grid of spectra and synthesizing) with their best line selection, and I repeated the analysis by first computing the microturbulence velocity and resolution for each star (which will compensate for errors in the reference macroturbulence velocity and rotation) while fixing the rest of the parameters to their reference values, and then calculating the iron abundance using the reference parameters plus the microturbulence and resolution found. I obtained $0.00\pm0.03$ and $-0.02\pm0.02$~dex for Grid and SPECTRUM respectively, which show lower dispersions than the values $-0.05\pm0.10$ and $-0.06\pm0.12$~dex obtained with all the parameters fixed.

Finally, when considering derived abundances instead of line-by-line differential ones (i.e. the derived abundance for a particular absorption line and target star minus the derived abundance for the same absorption line in the reference star, here the Sun), the accuracy of the results and level of agreement between codes significantly worsens. Line-by-line differential analysis helps to reduce systematics, such as the ones presented in Section~\ref{s:line_selection_results}, although the more different the target star is from the reference star, the less effective this strategy is.

\subsubsection{Limited Gaia Benchmark Stars data set}
\label{s:abundances_limited}

\begin{figure}
 \includegraphics[width=\columnwidth]{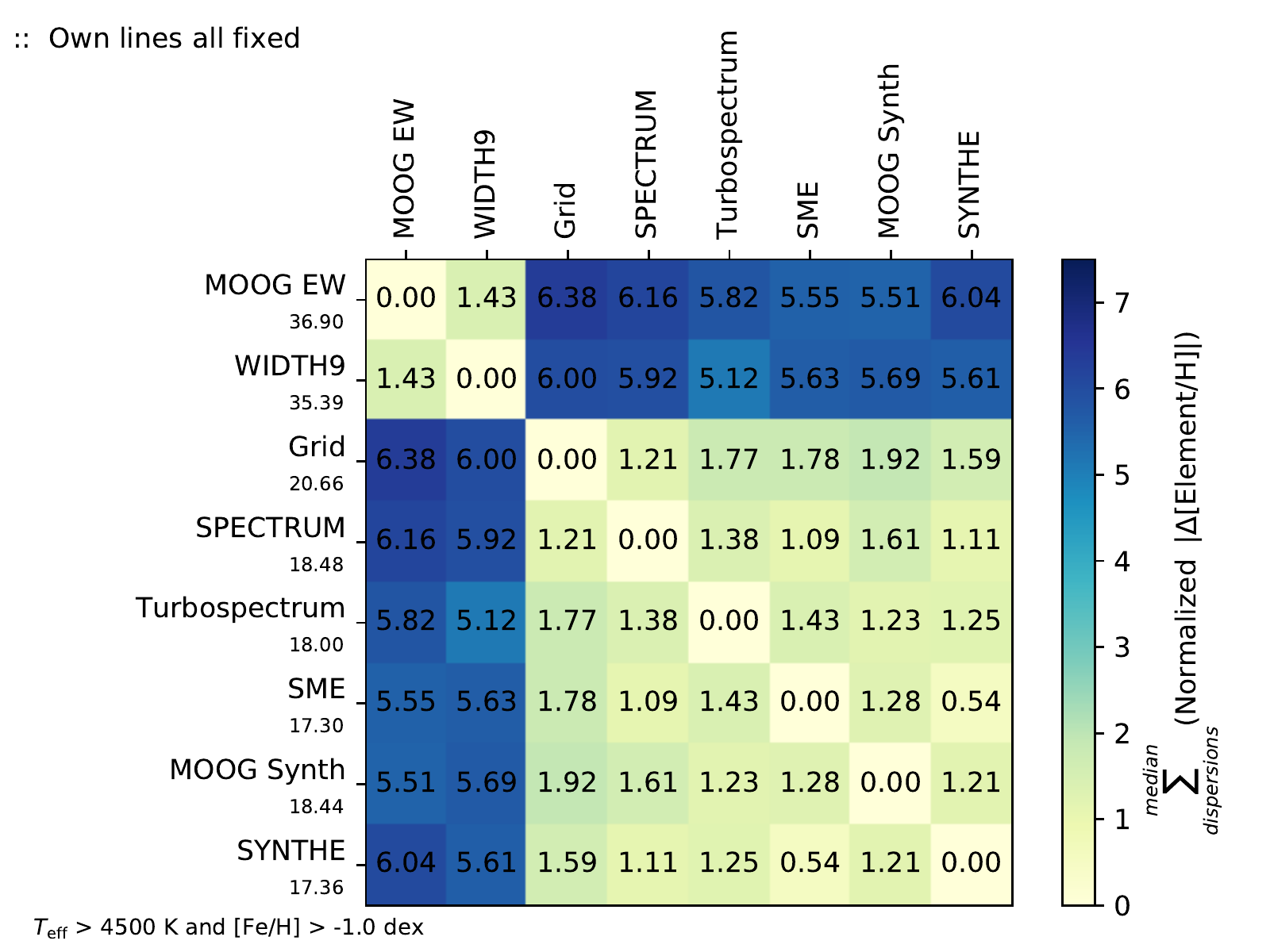}
 \caption{As Fig.~\ref{fig:best_abundance_precision}, but considering only Gaia Benchmark Stars with effective temperatures greater than 4\,500~K and metallicities greater than -1.0~dex.}
 \label{fig:best_abundance_precision_ew_range_limited}
\end{figure}

Similarly to in Section~\ref{sec:impact_on_ap_limited}, in order to explore a more limited region of the parameter space than the Gaia Benchmark Stars cover, I selected individual chemical abundances for stars with reference effective temperature higher than 4\,500~K and metallicity greater than -1.0~dex. Then I added the normalized median difference and robust standard dispersion between codes for all the analysed elements, as shown in Fig.~\ref{fig:best_abundance_precision_ew_range_limited}. Filtering out stars that are less convenient for the equivalent-width method does not erase the systematic differences already observed in Fig.~\ref{fig:best_abundance_precision_ew_range_limited} and described in Section~\ref{s:abundances_full}.

\begin{figure*}
 \includegraphics[width=\columnwidth]{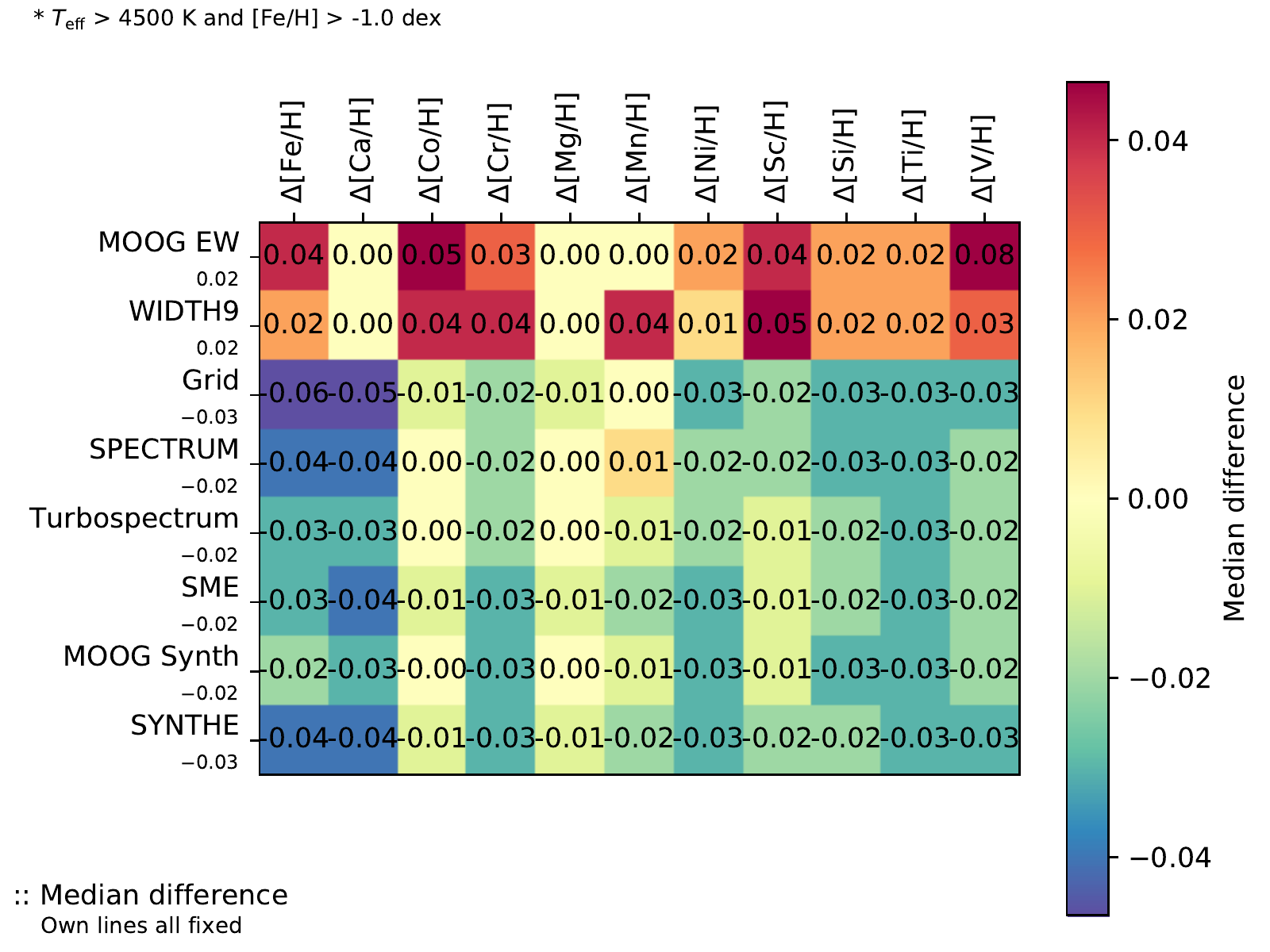}
 \includegraphics[width=\columnwidth]{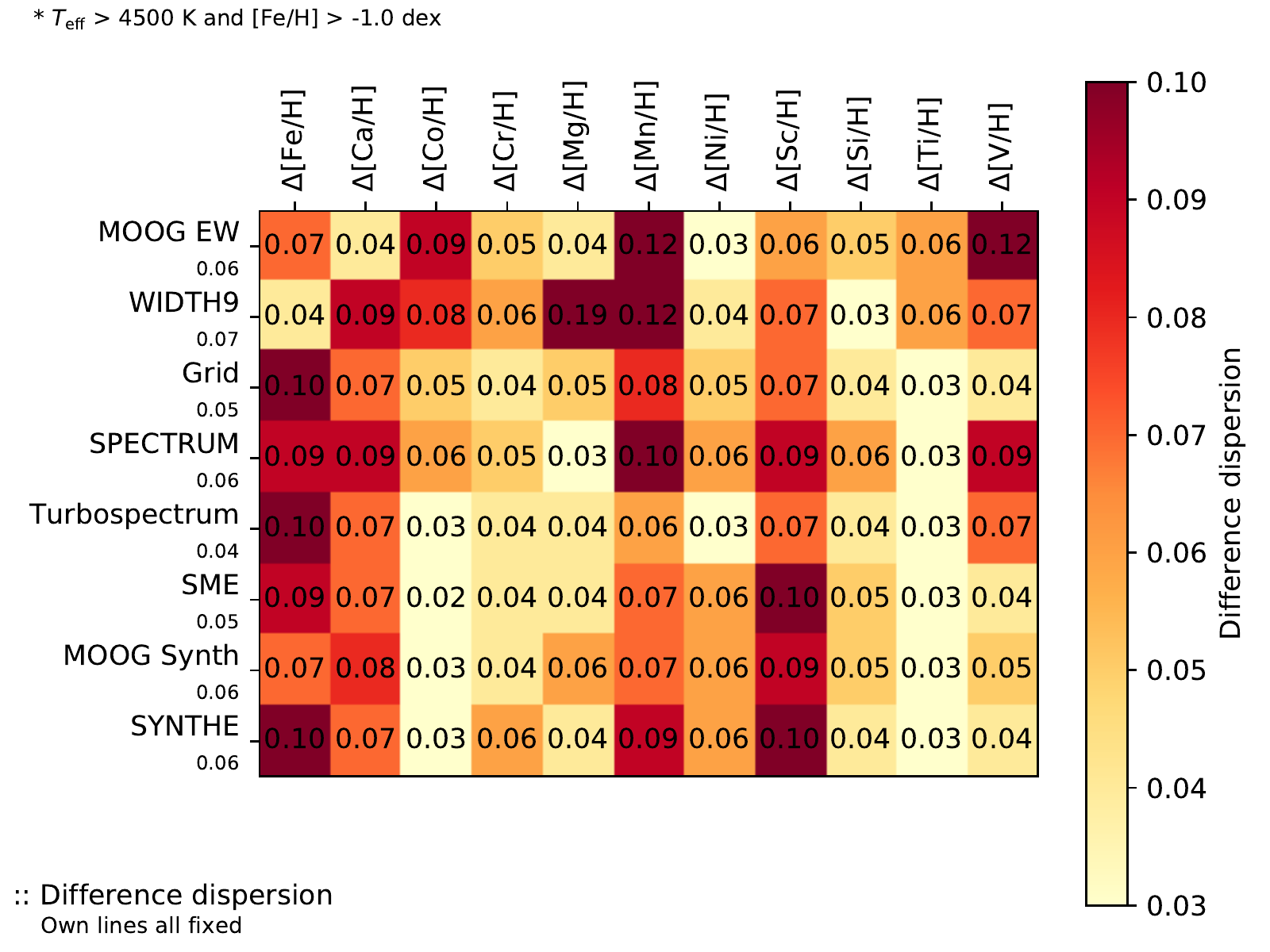}
 \caption{As Fig.~\ref{fig:accuracy_abundances} but considering only Gaia Benchmark Stars with effective temperatures greater than 4\,500~K and metallicities greater than -1.0~dex.}
 \label{fig:accuracy_abundances_ew_range_limited}
\end{figure*}

When comparing the results to the reference values as shown in Fig.~\ref{fig:accuracy_abundances_ew_range_limited}, the median differences and robust standard deviations improve for all the codes compared with Fig.~\ref{fig:accuracy_abundances}, but more significantly for the equivalent-width methods, with MOOG EW obtaining better results than WIDTH9.

\subsection{The non-observed data set experiment}
\label{s:non_observed_dataset_results}

\begin{figure*}
 \includegraphics[width=\columnwidth]{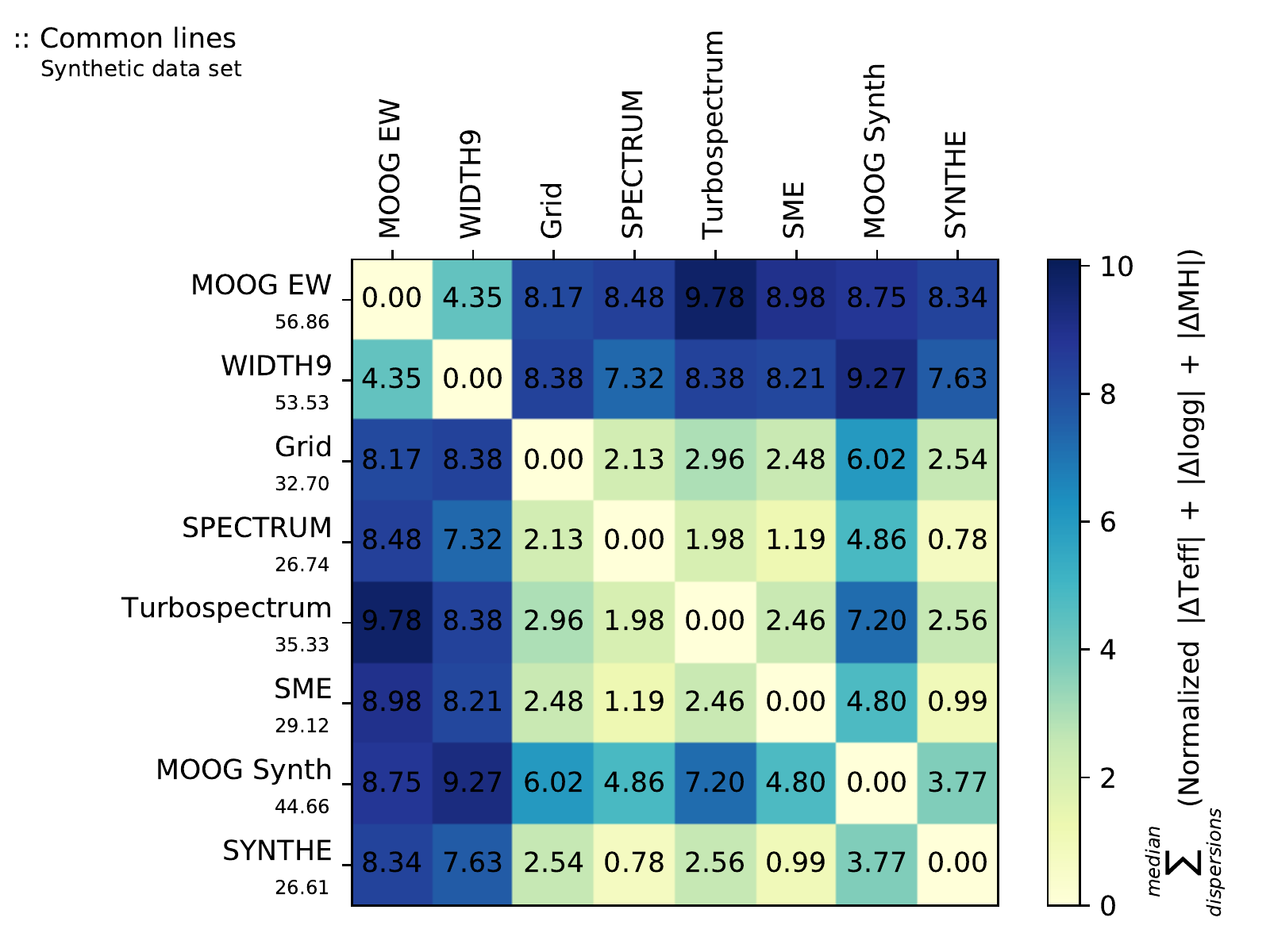}
 \includegraphics[width=\columnwidth]{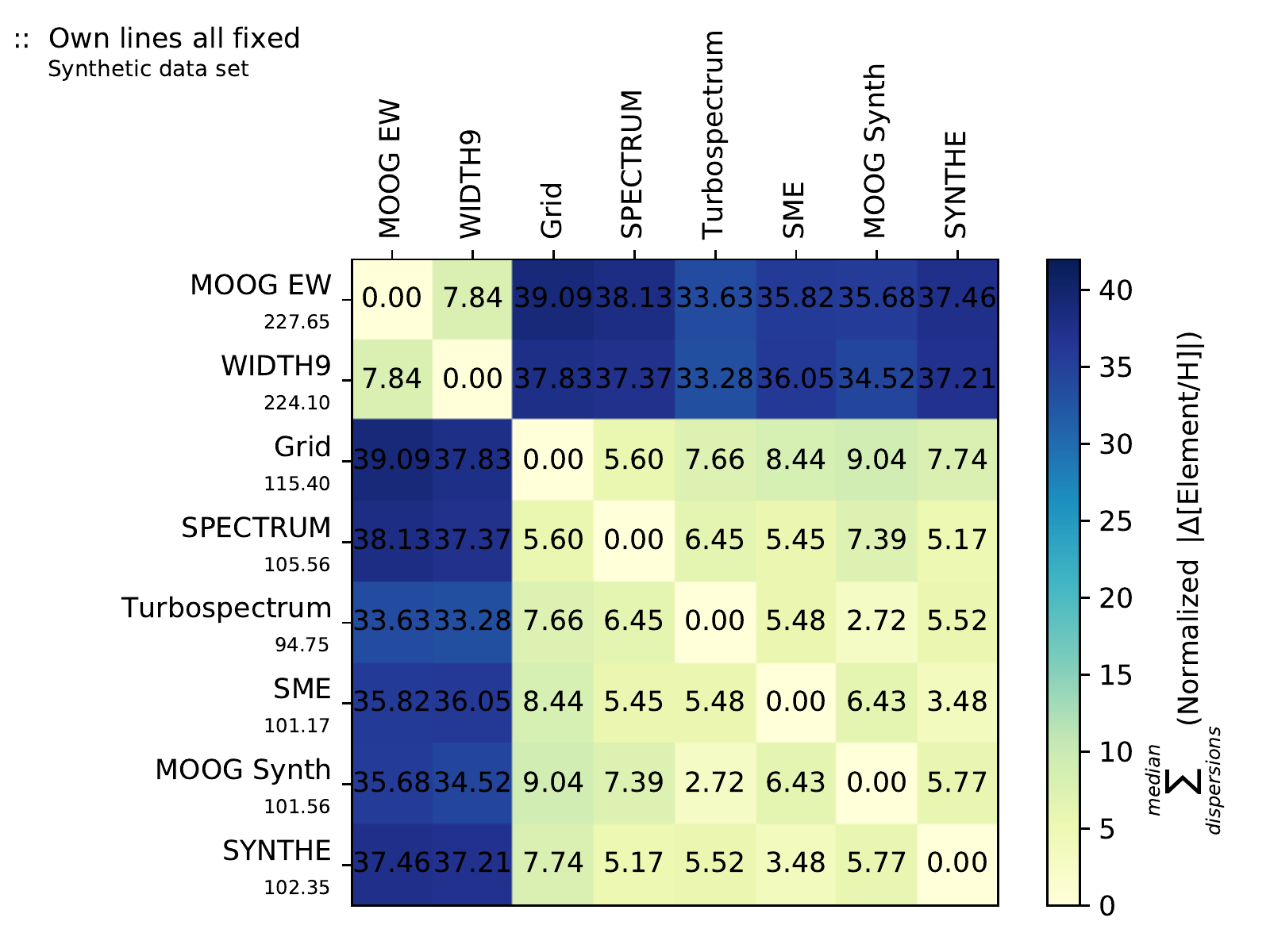}
 \caption{As Fig.~\ref{fig:best_precision} and \ref{fig:best_abundance_precision} (left and right plots, respectively), but using the results from the analysis of the synthetic spectra built for the non-observed data set experiment.}
 \label{fig:best_non_observed_dataset}
\end{figure*}

The analysis of the data set of 672 normalized synthetic spectra (112 spectra synthesized with the five codes included in this analysis plus 112 interpolated with the pre-computed grid) described in Section~\ref{s:non_observed_dataset_experiment} represents a major computational effort, involving hundreds of CPU hours. Apart from the determination of atmospheric parameters, it has led to the determination of more than two million abundances from more than 60\,000 absorption lines analysed with each method and code.

The outcome of this experiment, summarized in Fig.~\ref{fig:best_non_observed_dataset}, reproduces very closely the same main results as obtained using the observed data set. This ensures that the conclusions from this work are not affected by observational biases such as instrument effects, night conditions, raw data processing, or the inability of models to perfectly reproduce physical processes.

\subsection{The one variable at a time experiment}
\label{s:one_variable_at_a_time_results}

The zero-point of this experiment (described in Section~\ref{s:one_variable_at_a_time_experiment}) is set by the atmospheric parameters of the Sun, the MARCS model atmosphere, the GES atomic line list, and the code MOOG for the equivalent-width approach and SPECTRUM for synthesis approach. In all the figures referenced in this section, the zero-point is found at the intersection of the horizontal axis and the thick light grey vertical line. The $x$-axis corresponds to the independent variable being changed (e.g. effective temperature, electron density), and the $y$-axis is the outcome being evaluated, which is very simply computed as

\begin{equation}
f(x) = \textrm{median}\left(v - v_{\textrm{ref}}\right)
\end{equation}

where $v$ is the abundance for equivalent-width codes or the depth for synthesis codes, and $v_{\textrm{ref}}$ is the abundance or depth of reference. The dispersion is computed using the absolute median deviation

\subsubsection{Atmospheric parameters}

\begin{figure}
 \includegraphics[width=\columnwidth]{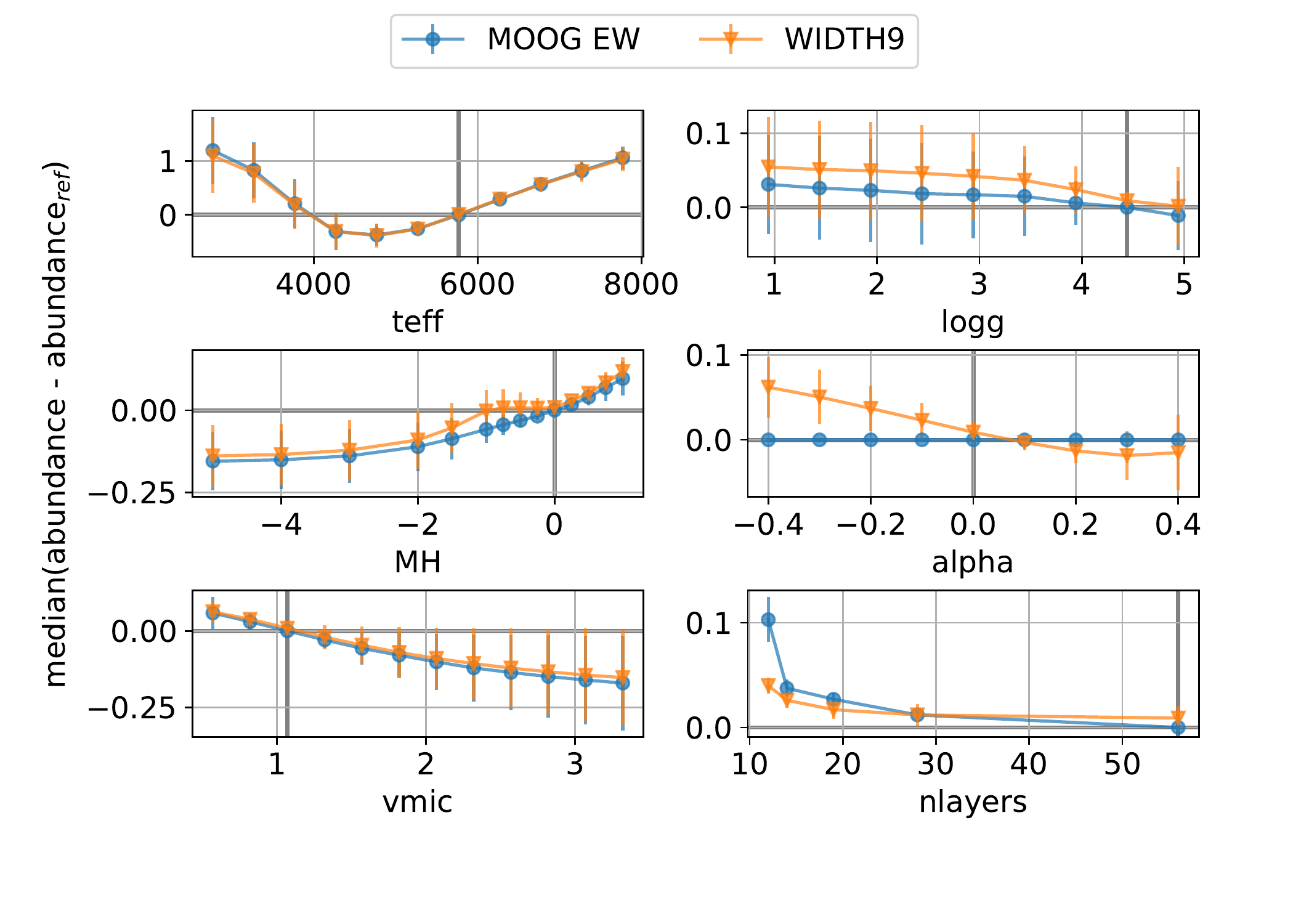}
 \caption{Median and absolute median deviation (error bars) difference in abundance over reference abundance (i.e. MOOG abundances for the Sun, zero-point represented by the intersection of the two thick grey lines) when varying selected atmospheric parameters and setting the rest to the reference values for the Sun. The lower right plot shows the results when all the atmospheric parameters are set to the solar reference and the number of layers in the atmospheric model is reduced.}
 \label{fig:solar_ew_abundances}
\end{figure}

As shown in Fig.~\ref{fig:solar_ew_abundances}, the equivalent-width codes show a systematic discrepancy for lower surface gravities, and the different pattern shown for metallicities between -2.0 and 0.0~dex is particularly puzzling. The effect of the microturbulence velocity shows a small but increasing systematic when going towards higher values (i.e. it is not just a constant offset between codes). Alpha enhancement does not have an impact on MOOG results, probably because MOOG does not re-compute the electron density but directly uses the values from the model atmosphere (i.e. changing the alpha parameter does not have an effect unless a full new model atmosphere is computed with the changed alpha parameter). WIDTH9 leads to more stable results, despite significantly reducing the number of layers in the model atmosphere, while MOOG median abundances show an offset of 0.10~dex when only 12 layers are used instead of the original 56

\begin{figure}
 \includegraphics[width=\linewidth]{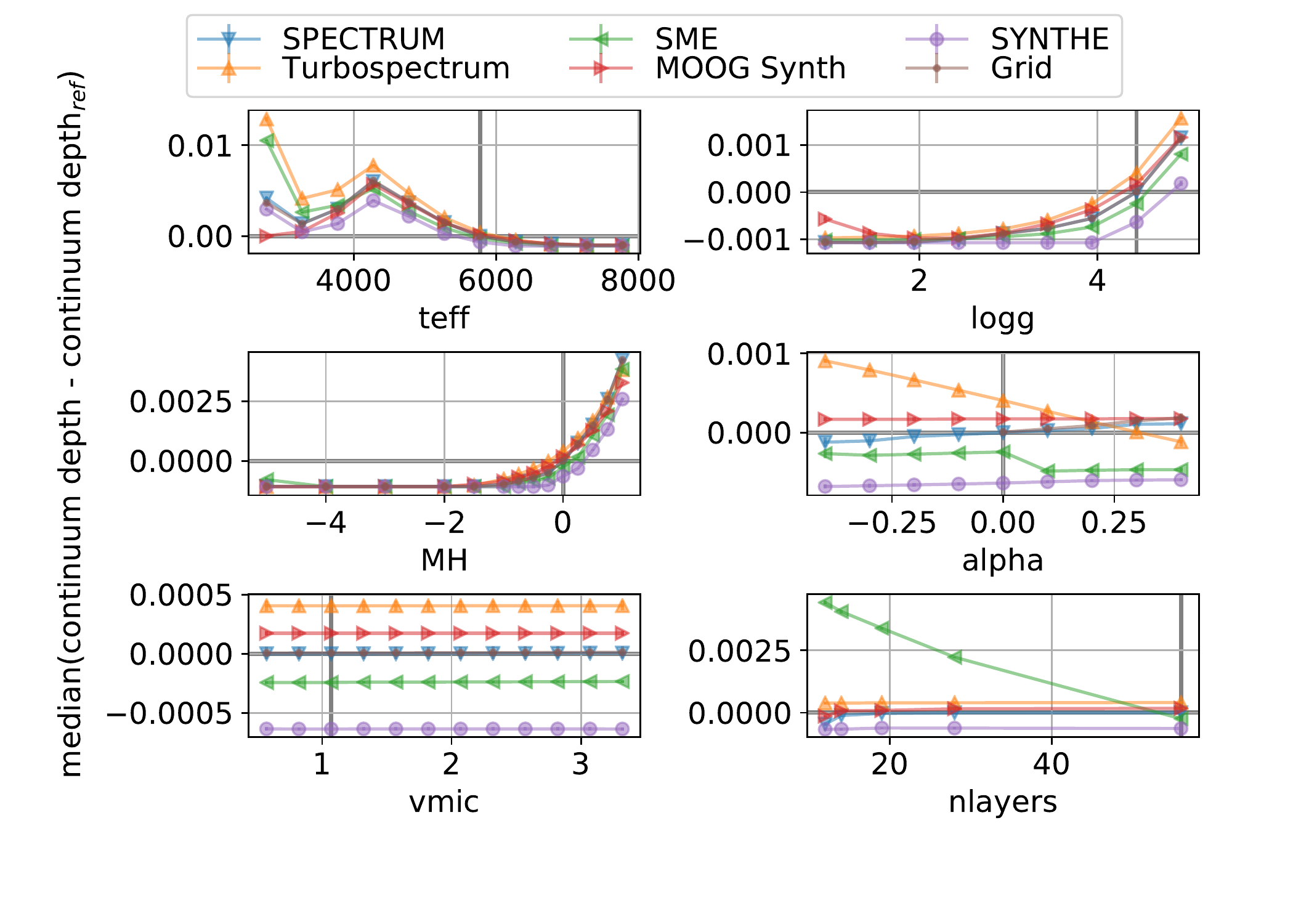}
 \caption{Median difference in continuum depth around 556.45 nm over the reference continuum depth (i.e. SPECTRUM continuum depth for the Sun, zero-point represented by the intersection of the two thick grey lines) when varying selected atmospheric parameters and setting the rest to the reference values for the Sun. The lower right plot shows the results when all the atmospheric parameters are set to the solar reference and the number of layers in the atmospheric model is reduced.}
 \label{fig:solar_synth_flux_continuum}
\end{figure}

\begin{figure}
 \includegraphics[width=\linewidth]{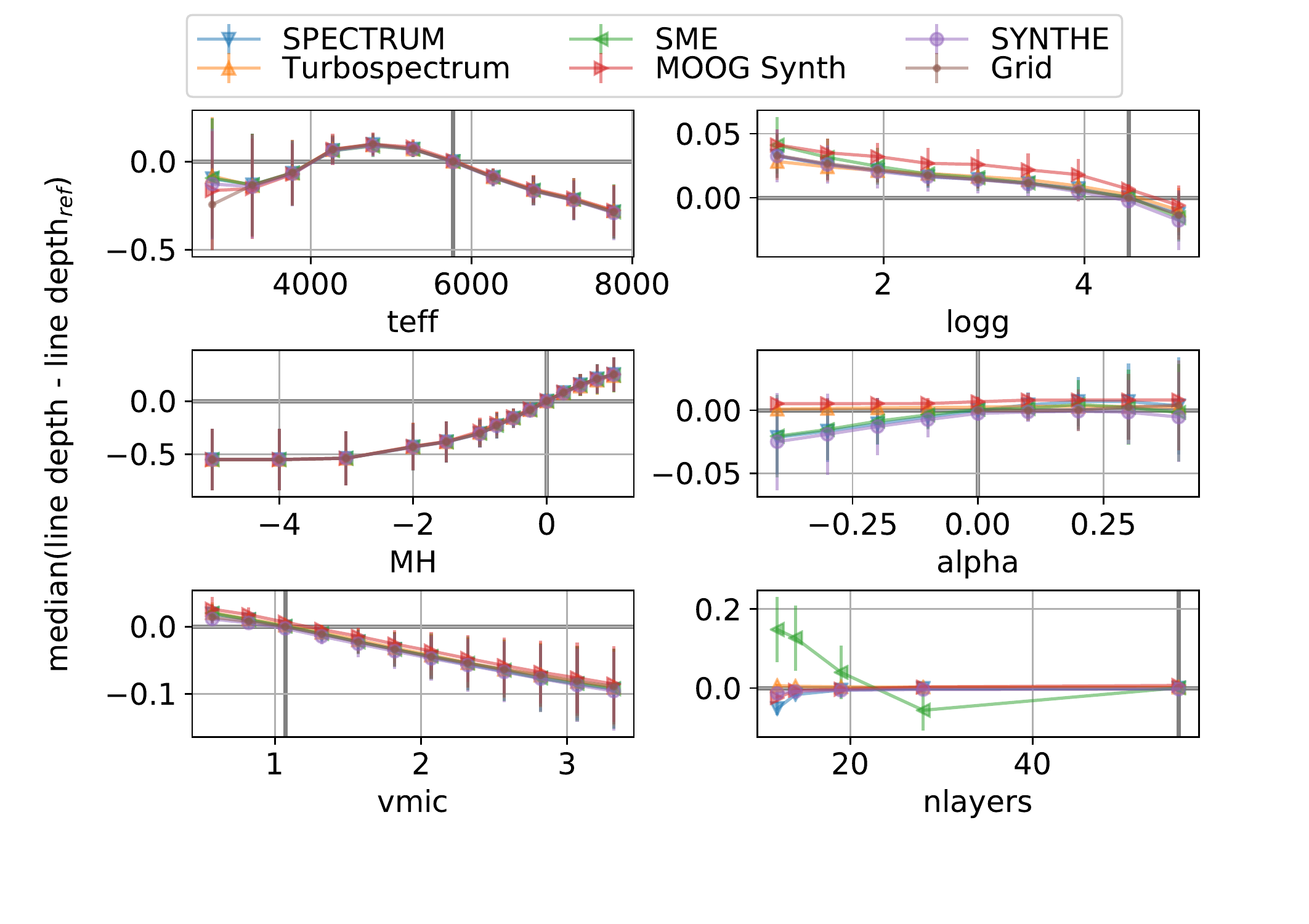}
 \caption{As Fig.~\ref{fig:solar_synth_flux_continuum}, but considering the depth at the line peaks for the common selection.}
 \label{fig:solar_synth_flux_line_cores}
\end{figure}

Regarding synthesis (see Figs~\ref{fig:solar_synth_flux_continuum} and \ref{fig:solar_synth_flux_line_cores}), lower effective temperatures lead to increasing discrepancies in the continuum and absorption line core depths. At the coolest end (2\,800~K), Turbospectrum and SME are relatively close, with the deepest continuum among all the codes. MOOG is at the other end of the range, with a continuum depth similar to the Sun. In terms of surface gravity, all the codes compute a continuum depth that is generally in better agreement for giant stars, while line core depths show the opposite pattern, with the exception of MOOG, which is systematically deeper than the rest of the codes (the opposite is true for line depths). Changes in metallicity show a high level of agreement for absorption line depths, although small discrepancies in continuum depth are present for solar values. Interestingly, variations in alpha abundances lead to very different continuum depth patterns among the codes. The highest agreement is found for absorption line depths with solar alpha abundances, while MOOG and Turbospectrum deviate from the rest for negative values. Microturbulence effects are only evident on the continuum depths, where all the codes keep an extremely small constant difference for all tested values. Finally, when the number of layers in the model atmosphere is reduced, SME is the code that deviates the most from the reference point.

\subsubsection{Model atmosphere}

\begin{figure}
 \includegraphics[width=\columnwidth]{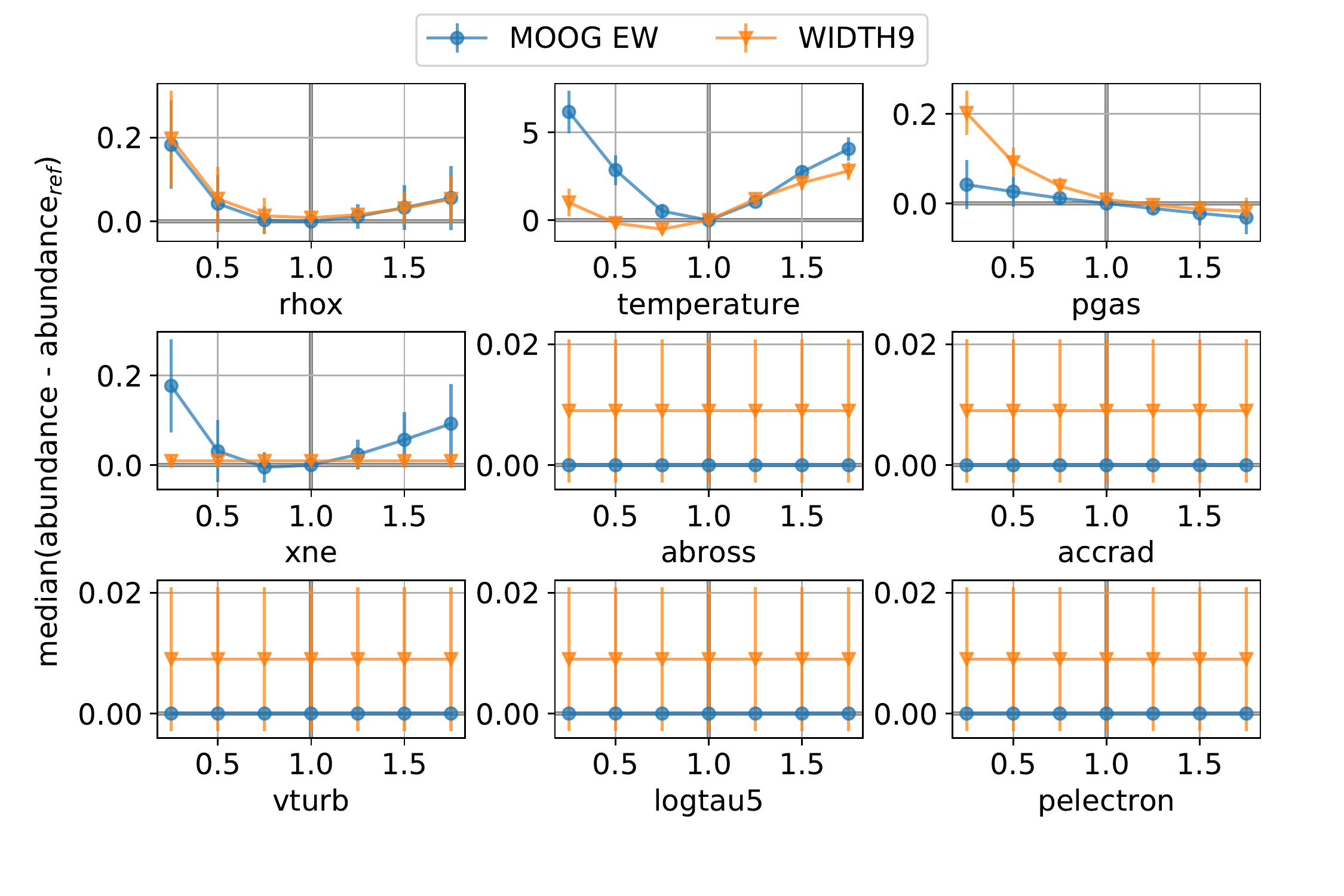}
 \caption{Median and absolute median deviation (error bars) difference in abundance over the reference abundance (i.e. MOOG abundances for the Sun, zero-point represented by the intersection of the two thick grey lines) when varying selected values of the solar model atmosphere.}
 \label{fig:atm_variations_ew_abundances}
\end{figure}

From the model atmosphere, equivalent-width codes are only affected by changes in column mass, temperature, and gas pressure, plus electron density in the case of MOOG, as shown in Fig.~\ref{fig:atm_variations_ew_abundances}. The highest disagreements occur with variation of temperature values, followed by reductions in gas pressure.

\begin{figure}
 \includegraphics[width=\linewidth]{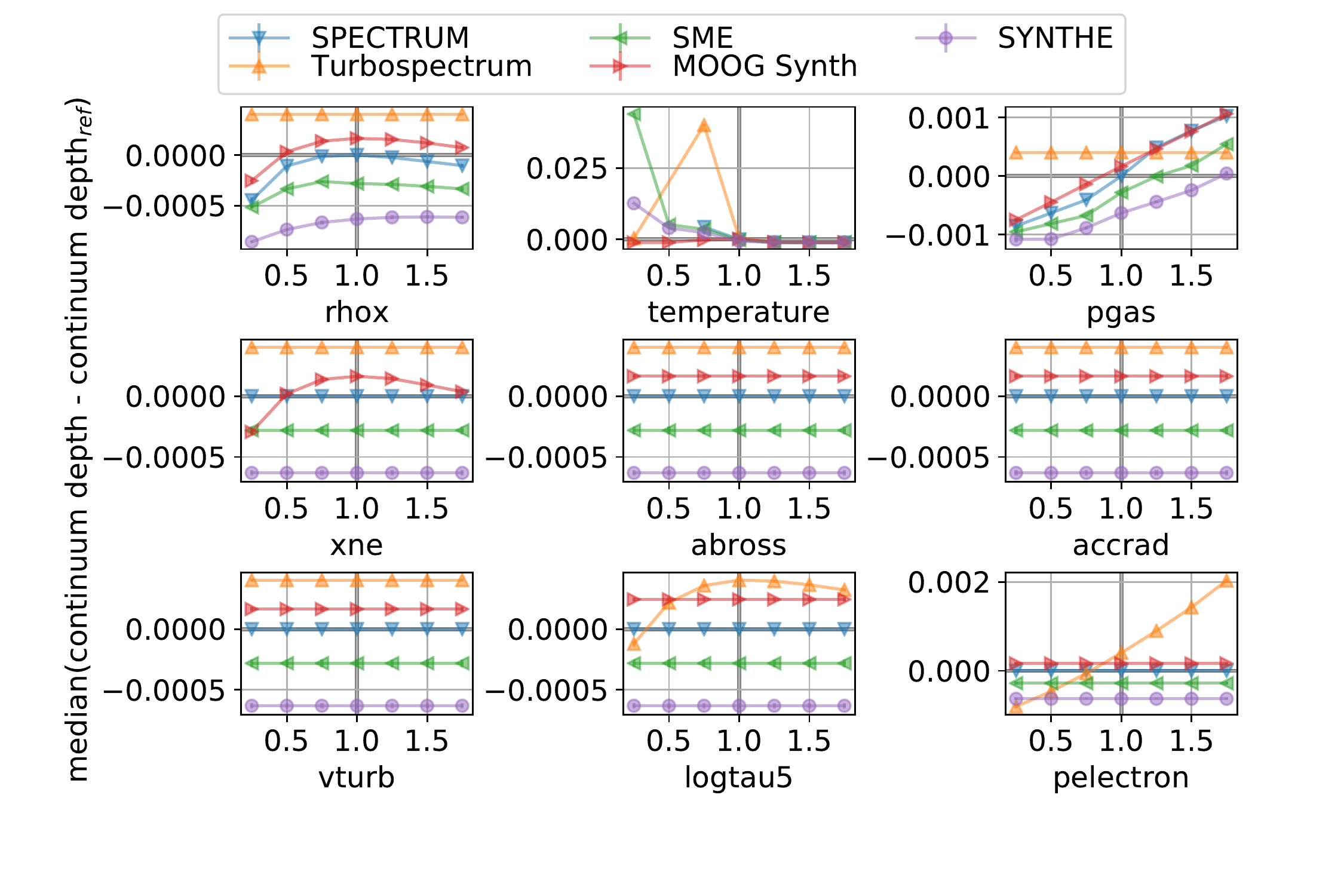}
 \caption{Median and absolute median deviation (error bars) in continuum depth around 556.45 nm over the reference continuum depth (i.e. SPECTRUM continuum depth for the Sun, zero-point represented by the intersection of the two thick grey lines) when varying different values of the solar model atmosphere.}
 \label{fig:atm_variations_synth_flux_continuum}
\end{figure}

\begin{figure}
 \includegraphics[width=\linewidth]{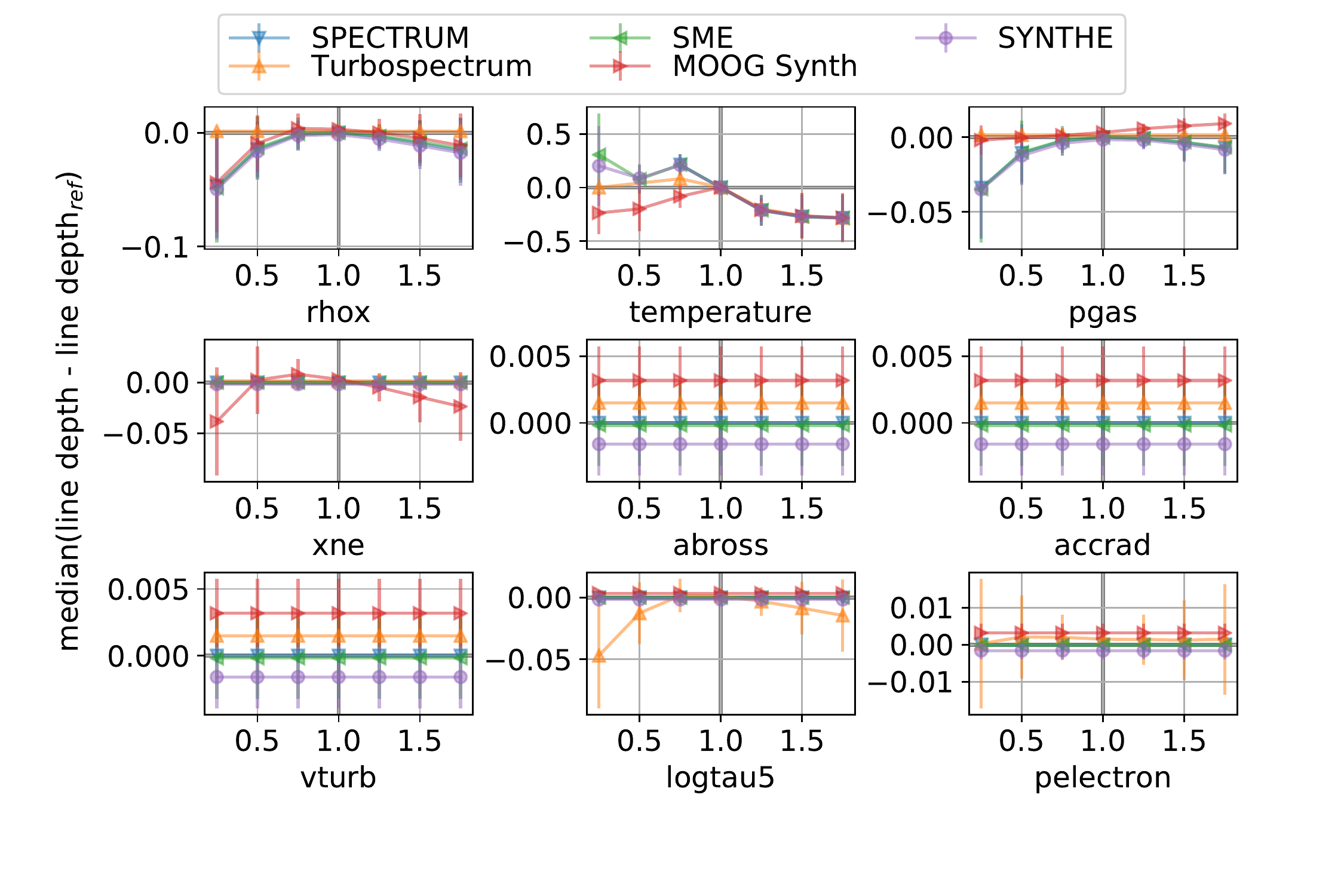}
 \caption{As Fig.~\ref{fig:atm_variations_synth_flux_continuum}, but considering the depth at the line peaks for the common selection.}
 \label{fig:atm_variations_synth_flux_line_cores}
\end{figure}

Regarding the results from synthesis codes, shown in Figs~\ref{fig:atm_variations_synth_flux_continuum} and \ref{fig:atm_variations_synth_flux_line_cores}, no effects are observed when changing the Rosseland mean absorption coefficient, radiation pressure, and microturbulence velocities, plus electron density except in the case of MOOG. Turbospectrum is the only code that uses optical depth instead of column mass, and electron pressure instead of gas pressure. In general terms, all the codes show differences that remain constant, with the exception of the variation of temperature and gas pressure.

\subsubsection{Atomic line list}
\label{s:non_observed_dataset_results_atomic_linelist}

\begin{figure}
 \includegraphics[width=\columnwidth]{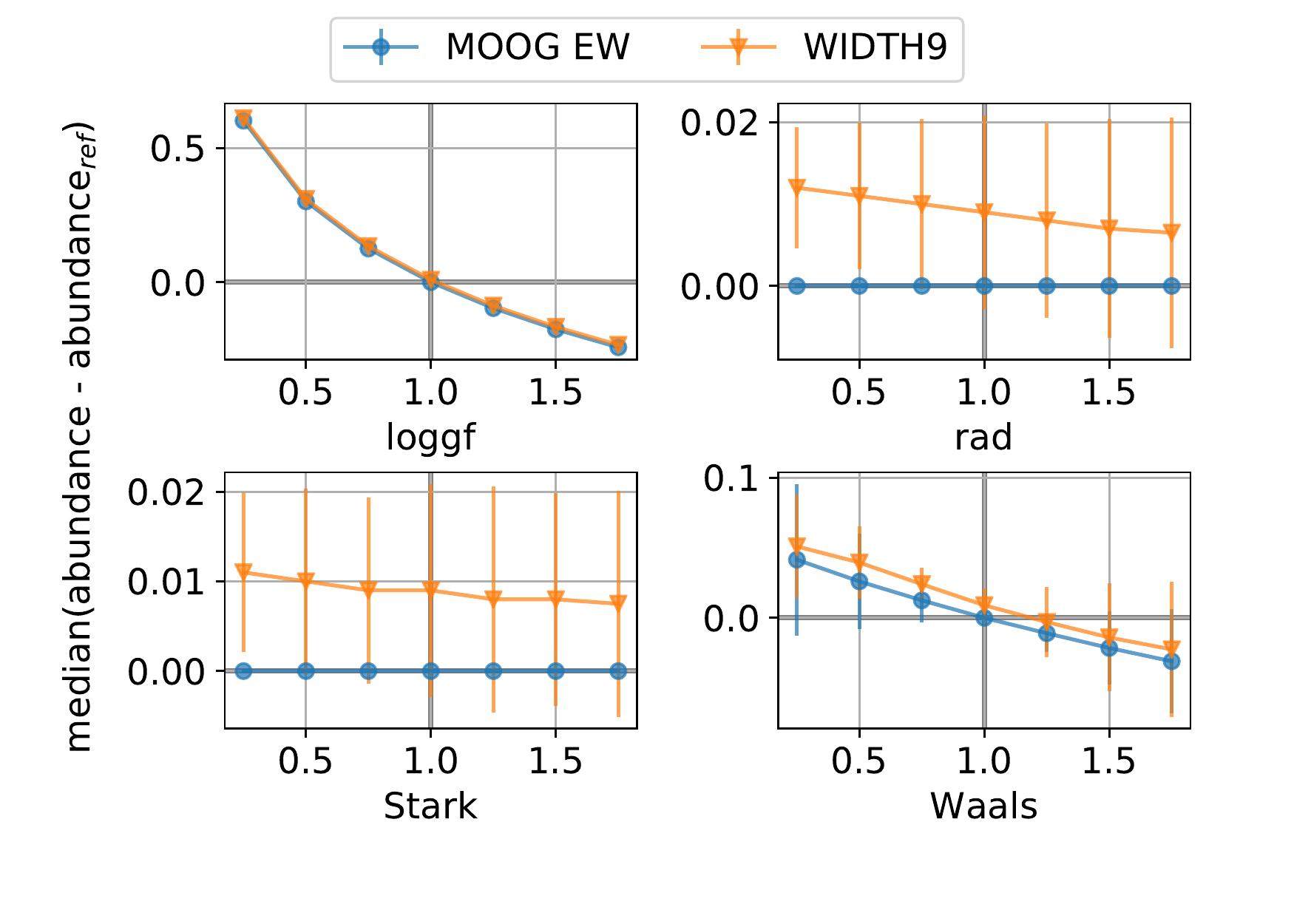}
 \caption{Median and absolute median deviation (error bars) difference in abundance over the reference abundance (i.e. MOOG abundances for the Sun, zero-point represented by the intersection of the two thick grey lines) when varying selected values of the atomic line list.}
 \label{fig:linelist_variations_ew_abundances}
\end{figure}

All the equivalent-width codes (see Fig.~\ref{fig:linelist_variations_ew_abundances}) present very similar results when varying the oscillator strength, but discrepancies appear for the atomic broadening parameters. MOOG does not implement the Stark damping parameter, and hence no effect is expected when changing this value. However, MOOG also does not show any change when varying the radiative damping parameter. Regarding the van der Waals damping parameter, MOOG and WIDTH9 show an almost constant systematic.

\begin{figure}
 \includegraphics[width=\linewidth]{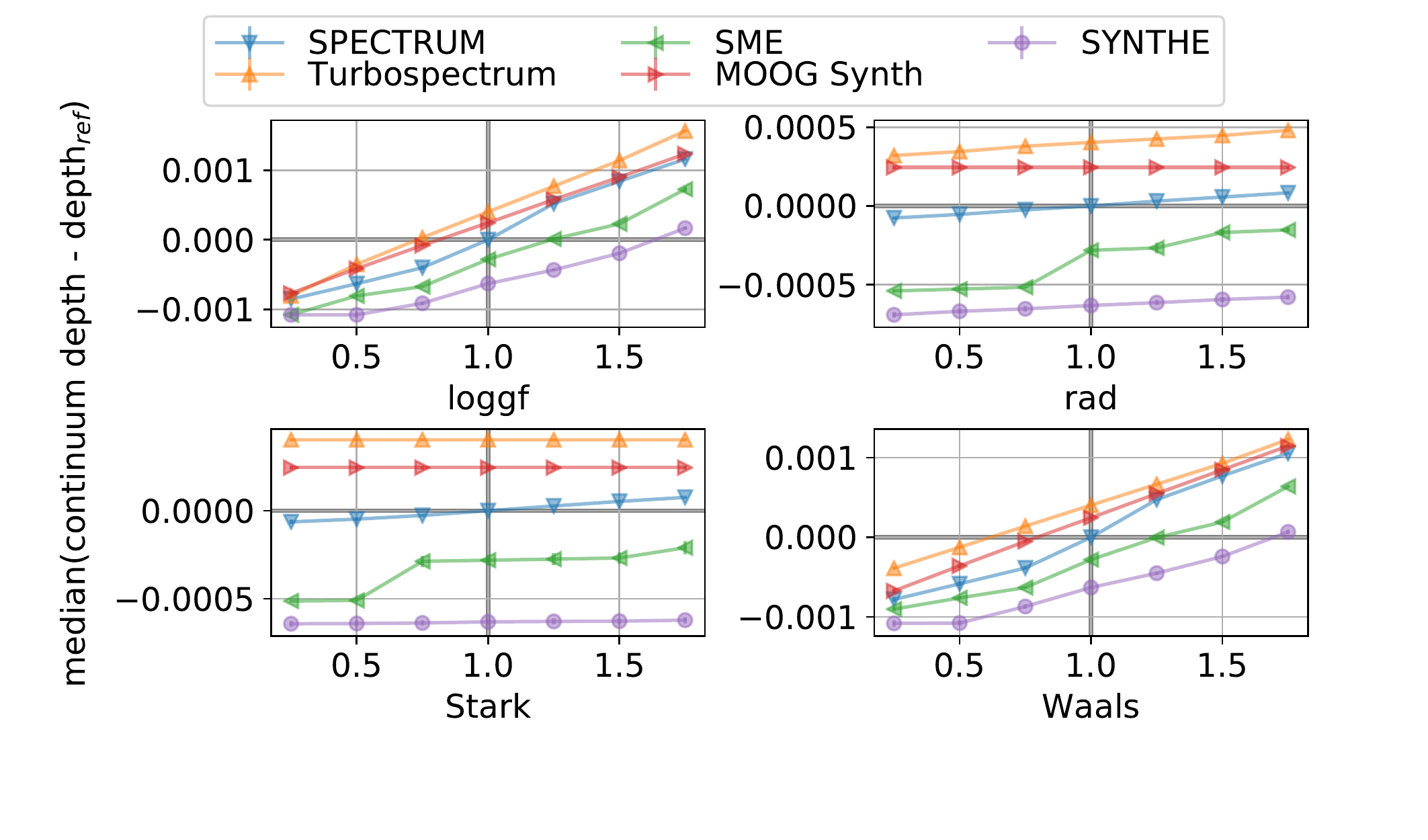}
 \caption{Median and absolute median deviation (error bars) difference in continuum depth around 556.45 nm over the reference continuum depth (i.e., SPECTRUM continuum depth for the Sun, zero point represented by the intersection of the two thick grey lines) when varying different values of the atomic line list.}
 \label{fig:linelist_variations_synth_flux_continuum}
\end{figure}

\begin{figure}
 \includegraphics[width=\linewidth]{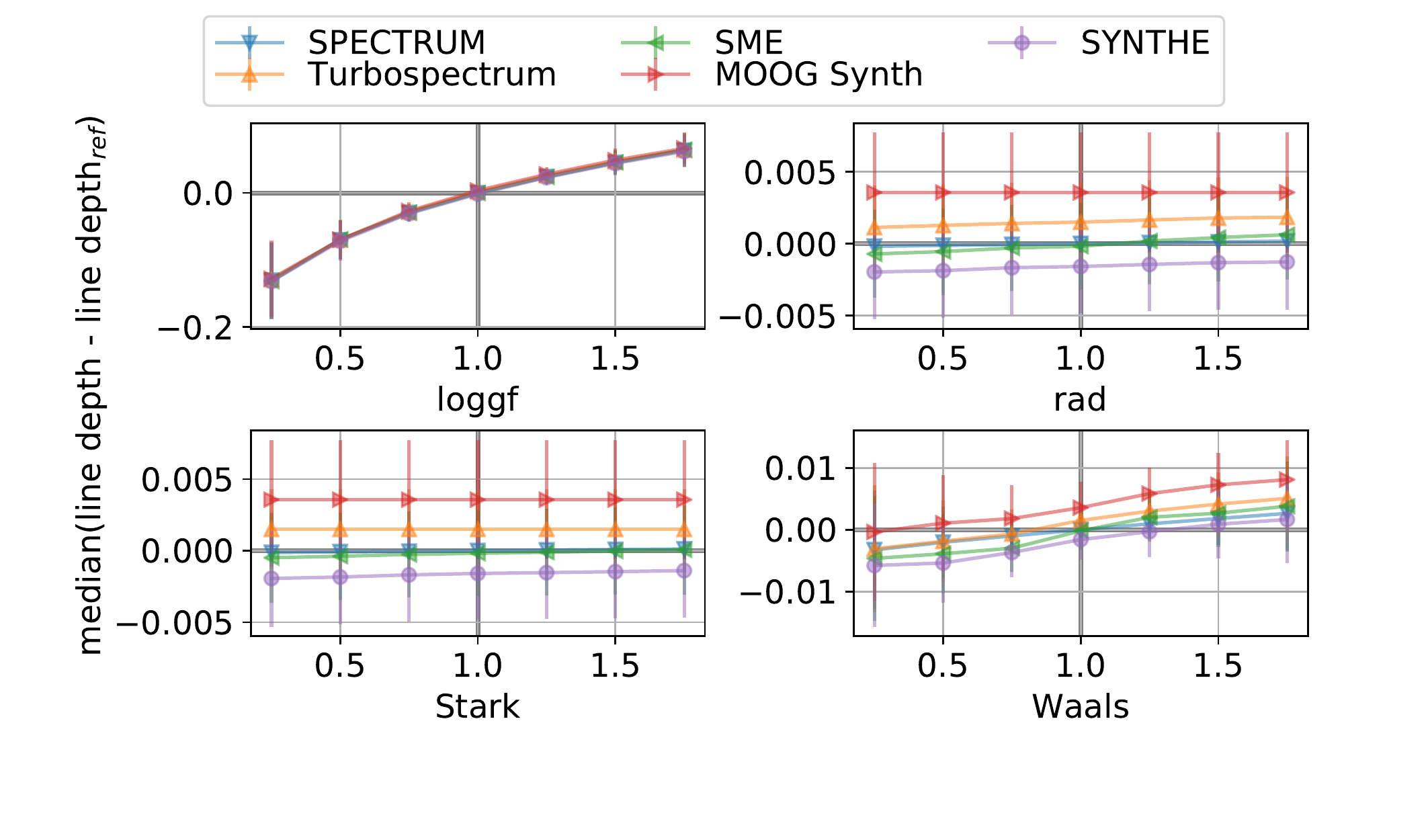}
 \caption{As Fig.~\ref{fig:linelist_variations_synth_flux_continuum}, but considering the depth at the line peaks for the common selection.}
 \label{fig:linelist_variations_synth_flux_line_cores}
\end{figure}

With respect to synthesis codes (see Figs~\ref{fig:linelist_variations_synth_flux_continuum} and \ref{fig:linelist_variations_synth_flux_line_cores}), the effect of varying oscillator strength parameters shows small changing continuum differences between all the codes. MOOG and Turbospectrum do not implement broadening using the Stark damping parameter, and thus no effect is observed when it is modified. MOOG does not show any effect when varying the radiative damping parameter either. The rest of the codes show very small variations in line depths, and changes are more noticeable in continuum depth, where different patterns emerge. These results reveal clear differences in the implementation of line-broadening effects for each code.

\section{Discussion}
\label{s:discussion}

From the 2496 absorption lines studied in the optical range (480-680 nm) of the solar spectrum, only an average of $\sim$300 abundances are within the range $\pm0.10$~dex when using the equivalent-width method, and within $\pm0.05$~dex when using the synthetic spectral-fitting technique. This represents less than 15 per cent of the total number of analysed absorption lines, despite the fact that the Sun is the star of reference that we know best and that the same observed data, normalization process, model atmosphere, solar abundances and atomic data were used in all the codes. When considering a larger margin of $\pm0.25$~dex, the percentage increases to 50 per cent, 65 per cent and 75 per cent for equivalent-width methods, synthesis, and interpolation from a pre-computed grid of spectra (i.e. Grid), respectively. The equivalent-width method suffers from blends, skewing the abundance distribution towards greater abundances (i.e. overestimating abundances from blended lines), and, consequently, presenting a lower number of lines for narrower margins. Grid is the strategy with which a higher number of lines are close to the reference abundance. The reasons for this are not obvious, and there could be a combination of causes, such as the following. (1) Model atmospheres are computed with a certain chemical composition (typically directly related to the solar abundances). Grid scales the metallicity and does not alter the individual chemical pattern, and thus its coherence with the model atmosphere is maximal compared with the synthesis codes. (2) For blended absorption lines, synthesis approaches have a wider range of allowed abundances for a particular element, while Grid has more constraints, given that all the abundances are being increased/decreased at the same time via the metallicity parameter. As shown in Section~\ref{s:line_selection_results}, line-by-line comparisons between all the codes show systematics for certain elements and disagreements that depend on the reduced equivalent width of each line (except for synthesis codes).

The number of good lines in common between equivalent-width codes is high, while the average percentage of matching for synthesis codes is around 85 per cent. As explained in Section~\ref{s:model_atmosphere} and \ref{s:atomic_data}, using the same model atmosphere and atomic line list does not imply that the synthetic spectrum is going to be computed using the same data, because each code may use different values (also discussed in Section~\ref{s:one_variable_at_a_time_results}), there are differences in the implementation of the line-broadening parameters (see Section~\ref{s:non_observed_dataset_results_atomic_linelist}), and in the case of synthesis the effect of these differences increases because blends are also taken into account (e.g. nearby lines can negatively influence the results if the atomic data are inaccurate). In addition, it is remarkable how much the average percentage of matching lines decreases when comparing results between an equivalent-width code and a synthesis one ($\sim$50 per cent). The reason for this can again be related to the intrinsic difference already discussed between each approach.

For the different set-ups evaluated in this work, the codes that lead to more similar results are SME, SYNTHE and SPECTRUM, as shown in Figs~\ref{fig:best_precision} and \ref{fig:best_abundance_precision}, although the first one performs worse when using the wings of H-$\alpha$/$\beta$ and/or the Mg triplet, as described in Section~\ref{sec:impact_on_ap}. For studies in which only the wings of hydrogen lines are used to determine effective temperature, SPECTRUM and SYNTHE produce the results with the smallest systematics, and, for all the codes, it is preferable to use H-$\alpha$ over H-$\beta$ because the latter is not well reproduced. In general terms, the most accurate results are obtained with SPECTRUM closely followed by SYNTHE, as seen in Fig.~\ref{fig:best_accuracy}, but the former is faster in terms of computation time. Furthermore, it is worth remembering that all the radiative transfer codes considered in this work use 1D models and assume LTE: overviews of the present and future prospects of 3D non-LTE models can be found \cite{2016A&ARv..24....9B, 2018A&ARv..26....6N, 2018arXiv181108041J}.

Interpolating from a grid of pre-computed synthetic spectra (i.e. Grid) also provides a remarkably good accuracy, despite being slightly lower than in SPECTRUM and SYNTHE. It is worth considering, given that for a single spectrum it can take an average of 30 min to derive atmospheric parameters with synthesis and only 5 min with interpolation. Grid requires having an existing pre-computed grid, which is computational work that only has to be done once, and a grid of $\sim$30\,000 spectra covering the region from 480 to 680 nm can be completed in less than a day on a modern machine with 32 CPUs.

When comparing all the codes in an even more controlled experiment (see Sections~\ref{s:non_observed_dataset_results} and \ref{s:one_variable_at_a_time_results}), differences in abundance determinations (equivalent-width methods) or synthetic flux computation emerge when any of the input parameters (atmospheric parameters, model atmosphere or atomic data) are varied. This is because of implementation differences such as continuum calculations and line-broadening effects, and it shows how varying discrepancies appear when using the same method (i.e. equivalent-width or synthetic spectral-fitting) but different codes.

In addition to individual differences between codes, we can also note that there are some clear systematics between equivalent-width methods and the synthetic spectral-fitting technique. Limiting the analysed data set by discarding the cooler and metal-poorer stars, which are very challenging types of stars for the equivalent-width method, does not fully erase the systematics, as shown in Figs~\ref{fig:best_precision} and \ref{fig:best_precision_ew_range_limited}. MOOG EW and WIDTH9 lead to extremely similar results when they use the common line selection, but they deviate when using their own line selection. It is particularly interesting that depending on whether the full or the limited data set of Gaia Benchmark Stars is considered, it is WIDTH9 or MOOG EW that obtains results more similar to synthesis codes. This shows that the systematics between codes and methods is not independent from the stellar type, and discourages the practice of blindly combining results from multiple methods and/or codes.

In general terms, the equivalent-width method can only compete with the synthetic spectral-fitting technique when the spectral range is limited and the cooler and metal-poorer stars (higher number of blends) are not considered, with MOOG EW leading to better results than WIDTH9. The analysis of one spectrum using equivalent widths can take less than a minute, and this constitutes a strong argument for this approach if the target stars are expected to be in the optimal range of parameters.

It is worth mentioning that the high level of precision between codes shown in this work is higher than the level of agreement that one could find between studies from different authors. The tests presented here used a line selection that was executed in a very homogeneous way for all the codes, and it can be expected that heterogeneous line selections lead to higher differences between codes. Moreover, the strictly line-by-line differential analysis for the determination of individual chemical elements had a significant effect, increasing the precision among codes and methods. This work also shows how the selection of absorption lines and other spectral regions plays a very important role in terms of accuracy, and this selection should be done using the same pipeline and criteria that are going to be used for the target spectra. Re-using line selection done by other authors using different model atmospheres, atomic data, radiative transfer codes and normalization processes will not guarantee the best results. These results also question the practice of blindly using the full spectrum to derive atmospheric parameters, given that with the existing models it is not possible to completely reproduce even the Sun, the star that we know the best. 

The dependence of the microturbulence velocity on the spectroscopic method seen in this work clearly suggests that re-using microturbulence results from different methods should not be a recommended practice. In the case of synthesis, the same recommendation could be extended to the macroturbulence velocity, for which different recipes exist (e.g. Gaussian and radial-tangential broadening), and it tends to be degenerated together with the projected rotational velocities and the resolution. Moreover, it is sometimes common practice to fix some of the atmospheric parameters to values obtained by other independent (occasionally more accurate) means, but, for instance, forcing the model to use a certain effective temperature can lead to biases in the rest of the free parameters, as described in Section~\ref{s:abundances}. There are advantages in fixing parameters to values derived by more accurate methods, but the effects of this practice should be carefully assessed and controlled.

Given the results presented in this work, the situation can be illustrated with the following analogy. We consider a given selection of model atmospheres, solar abundances, atomic data, radiative transfer codes, normalization procedures, general data treatments and spectral wavelength ranges as a ruler in centimetres (for instance) with a precision of 0.1 cm, while a different selection would correspond to a ruler in inches with a precision of 0.1 inches (i.e. 0.254 cm). If we wanted to compare people's heights, we would obviously not choose to measure them using rulers in different units and with different precisions: homogeneity is necessary unless we know how to transform one unit into another very accurately and the precision of each measurement is similar. We would not also gain much benefit from blindly combining measurements for each person with several different rulers of different units, especially if some of these rulers work worse for a particular type of person (e.g. because of their body structure): who we measure, what/how we measure and how we combine the results may introduce biases that are not easy to account for. It is generally preferable to have a controlled and homogeneous measurement process.

The combination of results from different sources would be valuable if it were shown that one method is particularly good for a type of star for which the other methods are less reliable (e.g. as claimed in \citealt{2014A&A...570A.122S}). Then the difficult task of combining measurements obtained from different rulers may make sense, although the process should correctly account for the strengths of each approach, each method should be assessed using benchmark objects, the strategy should be described in detail, and the original non-combined individual values (spectrum per spectrum and line by line if it applies) should also be included in the publication. For instance, the current work does not provide any reason to support mixing results from the equivalent-width method and the synthetic spectral-fitting technique, because, apart from the identified systematics, the former is less reliable for stars with many blends while the latter leads to more robust results for a wider range of parameters. Thus, combining results would not lead to a better overall outcome but will introduce biases and complicate comparisons.

\section{Conclusions}
\label{s:conclusions}

In this work, I expanded the capabilities of iSpec to derive atmospheric parameters and abundances using several new radiative transfer codes. The user can choose between MOOG and WIDTH9 for equivalent-width methods, and between SPECTRUM, Turbospectrum, SME, MOOG and SYNTHE for the synthetic spectral-fitting technique. In addition, I included the possibility of interpolating from a grid of pre-computed/observed spectra (i.e. Grid), which reduces the analysis time for the determination of atmospheric parameters of a single spectrum down to 5 min or less with a modern computer.

By designing a completely automatic spectroscopic pipeline and analysing the Gaia Benchmark Stars, I executed an experiment that explores the key pitfalls of modern spectroscopy by comparing the spectroscopic results (atmospheric parameters and individual chemical abundances) obtained when using several radiative transfer codes and different spectroscopic techniques with multiple set-ups.

The results showed that the synthetic spectral technique has a higher accuracy when considering the full range of atmospheric parameters that the Gaia Benchmark Stars cover, while the equivalent-width method is competitive only when the data set is limited by discarding cooler and metal-poorer stars. When the right set-up is selected, interpolating from a grid of pre-computed spectra has an accuracy almost comparable to the results obtained when using synthesis with an interpolated model atmosphere. To obtain the best results, an appropriate line selection should be executed, using the same model atmosphere, atomic data, radiative transfer code and normalization process that will be applied to the target spectra. In the case of synthesis, including the wings of H-$\alpha$/$\beta$ and the Mg triplet can help to increase the accuracy when using Grid, SPECTRUM and SYNTHE codes.

Despite using a homogeneous process to select the best common absorption lines, there are clear systematics between the equivalent-width method and the synthetic spectral-fitting technique. In addition, for the latter, there are also differences when interpolating from a pre-computed grid of spectra or synthesizing with an interpolated atmospheric model. This demonstrates that blindly combining atmospheric parameters and chemical abundances measured using heterogeneous set-ups and methods is not a recommended procedure, especially when a high precision is key for the scientific goals of the study.

This study has uncovered code-to-code differences that can affect the scientific interpretation of spectroscopic analysis; this suggests that it would be a good practice to assess if the conclusions still hold when deriving atmospheric parameters and/or abundances with different codes.

There are plenty of models and tools freely accessible today to analyse the growing number of high-quality spectra available in the public archives or to execute our own observations and studies. This ease of access represents an unprecedented opportunity in the history of stellar spectroscopy to investigate the wonders of the stars, but it carries with it the need to consider carefully the caveats of modern spectroscopy exposed here.

\section*{Acknowledgements}

This research has made use of NASA's Astrophysics Data System. This work would not have been possible without the invaluable contribution from all the authors that developed the considered radiative transfer codes: Richard O. Gray, Robert L. Kurucz, Luca Sbordone, Nikolai Piskunov, Jeff A. Valenti, Bertrand Plez, Chris Sneden plus any other major contributor that I may have forgotten. Likewise, Bengt Gustafsson and contributors should be acknowledged for their major contribution to stellar astrophysics with their MARCS model atmosphere. Additionally, I am very grateful to Thomas Nordlander, Ulrike Heiter, Paula Jofr\'e, Thomas Masseron, Laia Casamiquela, Hugo M. Tabernero, Andrew R. Casey, Jorge Mel\'endez, and Ivan Ram\'irez for their help integrating and testing all these radiative transfer codes in iSpec. I thank the anonymous referee, whose great ideas and suggestions significantly improved the quality of this work.




\bibliographystyle{mnras}
\bibliography{stellar_spectroscopy} 



\appendix

\section{Model atmosphere and atomic data}


\begin{table*}
    \begin{center}
\resizebox{\textwidth}{!}{%
        \begin{tabular}{l|c|c|c|c|c}

    &   SPECTRUM    &   Turbospectrum   &   SME &   MOOG    &   WIDTH9/SYNTHE   \\
\hline                                          
Column mass above each point [g cm${^-2}$]  &   \checkmark  &       &   \checkmark  &   \checkmark  &   \checkmark  \\
\hline                                          
Temperature [K] &   \checkmark  &   \checkmark  &   \checkmark  &   \checkmark  &   \checkmark  \\
\hline                                          
Gas pressure [dyn cm$^{-2}$]    &   \checkmark  &   \checkmark  &   \checkmark  &   \checkmark  &   \checkmark  \\
\hline                                          
Electron density [cm$^{-3}$]    &   \checkmark  &       &   \checkmark  &   \checkmark  &   \checkmark  \\
\hline                                          
Rosseland mean absorption coefficient [cm$^{2}$ g$^{-1}$]   &   \checkmark  &       &       &       &   \checkmark  \\
\hline                                          
Radiation pressure [dyn cm$^{-2}$]  &   \checkmark  &       &       &       &   \checkmark  \\
\hline                                          
Microturbulence velocity [m s$^{-1}$]   &   \checkmark  &       &       &       &   \checkmark  \\
\hline                                          
Optical depth [$\log \tau$ at 5000~{\AA}] &       &   \checkmark  &       &       &       \\
\hline                                          
Depth [cm]  &       &   \checkmark  &   \checkmark  &       &       \\
\hline                                          
Electron pressure [cm$^{2}$ g$^{-1}$]   &       &   \checkmark  &       &       &       \\
\hline                                          
Turbulence pressure [dyn cm$^{-2}$] &       &   \checkmark  &       &       &       \\

        \end{tabular}
}
    \end{center}
    \caption{Model atmosphere fields required as input values for each radiative transfer code. This list respects the expected input order as required by the used version of iSpec.}
    \label{tab:input_values_model_atmospheres}
\end{table*}

\begin{table*}
    \begin{center}
\resizebox{\textwidth}{!}{%
        \begin{tabular}{l|c|c|c|c|c}

    &   SPECTRUM    &   Turbospectrum   &   SME &   MOOG    &   WIDTH9/SYNTHE   \\
\hline                                          
Element name    &   \checkmark  &   \checkmark  &   \checkmark  &   \checkmark  &       \\
\hline                                          
Wavelength [\AA]    &   \checkmark  &   \checkmark  &   \checkmark  &   \checkmark  &       \\
\hline                                          
Wavelength [nm] &       &       &       &       &   \checkmark  \\
\hline                                          
loggf   &   \checkmark  &   \checkmark  &   \checkmark  &   \checkmark  &   \checkmark  \\
\hline                                          
Lower state [eV]    &       &   \checkmark  &   \checkmark  &   \checkmark  &       \\
\hline                                          
Lower state [cm$^{-1}$] &   \checkmark  &       &       &       &   \checkmark  \\
\hline                                          
Lower j or total angular momentum quantum number    &       &       &       &       &   \checkmark  \\
\hline                                          
Upper state [eV]    &       &       &       &       &       \\
\hline                                          
Upper state [cm$^{-1}$] &       &       &       &       &   \checkmark  \\
\hline                                          
Upper j or total angular momentum quantum number    &       &       &       &       &   \checkmark  \\
\hline                                          
Upper g or statistical weight   &       &   \checkmark  &       &       &       \\
\hline                                          
Lower Land\'e g-factor  &       &       &       &       &   \checkmark  \\
\hline                                          
Upper Land\'e g-factor  &       &       &       &       &   \checkmark  \\
\hline                                          
Transition type:   &   \checkmark  &       &       &       &   \checkmark  \\
\hline                                          
10 to the power of the radiative damping parameter  &       &   \checkmark  &       &   \checkmark  &       \\
\hline                                          
Radiative damping parameter &   \checkmark  &       &   \checkmark  &       &   \checkmark  \\
\hline                                          
Stark damping parameter &   \checkmark  &       &   \checkmark  &       &   \checkmark  \\
\hline                                          
van der Waals damping parameter ($\sigma$.$\alpha$ format for AO theory)  &   \checkmark  &       &   \checkmark  &       &    \\
\hline                                          
van der Waals damping parameter (classic)  &       &       &       &   \checkmark  &   \checkmark  \\
\hline                                          
Fudge factor (common for the same atomic number and ion) \\
or van der Waals damping parameter (classic or AO theory) if present &       &   \checkmark  &       &       &       \\
\hline                                          
Fudge factor (always set to 1.0)    &   \checkmark  &       &       &       &       \\
\hline                                          
Lower orbital type  &       &   \checkmark  &       &       &       \\
\hline                                          
Upper orbital type  &       &   \checkmark  &       &       &       \\
\hline                                          
Line due to molecular absorption (True/False)  &   \checkmark  &   \checkmark  &   \checkmark  &   \checkmark  &   \checkmark  \\
\hline                                          
Isotope in spectrum format  &   \checkmark  &       &       &       &   \checkmark  \\
\hline                                          
Ion (e.g., 0 for neutral lines, 1 for ionized) &       &   \checkmark  &   \checkmark  &       &       \\
\hline                                          
Species code: "atomic number" + "." + "ion state - 1" &   \checkmark  &       &       &   \checkmark  &       \\
\hline                                          
Species code: "atomic number" + "." + "isotope code"  &       &   \checkmark  &       &       &       \\
\hline                                          
Species code: "atomic number" + ".0" + "ion state - 1"    &       &       &       &       &   \checkmark  \\

        \end{tabular}
}
    \end{center}
    \caption{Atomic line list fields required as input values for each radiative transfer code. Transition type indicates whether the $\alpha$ and $\sigma$ parameters used in the Anstee and O'Mara broadening theory are provided \citep[][coded as AO type;]{1991MNRAS.253..549A, 1995MNRAS.276..859A} or the classic van der Waals broadening should be used (GA type) as described in SPECTRUM documentation. Fudge factors are arbitrary non-physical values used to increase the line broadening to compensate for unknowns. This list respects the expected input order as required by the used version of iSpec.}
    \label{tab:input_values_atomic_linelist}
\end{table*}


\bsp	
\label{lastpage}
\end{document}